%% file: paper_2026.tex
\documentclass[]{fairmeta}

\input{imports}

\usepackage{tikz}
\usepackage{multirow}
\usepackage{booktabs}
\usepackage{adjustbox}
\usepackage{textcomp}
\usetikzlibrary{shapes.geometric, arrows}

\newcommand{\name}[0]{FastCSP}
\newcommand{\fullname}[0]{FastCSP}

\usepackage{listings}
\usepackage{changepage}

\lstdefinelanguage{ini}{
    basicstyle=\ttfamily\small,
    morecomment=[l]{;},
    morecomment=[l]{\#},
    morestring=[b]",
    alsoletter={=},
    escapeinside={(*}{*)},
}

\tikzstyle{startstop} = [rectangle, rounded corners, 
minimum width=3cm, 
minimum height=1cm,
text centered, 
draw=black, 
fill=red!30]

\tikzstyle{io} = [trapezium, 
trapezium stretches=true,
trapezium left angle=70, 
trapezium right angle=110, 
minimum width=3cm, 
minimum height=1cm, text centered, 
draw=black, fill=blue!30]

\tikzstyle{process} = [rectangle, 
minimum width=3cm, 
minimum height=1cm, 
text centered, 
text width=3cm, 
draw=black, 
fill=orange!30]

\tikzstyle{decision} = [diamond, 
minimum width=3cm, 
minimum height=1cm, 
text centered, 
draw=black, 
fill=green!30]
\tikzstyle{arrow} = [thick,->,>=stealth]

\title{FastCSP: Accelerated Molecular Crystal Structure Prediction with Universal Model for Atoms}

\author[1,*, \dagger]{Vahe Gharakhanyan}
\author[2,*]{Yi Yang}
\author[1]{Luis Barroso-Luque}
\author[1]{Daniel S. Levine}
\author[1]{Sushree Jagriti Sahoo}
\author[1]{Brandon M. Wood}
\author[1]{Kyle Michel}
\author[1]{Muhammed Shuaibi}
\author[3]{Gregory J. O. Beran}
\author[4]{Viachaslau Bernat}
\author[1]{Misko Dzamba}
\author[1]{Xiang Fu}
\author[1]{Meng Gao}
\author[4]{Xingyu Liu}
\author[1]{Benjamin K. Miller}
\author[4]{Keian Noori}
\author[4]{Lafe J. Purvis}
\author[4]{Tingling Rao}
\author[1]{Ammar Rizvi}
\author[1]{Matt Uyttendaele}
\author[4]{Andrew J. Ouderkirk}
\author[4]{Chiara Daraio}
\author[1]{C. Lawrence Zitnick}
\author[4, \dagger]{Arman Boromand}
\author[2,5,6, \dagger]{Noa Marom}
\author[1, \dagger]{Zachary W. Ulissi}
\author[1]{Anuroop Sriram}

\affiliation[1]{Fundamental AI Research at Meta}
\affiliation[2]{Department of Materials Science and Engineering, Carnegie Mellon University, Pittsburgh, PA, USA}
\affiliation[3]{Department of Chemistry, University of California Riverside, Riverside, CA, USA}
\affiliation[4]{Reality Labs Research at Meta}
\affiliation[5]{Department of Physics, Carnegie Mellon University, Pittsburgh, PA, USA}
\affiliation[6]{Department of Chemistry, Carnegie Mellon University, Pittsburgh, PA, USA}

\contribution[*]{Equal contribution}
\contribution[\dagger]{Corresponding authors}

\abstract{Molecular crystal structure prediction (CSP) is essential for applications in pharmaceuticals and organic electronics. However, CSP remains challenging and computationally intensive due to the need to explore a large search space with sub-kJ/mol accuracy to distinguish between competing polymorphs. While dispersion-inclusive density functional theory (DFT) offers the necessary precision, its computational cost is impractical for a large number of putative structures. Here, we present FastCSP, an open-source, end-to-end CSP workflow driven entirely by a single pretrained universal machine learning interatomic potential (MLIP), the Universal Model for Atoms (UMA), without any system-specific fine-tuning or DFT calculations. FastCSP integrates conformer generation, random structure generation via Genarris 3, geometry optimization, free energy evaluation, and conformer energy corrections, all powered by UMA. Benchmarked on 28 semi-rigid and 10 flexible molecules spanning 74 experimental polymorphs, FastCSP reliably recovers all known structures, ranking them within 9 kJ/mol of the global minimum. UMA reproduces dispersion-inclusive DFT results with high fidelity across chemically diverse compounds. Conformer corrections are particularly beneficial for flexible compounds with conformational polymorphism, such as ROY. UMA's accuracy, transferability, and computational cost thus eliminate the need for classical force fields in early-stage screening and DFT-based re-ranking in CSP workflows. The open-source release of the entire FastCSP workflow lowers the barrier to accessing CSP, enabling both pharmaceutical-grade and high-throughput polymorph screening within practical computational reach.}

\metadata[Code]{\url{https://github.com/facebookresearch/fairchem/tree/main/src/fairchem/applications/fastcsp}}
\correspondence{V.G. (\email{vaheg@meta.com}), Z.W.U. (\email{zulissi@meta.com}), N.M. (\email{nmarom@andrew.cmu.edu}), A.B. (\email{aboromand@meta.com})}

\begin{document}

\maketitle

\section{Introduction}
Crystal structure prediction (CSP) of molecular crystals is a fundamental challenge in materials science~\cite{beran2023frontiers}, with profound implications across several industries including pharmaceuticals~\cite{price2016can, yang2020prediction}, organic electronics~\cite{campbell2017predicted, bhat2023computational, tom2023inverse}, 
energetic materials~\cite{bier2021crystal, arnold2023crystal, o2023ab}, pigments~\cite{panina2007crystal}, and agrochemicals~\cite{tremayne2004characterization, yang2017ddt}. Many molecules can form different crystal structures, known as polymorphs, depending on crystallization conditions~\cite{lee2011crystal, cruz2015facts, neumann2015combined, bernstein2020polymorphism}.
Different polymorphs of the same compound can exhibit widely varying physical properties~\cite{yu2003scientific, coropceanu2007charge, censi2015polymorph, chung2016polymorphism, liu2018polymorphism}.
Therefore, it is desirable to explore the configuration space of structure and properties to optimize the performance of products based on molecular crystals. Computer simulations can greatly accelerate this process compared to experimental polymorph screening and steer crystal growth efforts towards potentially stable polymorphs with desirable properties~\cite{neumann2008major, rice2018computational, yang2018large, hoja2019reliable, schmidt2021computational, tom2023inverse, faruque2024high, johal2025exploring}.

Different polymorphs are often separated by only a few kJ/mol per molecule in energy~\cite{nyman2015static, cruz2015facts}. This necessitates very accurate evaluation of the thermodynamic stability of putative structures in order to determine the most likely form(s) of a given compound. CSP is thus challenging because it requires exploring the vast configuration space of possible crystal structures with very high accuracy. This challenge is represented by the CSP blind tests organized by the Cambridge Crystallographic Data Centre (CCDC), which have been conducted periodically for over two decades to assess the progress of the field~\cite{lommerse2000test, motherwell2002crystal, day2005third, day2009significant, bardwell2011towards, reilly2016report, hunnisett2024seventhgen, hunnisett2024seventhrank}. In these blind tests, participants aim to predict the crystal structure of target compounds with a varying degree of complexity starting from a chemical ``stick diagram''. Over the years, supercomputing capabilities and CSP methods have advanced hand in hand. In the early blind tests only classical force fields were used~\cite{lommerse2000test, motherwell2002crystal}, whereas in more recent blind tests the use of dispersion-inclusive density functional theory (DFT) for final stability ranking has become an established best practice~\cite{bardwell2011towards, reilly2016report}. Unfortunately, the high computational cost of dispersion-inclusive DFT methods limits the scale at which they can be applied. In practice, a hierarchical CSP approach is often used, wherein classical force fields and/or semi-empirical methods filter structures in the initial steps of a multi-stage workflow before final DFT refinement and ranking. 

DFT lattice energies are sometimes supplemented with additional corrections to improve the reliability of the final ranking in CSP workflows. Under realistic conditions of finite temperature and pressure, the stability of crystals is governed by free energy rather than 0 K lattice energy. To account for this, vibrational free energy contributions may be added via the harmonic approximation  (HA) or the quasi-harmonic approximation (QHA), which also considers the effect of thermal expansion. These terms can change the predicted stability order, especially for large flexible molecules~\cite{heitPredictingFinitetemperatureProperties2015, nyman2016modelling, heitHowImportantThermal2016, hoja2019reliable, o2022performance}. In the ranking stage of the seventh CSP blind test, several groups performed such corrections~\cite{hunnisett2024seventhrank}. However, it was found that free energy corrections did not necessarily improve the ranking of the experimental forms.

A second correction addresses the balance between the stabilization contributed by intermolecular interactions and the destabilization caused by changes in molecular conformation. Semi-local approximations to the exchange-correlation functional of DFT, such as the generalized gradient approximation (GGA), suffer from the delocalization error~\cite{bryentonDelocalizationErrorGreatest2023}, which may distort the relative energies of different intramolecular conformations and bias the ranking of conformational polymorphs~\cite{greenwell2020inaccurate, rana2023correcting,midgley2026dft}. To address this problem, the intramolecular part of the lattice energy can be replaced with a higher-level calculation, such as a correlated wave-function method or a hybrid exchange-correlation functional, while retaining the (dispersion-inclusive) semi-local description of the intermolecular interactions~\cite{beran2023frontiers, chattopadhyay2025lattice}. In the seventh blind test Group 2 employed this strategy, which was found to make the predictions less sensitive to spurious self-interaction errors~\cite{hunnisett2024seventhrank, cook2024contrasting}. 

Machine learning interatomic potentials (MLIPs) have revolutionized atomistic modeling by enabling accurate predictions of energies and forces at a fraction of the cost of quantum mechanical methods such as DFT~\cite{wines2025chips,loew2025universal,Batatia2022mace,esen,schutt2017schnet,sriram2022towards,gasteiger2021gemnet,gasteiger2022graph,schutt2021equivariant,passaro2023reducing,liao2023equiformerv2}. 
Historically, MLIPs were trained almost exclusively on large datasets of DFT calculations of either crystalline inorganic materials or isolated molecules~\cite{omol, odac23, omat24, smith2017ani, smith2020ani, ani2x, eastman2023spice, deng2023chgnet}, leaving molecular crystals largely outside the training domain of general-purpose models. Early machine learning approaches to molecular CSP therefore relied on system-specific potentials, retrained or fine-tuned for each target~\cite{musilMachineLearningStructure2018a,glycine_pbe0_free,egorovaMultifidelityStatisticalMachine2020,wengertDataefficientMachineLearning,butlerMachineLearnedPotentialsActive2024,piaAccurateEfficientMachine2025,bahramiPredictionCrystalStructure2026}. In the seventh blind test, one team (Group 16) developed system-specific AIMNet2 MLIPs~\cite{nayal_efficient_2025}. These MLIPs were trained on DFT data for small clusters of molecules extracted from putative crystal structures of each target. 
They were used for structure relaxations, lattice energy evaluations, and free energy calculations in both the structure generation phase and the ranking phase. In the structure ranking phase, two other teams applied machine learning methods. Group 12 used Gaussian process regression trained on density functional tight binding (DFTB) data and Group 15 used ANI-2x MLIPs fine-tuned by transfer learning on the lists of crystal structures provided by the CCDC. Group 16's system-specific AIMNet2 MLIPs delivered the best performance, on par with dispersion-inclusive DFT. However, this strategy does not scale to high-throughput screening, since a new model must be trained for every compound of interest. 

This landscape has shifted with the Open Molecular Crystals 2025 (OMC25) dataset~\cite{omc25}, a large open collection of dispersion-inclusive DFT calculations spanning chemically diverse molecular crystals. OMC25 has enabled the first generation of general-purpose MLIPs that treat periodic organic solids on the same footing as inorganic materials and gas-phase molecules~\cite{uma}.
Leveraging such large-scale training, several groups have begun applying pretrained, general-purpose MLIPs to molecular CSP without per-system retraining~\cite{zhou2025robust,weber2025efficient, taylor2025predictive, price2025one, glick2025toward, hafiziFoundationModelsTrained, midgley2026dft}. Questions nevertheless remain about the reliability and transferability of these models across chemical space. In practice, most reported workflows still rely on a final DFT re-ranking step, which limits the speed and scalability that motivate the use of MLIPs in the first place. Whether a single pretrained potential can support an end-to-end molecular CSP workflow with DFT-level accuracy and without system-specific fine-tuning remained an open question.

In this work, we present the \fullname~framework, an open-source end-to-end crystal structure prediction workflow that leverages a single universal machine learning potential, Universal Model for Atoms (UMA)~\cite{uma}, for geometry relaxations, free-energy evaluations, and conformer energy corrections of random structures generated by Genarris 3~\cite{genarrisv3}. UMA is built on the eSEN architecture~\cite{esen}, an equivariant graph neural network model, and is extended with Mixture of Linear Experts (MoLE) layers to scale to large model sizes without compromising inference speed. 
Unlike most existing MLIPs, which are typically trained on a single domain (e.g., molecules or inorganic crystals), UMA is trained jointly across multiple domains using shared representations and separate dataset-specific tasks.

Here, we primarily use the Open Molecular Crystals (OMC) task of UMA for geometry relaxations and free-energy evaluations. The OMC25 dataset~\cite{omc25} includes over 25 million configurations extracted from relaxation trajectories of thousands of putative molecular crystal structures generated by Genarris 3 from 50,000 chemically diverse compounds. The crystal structures in OMC25 are labeled with energies, forces, and stress tensors, calculated using dispersion-inclusive semi-local DFT.
This large, chemically diverse dataset enables UMA to generalize across a wide range of compounds, comprising different elements and capable of forming a variety of intermolecular interactions and crystal packing motifs, without requiring molecule-specific fine-tuning. 

The Open Molecules (OMol) task of UMA is used in the conformer generation stage and to evaluate a conformer energy correction that rebalances intramolecular energies against the lattice contribution. The OMol25 dataset~\cite{omol} comprises approximately 140 million single-point and partial optimization calculations of molecules and clusters sampled from equilibrium and non-equilibrium geometries. These were computed using a range-separated hybrid meta-GGA functional with nonlocal correlation using Gaussian basis sets without pseudopotentials. This level of theory is substantially more accurate for both inter- and intramolecular conformational energies than the dispersion corrected semi-local DFT approach used in OMC25. The key distinction between the two tasks is their support for periodic systems: OMC is trained on periodic molecular crystal configurations and optimized for crystal packing energetics, whereas OMol is trained on finite systems and does not include unit cell stress values. This complementarity is exploited in our conformer-energy correction scheme, which combines the OMC task's accuracy for lattice energy with the OMol task's superior description of monomer conformational preferences.

We demonstrate the effectiveness of our approach on two complementary benchmark subsets: a Semi-Rigid subset of 28 mostly rigid molecules with 37 experimental polymorphs in total, and a Flexible subset of 10 conformationally rich molecules with multiple rotatable degrees of freedom and 37 experimental polymorphs in total. Our FastCSP workflow predicts the known experimental structures for all of these molecules. For single-polymorph systems, the experimentally observed structure is ranked as the lattice energy global minimum in 14 out of 26 cases, and within the top 5 in 9 additional cases. For multi-polymorph systems, one of the experimentally known polymorphs is ranked as the global minimum in 9 out of 12 cases, and at least one polymorph is within the top 5 for the remaining 3 cases. All experimentally observed polymorphs fall within 9 kJ/mol of the predicted global minimum. We further show that the performance of UMA for geometry optimization, ranking, and free-energy calculations is on par with dispersion-inclusive DFT. Notably, for some flexible molecules, including the well-known ROY, an optional OMC$\to$OMol conformer-energy correction recovers the correct stability ordering of conformational polymorphs without  requiring post-hoc high-level DFT recalculation.

Thus, the speed and accuracy of UMA obviates the need for classical force field pre-screening or DFT-based (re)ranking and enables a CSP workflow that relies exclusively on MLIPs, unlocking results within hours to several days with tens of modern NVIDIA H200 GPUs available. 
Our entire workflow is released open-source, with the data used to train the models released in the companion papers~\cite{omc25, omol}. This combination of accuracy, speed, and accessibility represents a significant step toward scalable, transparent, and reliable molecular CSP workflows. 

\section{Methods}
\subsection{Overview of FastCSP}

\begin{figure}
    \centering
    \includegraphics[trim={0 1.25cm 0 3.23cm},clip,width=\linewidth]{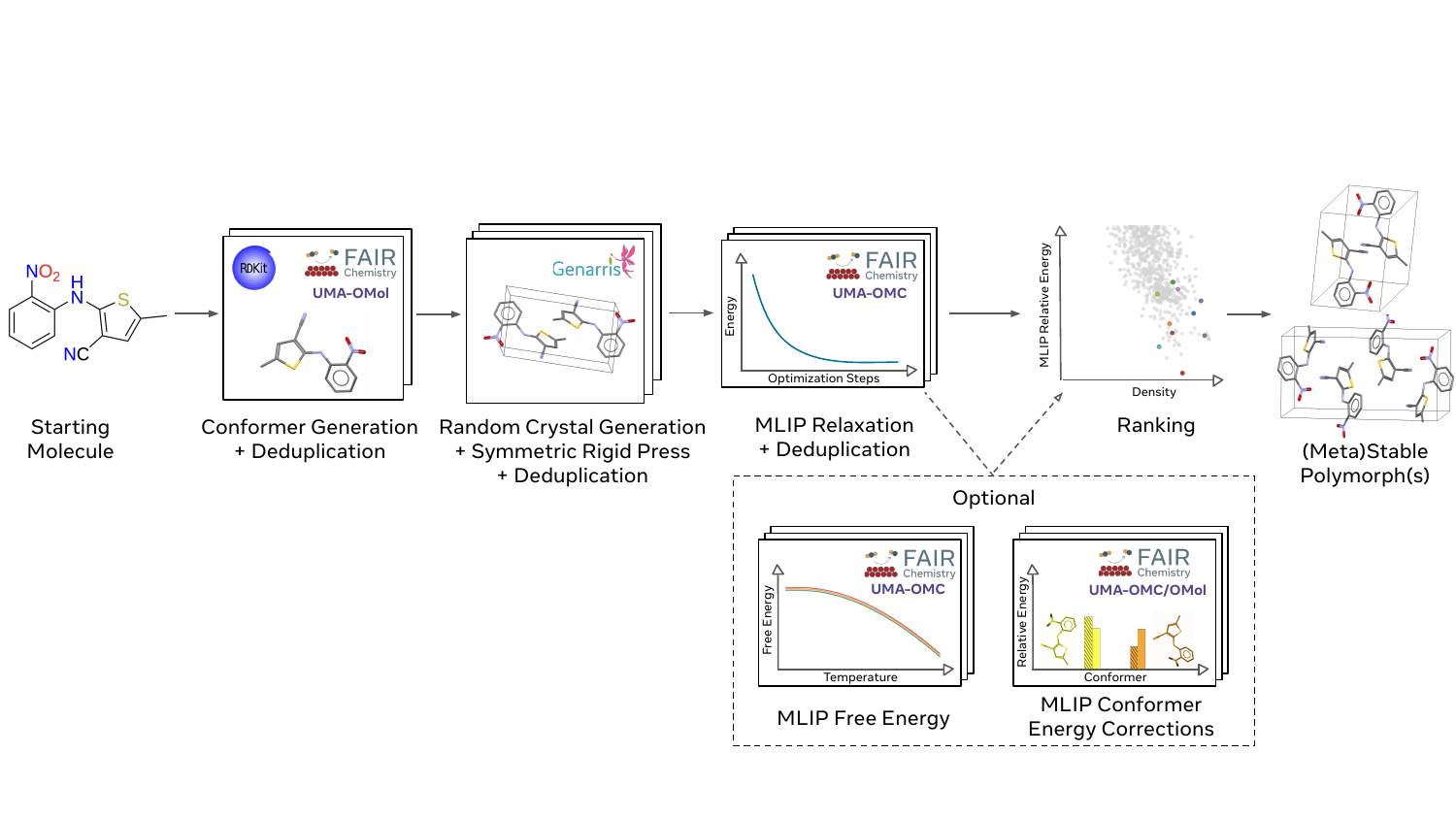}
    \caption{Overview of the \name~workflow. Starting from a target molecule, an initial pool of conformers is generated with RDKit and relaxed with the OMol task of the UMA model. Each retained conformer is then passed to Genarris 3, which generates a diverse set of random crystal structures. The generated structures undergo optimization with Rigid Press feature of Genarris 3 followed by deduplication. The remaining structures are fully relaxed using the OMC task of UMA and undergo another round of deduplication. The final ranking is based on UMA lattice energy. Optionally, a free energy or an OMC$\to$OMol conformer energy correction, also calculated using UMA, can be applied.}
    \label{fig:workflow}
\end{figure}

FastCSP is an end-to-end, high-throughput crystal structure prediction (CSP) workflow built around a single universal machine-learning interatomic potential (MLIP). The full workflow, illustrated in Figure \ref{fig:workflow}, comprises three main and two optional stages: (i) generation of molecular conformers using RDKit~\cite{rdkit} and the Universal Model for Atoms (UMA)~\cite{uma}, (ii) random crystal structure generation using Genarris 3~\cite{genarrisv3}, (iii) geometry relaxation and ranking with UMA, (iv) optional finite-temperature free-energy evaluation with UMA, and (v) an optional energy correction with UMA that rebalances intramolecular conformer energies against the lattice energy. The entire workflow is driven by UMA. Dispersion-inclusive DFT is only used here to benchmark the MLIP performance.

For a given target compound, \textsc{RDKit} is used to seed an initial conformer pool that is relaxed with the OMol task of UMA-Small version 1.1 (UMA-S-1.1 [OMol]), and clustered down to a small set of low-energy conformers. Each selected conformer is then passed to Genarris 3, which generates random molecular crystal structures in all space groups compatible with the molecular point group and the requested number of molecular formula units per unit cell ($Z$), including cases in which molecules occupy special Wyckoff positions ($Z'\leq1)$. The initial structures are compressed with the Rigid Press feature of Genarris 3 to achieve close packing while preserving the space group symmetry. Near-identical structures are subsequently removed with \textsc{Pymatgen}'s \textsc{StructureMatcher}~\cite{ong2013python}.

The remaining structures are fully relaxed using the OMC task of UMA-S-1.1 (UMA-S-1.1 [OMC]). The use of a general-purpose MLIP allows us to scale relaxation across thousands of candidate structures, making the workflow suitable for high-throughput CSP campaigns. Structures whose relaxation fails to converge and those whose molecular connectivity changed during relaxation are discarded. Then, deduplication is performed again to eliminate redundant relaxed minima. UMA can optionally be used to compute Helmholtz and Gibbs vibrational free energies at finite temperatures to reflect the relative thermodynamic stability of candidate structures under real conditions. OMC$\to$OMol (monomer) conformer energy corrections can be applied to rebalance the contributions of intramolecular conformer energies and intermolecular interactions to the lattice energy. Full details of all workflow parameters are provided in SI Section~\ref{app:hyperparams}.

The use of a single pretrained MLIP for all stages yields substantial computational savings compared to DFT-based or system-specific MLIP approaches. Conformer generation completes in minutes on a single GPU. Crystal structure generation with Genarris~3 requires on the order of 100 CPU-core hours per conformer (depending on the sampled search space), and MLIP relaxation, which dominates the GPU cost, averages approximately 2.3~minutes per structure for semi-rigid molecules and 2.9~minutes per structure for flexible molecules on an NVIDIA V100 GPU. 
Even at this conservative configuration without inference optimizations, each MLIP relaxation step is on average ${\sim}$5,600$\times$ (CPU-core hour per GPU hour) cheaper than a corresponding dispersion-inclusive DFT ionic step.
Transitioning to current-generation NVIDIA H200 GPUs and parallelized batch inference (processing multiple independent structures on a single GPU simultaneously) can increase the effective throughput by ${\sim}$6$\times$, reducing the amortized GPU-time per structure below 30~s. 
With this optimization on NVIDIA H200 GPUs, end-to-end CSP results for a single system are achievable within hours to several days depending on the number of selected conformers and search space size. 

Conformer energy corrections are inexpensive, requiring only single-point UMA-OMC and UMA-OMol evaluations on each of the $Z$ constituent molecules extracted from each relaxed crystal. Free energy calculations within the quasi-harmonic approximation are considerably more expensive, as they require phonon calculations on supercells at multiple volumes for each retained candidate structure. The primary knobs for controlling computational cost are the number of conformers forwarded to crystal structure generation, the space groups and $Z$ values sampled, the number of structures sampled in each space group, and the deduplication stringency prior to relaxation. Deeper discussion of runtime is presented in SI~Section~\ref{app:si_runtime} and per-system runtime breakdowns are provided in SI~Table~\ref{tab:runtime}.

\subsection{Conformer Generation}
\label{sec:main_methods_conformer_generation}
For each target, a pool of conformers was generated from the canonical, stereochemistry-aware SMILES using \textsc{RDKit}, building on the multi-sampler scheme of~\cite{zhou_deep_2023}. To reach a combined target pool size of 500 conformers, four generation passes are run: (I) ETKDGv3~\cite{etkdg,etkdgv3} distance-geometry embedding followed by MMFF force-field optimization~\cite{mmff94}, (II) the same with the embedding initialized from random Cartesian coordinates, (III) ETKDGv3 embedding without force-field optimization, and (IV) uniform random sampling of each rotatable torsion from a single ETKDGv3-embedded reference. The four passes contribute in a 4:4:1:1 ratio to the combined pool. The combined pool is clustered with the Butina algorithm~\cite{butina1999unsupervised}, which assigns all conformers as neighbors or not, on the basis of a given RMSD threshold, and then selects the set of representative structures with the most neighbors until all structures are assigned. Each cluster representative is relaxed with the OMol task of UMA-Small v1.1. The relaxed pool is re-clustered by Butina with a within-cluster energy gate of 1.5~kJ/mol, then truncated to a 40~kJ/mol window above the lowest-energy survivor. From this energy-sorted pool, we forward the lowest-energy conformers to Genarris: one per target for the Semi-Rigid subset, twenty per target for the Flexible subset, and five for nicotinamide because of its rotatable amide group. Glycine is a special case: relaxation with UMA-Small v1.1 starting from the zwitterionic SMILES invariably results in proton-transfer to neutral glycine (the expected gas-phase form~\cite{glycine_exp_data,genarrisv2}), such that every UMA relaxation fails the connectivity check. The workflow then reverts to the unrelaxed RDKit geometries (ranked by their UMA single-point energy) and forwards the single lowest-energy conformer to structure generation.

A complete evaluation of conformer generation quality is provided in SI Section~\ref{app:conf_eval}. Therein, we show that the generated conformer pools closely reproduce the molecular conformations observed in the experimental crystal structures. We also note that seeding the workflow with conformers that match the crystal geometry does not, in and of itself, guarantee successful CSP because intermolecular interactions ultimately determine the final relaxed molecular structure in the crystal. This means that a somewhat different initial conformation may still relax into the correct experimentally observed structure. Therefore, initiating CSP with a conformer ensemble generally improves the chances of generating the experimental structure(s).

\subsection{Structure Generation}
For each selected conformer, Genarris 3~\cite{genarrisv3} was used to generate a large number of putative crystal structures. Genarris generates structures with up to one molecule in the asymmetric unit ($Z' \leq 1$) in all space groups compatible with the requested number of molecules per unit cell ($Z\ \in$ \{1, 2, 3, 4, 6, 8\}) and the molecular point group symmetry, including cases where molecules occupy special Wyckoff positions~\cite{genarrisv2}. 
For the Semi-Rigid molecules subset, 500 structures were generated in each compatible space group for each conformer.
For the Flexible molecules subset, to balance computational cost with sampling quality, sampling was restricted to the top 10 most common space groups in the CSD. These 10 space groups jointly cover over 90\% of organic crystals in the CSD~\cite{csd}. Crucially, the experimental space group was also included if it fell outside the top 10. This inclusion was only necessary for two structures: the $\alpha _{1}$ polymorph of piroxicam ($Pca2_{1}$, No. 29; the 11th most frequent space group in the CSD) and the $\beta$ polymorph of chlorpropamide ($Pbcn$, No. 60; the 13th most frequent space group), where no top-10 space group produced a match. Consequently, a strict top-10 limit would have automatically resulted in two missed structures.
For each conformer, 1,500 structures were generated per space group. For olanzapine, Forms I and IV were not generated with these settings, and structure generation was repeated for the five lowest-energy conformers with 3,000 structures per space group. The larger pool for the Flexible subset reflects the larger accessible configuration space, which requires more sampling to recover the experimental packing arrangement(s).

Genarris 3 generates random molecular crystal structures by sampling unit cell volumes from a Gaussian distribution around a target volume estimated by the built-in PyMoVE model~\cite{bier2020machine}. The estimated volume is initially scaled by a factor of 1.5 to facilitate molecule placement. 
The first molecule is randomly placed within the unit cell, and the remaining molecules are placed according to the symmetry operations of the selected space group. When the initial placement occurs at a special Wyckoff position, the molecule is aligned to match the site symmetry~\cite{genarrisv2}. Once a molecular crystal structure is generated, it undergoes validation to ensure that interatomic distances $d_{ij}$ between atoms i and j on different molecules are physically reasonable: $d_{ij}$ must exceed $s_r \times (r_i^{vdW} + r_j^{vdW})$, where $r_{i/j}^{vdW}$ are the atomic van der Waals radii and $s_r$ is a user-defined fraction with a default of 0.95. Specific intermolecular distance thresholds are assigned for strong hydrogen bonds~\cite{genarrisv2}.

The initial generation of molecular crystal structures proceeds until the number of structures requested by the user is reached, after which
geometry optimization with the \emph{Rigid Press} feature of
Genarris 3 is performed. A regularized hard-sphere potential is employed to compress the unit cell while keeping the molecules' internal geometry rigid. This brings the molecules as close to each other as possible without breaking the space group symmetry to achieve close packing. We then deduplicate the generated structures with \textsc{Pymatgen}'s \textsc{StructureMatcher}~\cite{ong2013python} to remove similar crystal structures. A substantial fraction of the generated pool is removed at this step (an average of 67\% for the Semi-Rigid and 43\% for the Flexible molecules), lowering the cost of the relaxation stage downstream. We consider this as an indication that the configuration space was sampled exhaustively.

\subsection{Structure Relaxations} \label{sec:mlip_relaxation}
UMA-Small version 1.1 model with the OMC task~\cite{uma} was used for optimizing the atomic positions and lattice parameters of all generated structures. Structure relaxations used the Broyden-Fletcher-Goldfarb-Shanno (BFGS) optimizer as implemented in the Atomic Simulation Environment (\textsc{ASE})~\cite{ase}, with a 0.01 eV/\AA\ force threshold, under periodic boundary conditions and without symmetry constraints. Optimizations were limited to 1,000 steps. Structures whose relaxation failed to converge and those whose intramolecular covalent connectivity changed during relaxation were discarded. For each target, we retained all structures within 10 kJ/mol of the global minimum.
A second \textsc{StructureMatcher} deduplication pass was then applied to the entire relaxed pool to remove duplicates generated from different conformers, $Z$ values, or space groups that relaxed into the same minimum. Within each deduplication cluster, the lowest-energy structure is retained as the representative. For the free-energy calculations and the DFT re-ranking validation for the Semi-Rigid subset, we performed the aforementioned deduplication after filtering structures within a 5 kJ/mol window of the global minimum. For downstream analysis, all deduplicated structures were retained.

\subsection{Free Energy Calculations} \label{sec:free_energy}
Free energy calculations were conducted for putative structures with relative lattice energies within 5 kJ/mol (Semi-Rigid) or 10 kJ/mol (Flexible) of the global minimum of each target to account for finite temperature and pressure thermodynamic effects. To obtain Helmholtz free energies at constant volume, we performed phonon calculations within the harmonic approximation (HA), using the supercell method implemented in \textsc{Phonopy}~\cite{phonopy-phono3py-JPCM, phonopy-phono3py-JPSJ}. To account for thermal expansion, we calculated the Gibbs free energies within the quasi-harmonic approximation (QHA). The Gibbs free energy is approximated by calculating Helmholtz free energies for volume-constrained relaxed supercells at a range of fixed volumes. Here, we used volumes ranging between $\pm$6\% of the zero-temperature equilibrium volume obtained from our preceding MLIP relaxations. 
Subsequently, an expression for the Helmholtz free energy as a function of volume at a fixed temperature is obtained by fitting a Vinet equation of state~\cite{vinet1989universal} to the calculated free energies at each volume. Finally, we determined the Gibbs free energy of a structure at a given temperature and pressure by identifying the minimum value of the fitted Helmholtz free energy plus the hydrostatic work (PV).

\subsection{Conformer Energy Corrections} \label{sec:energy_corrections}
The OMC task of UMA-S-1.1 is trained on dispersion-inclusive DFT data for periodic molecular crystals~\cite{omc25}, acquired using the generalized gradient approximation of Perdew, Burke, and Ernzerhof (PBE)~\cite{pbe} with the Grimme D3~\cite{dftd3} dispersion correction. Therefore, UMA-OMC is accurate for intermolecular interactions in extended systems. The OMol task was trained on the OMol25 dataset~\cite{omol}, comprising non-periodic molecule and cluster reference data acquired at a higher level of theory using the $\omega$B97M-V range-separated hybrid meta-GGA functional with the VV10 non-local treatment of dispersion interactions~\cite{mardirossian2016omegab97m, hellweg2015development, rappoport2010property}, which is significantly more accurate than PBE-D3. In addition, the OMol25 dataset was acquired using Gaussian basis sets and without pseudopotentials of light atoms, although this is anticipated to have a minor effect relative to the accuracy of the exchange-correlation functional and dispersion model. OMol25 models have been shown to reproduce conformer energy differences within chemical accuracy for diverse molecular structures~\cite{omol}. Therefore, the OMol task is expected to be more accurate than the OMC task for intramolecular conformational energies. 

To account for the contributions of crystal packing versus conformational changes to the stabilization of molecular crystals, a post-relaxation OMC$\to$OMol conformer correction can be applied.  The relaxed crystal is decomposed into its $Z$ molecular fragments (monomers/conformers) and the corrected crystal energy is given by:
\begin{equation}
    E^{\text{UMA$+\Delta$UMA}}_{\text{crystal}}
   \;=\; E^{\text{UMA-OMC}}_{\text{crystal}}
        \;-\; \sum_{i=1}^{Z} \left(E^{\text{UMA-OMC}}_{\text{conformer}_i}
        \;-\; E^{\text{UMA-OMol}}_{\text{conformer}_i}\right)\ ,
    \label{eq:conformer_correction}
\end{equation}
where all conformer energies are evaluated with the same UMA-S-1.1 checkpoint. This correction scheme combines the accuracy of UMA-OMC for molecular crystals with the accuracy of UMA-OMol for isolated molecules. The correction is most impactful for systems whose experimentally observed structure differs from the UMA-OMC global minimum structure primarily due to intramolecular conformation energy, as discussed in detail below.

\subsection{DFT Validation} \label{sec:DFT}
To benchmark the accuracy of the UMA MLIP, we performed reference dispersion-inclusive DFT calculations for all deduplicated relaxed structures ranked by UMA within 5 kJ/mol (Semi-Rigid) or 10 kJ/mol (Flexible) of the global minimum of each target. We used the Vienna Ab initio Simulation Package (VASP)~\cite{vasp1,vasp2,vasp3} with the same computational settings as those employed in the generation of the OMC25 dataset~\cite{omc25}. The PBE exchange-correlation functional was paired with Grimme's D3~\cite{dftd3} pairwise dispersion correction. VASP 5.4 projector augmented-wave (PAW) PBE pseudopotentials were used with a plane-wave energy cutoff of 520 eV. PBE-D3 relaxations were started from the UMA-relaxed structures. Atomic positions and lattice vectors were relaxed until the maximum per-atom residual forces fell below 0.001 eV/\AA, or the relaxation process exceeded 1,500 steps.

To benchmark the accuracy of the free energy calculations with UMA, we performed DFT free energy calculations for the five polymorphs of glycine. The FHI-aims code~\cite{fhiaims1, fhiaims2, fhiaims3} was used with Tier 1 basis sets and \textit{light} numerical settings. Free energy calculations within the HA and QHA were performed at the PBE-D3 level using the finite-displacement supercell method implemented in \textsc{Phonopy}~\cite{phonopy-phono3py-JPCM, phonopy-phono3py-JPSJ}. We employed $2\times2\times2$ supercells and performed harmonic free energy calculations for six fixed volumes ranging from -6\% to +10\% of the 0~K equilibrium unit cell volume. All other computational settings matched those of the free energy calculations with UMA. The five polymorphs of glycine were further re-optimized with PBE+MBD using the Tier 2 basis sets and \textit{tight} numerical settings of FHI-aims, and re-ranked based on single-point energy evaluations with the PBE-based hybrid functional, PBE0~\cite{pbe0}, combined with the many-body dispersion (MBD)~\cite{mbd1, mbd2, libmbd} method using the Tier 2 basis sets and \textit{intermediate} numerical settings.

\section{Results}

\subsection{Benchmark Molecular Dataset} \label{sec:dataset} \begin{figure*}[!ht] \centering \includegraphics[trim={0cm 0cm 0cm 0cm},clip,width=\textwidth]{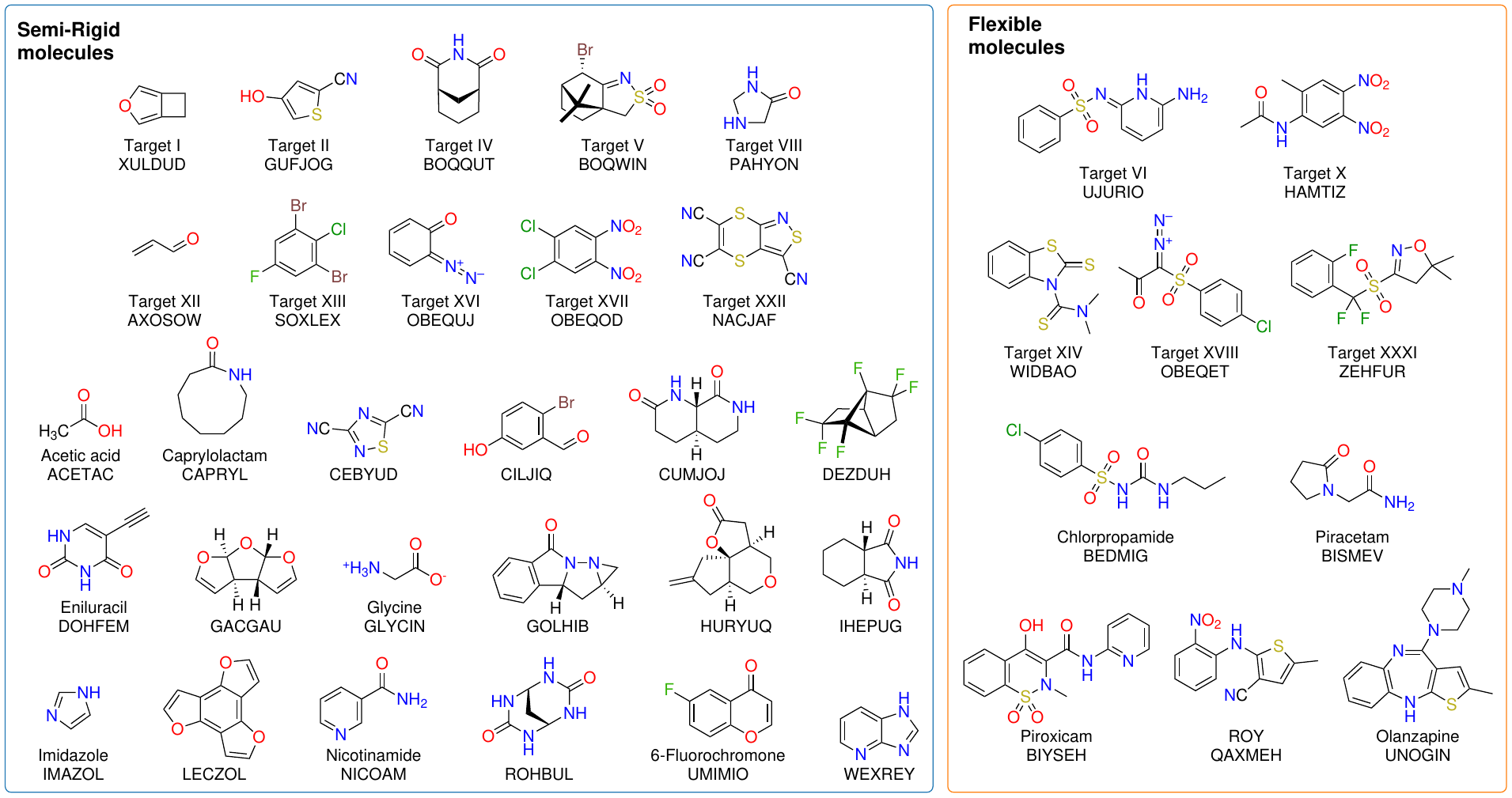} \caption{ Molecular diagrams, CSD reference codes, and common names of all 38 compounds considered as benchmark systems for FastCSP. More details can be found in SI Table~\ref{tab:all_molecules}. } \label{fig:molecules} \end{figure*} 

To validate FastCSP, we curated a chemically diverse benchmark set of 38 compounds (Figure~\ref{fig:molecules}), divided into two complementary subsets that together span 74 experimentally known crystal structures. 
The \textbf{Semi-Rigid molecules subset} comprises 28 compounds with at most one non-trivial rotatable degree of freedom and 37 experimental polymorphs. These compounds were selected to represent diverse space groups (with $Z$ values of 1, 2, 3, 4, 6, and 8 across 14 distinct space groups), chemically diverse intermolecular interactions (hydrogen bonds, halogen bonds, $\pi$-stacking), and elements beyond those commonly treated (including \ce{Br}). The subset includes 14 compounds from the ``Tier~1'' set of Ref.~\cite{zhou2025robust}, extended with 14 additional targets to cover space groups not among the 22 most common in the CSD~\cite{csd, csdspacegroups} and to incorporate molecules with more diverse bonding environments~\cite{taylor2025predictive, taniguchi2025crystal}. Five compounds (Target~I, CILJIQ, imidazole, nicotinamide, and glycine) have multiple known polymorphs, amounting to a total of 37 crystal structures for the 28 molecules. 

The \textbf{Flexible molecules subset} comprises 10 compounds with up to four rotatable bonds, up to 50 atoms, and 37 experimental polymorphs collectively. Nine targets come from the ``Tier~2'' set of Ref.~\cite{zhou2025robust} and one from Ref.~\cite{glick2025toward}. Five were selected from previous CSP blind tests (Targets VI~\cite{motherwell2002crystal}, X~\cite{day2005third}, XIV~\cite{day2009significant}, XVIII~\cite{bardwell2011towards}, and XXXI~\cite{hunnisett2024seventhgen, hunnisett2024seventhrank}) to provide a direct connection to established community benchmarks. The remaining five systems, for which we selected polymorphs with $Z' \leq 1$, were chosen as polymorph-rich systems where most ambient forms adopt distinct molecular conformations. These include ROY (10 polymorphs)~\cite{roy}, chlorpropamide (6 polymorphs)~\cite{cpa}, piroxicam (5 polymorphs)~\cite{yaoPolymorphismPiroxicamNew2020}, olanzapine (4 polymorphs)~\cite{reutzel-edensCrystalFormsPharmaceutical2020}, and piracetam (4 polymorphs)~\cite{nowellValidationSearchTechnique2005}. 

Jointly, the two subsets enable a thorough evaluation of FastCSP across increasing molecular flexibility, packing diversity, and polymorphic complexity. Some structures in the benchmark set are represented by multiple CSD entries. We selected the entries that correspond to the blind test reports, where available. For all other cases with multiple entries, we followed the procedure of Ref.~\cite{zhou2025robust} to select the most reliable crystal structure(s). Detailed crystallographic information for all selected structures, including CSD reference codes, space groups, $Z$ values, and polymorph designations, is provided in SI~Table~\ref{tab:all_molecules}.

\subsection{Structure Generation Results} 

\begin{figure*}[ht] \centering \includegraphics[width=0.8\textwidth]{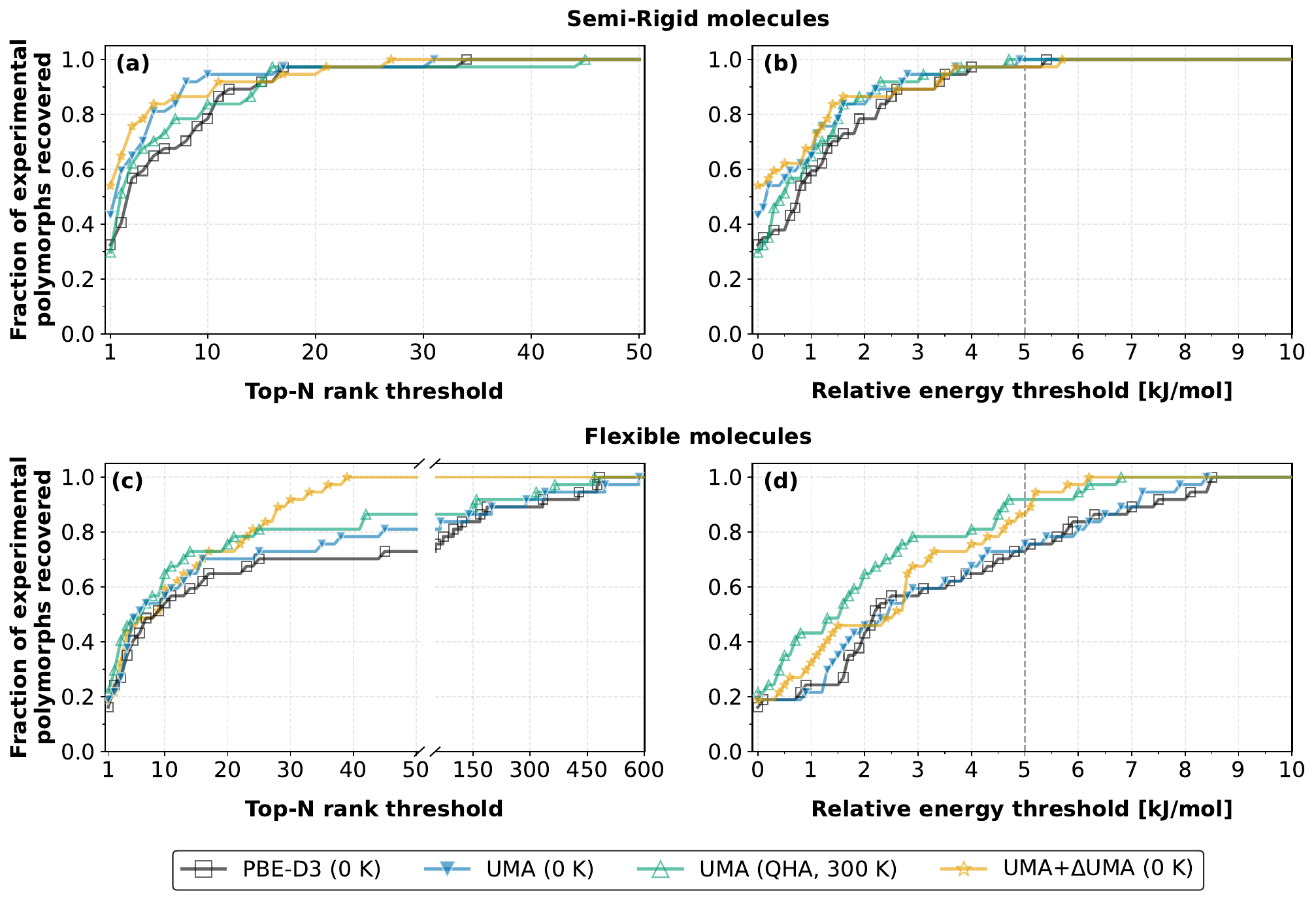} \caption{Cumulative recovery (recall) of experimental polymorphs. Recovery as a function of the top-N ranked predicted structures using ML (UMA MLIP at 0 K, quasi-harmonic approximation [QHA] at 300 K, and with conformer energy corrections) and DFT (PBE-D3 at 0 K) for (a) Semi-Rigid and (c) Flexible molecules. Recovery as a function of relative energy threshold (in kJ/mol) from the predicted minimum for (b) Semi-Rigid and (d) Flexible molecules. Recovery (recall) measures the fraction of known experimental structures that are matched within a given rank or energy threshold. All structures are recovered within 9 kJ/mol relative energy with all methods.} \label{fig:recall} \end{figure*} 

The \name\ workflow successfully generated the experimental structures of all compounds in the benchmark set. The only exception that was not generated with the default structure generation settings is \textbf{olanzapine} (\ce{C17H20N4S}, an antipsychotic drug~\cite{fultonOlanzapine1997} with over 60 reported solid forms~\cite{reutzel-edensCrystalFormsPharmaceutical2020}). Forms I and IV of olanzapine  were only generated when the number of structures generated by Genarris was increased from 1,500 to 3,000 per space group. The packing motif of all neat olanzapine forms comprises centrosymmetric dimers in which two molecules stack face-to-face~\cite{bhardwajExploringExperimentalComputed2013}. It is possible that the particular dimer stacking adopted by Forms I and IV may require more sampling to recover by random structure generation.
The recall analysis in Figure~\ref{fig:recall} shows that our UMA-based CSP workflow achieves over 94\% recall within the top~10 predictions and 100\% within 5~kJ/mol of the lattice energy (0~K) global minimum for semi-rigid molecules. For flexible molecules, FastCSP achieves over 81\% recall within the top~50 predictions and 100\% within 9~kJ/mol of the lattice energy (0~K) global minimum. 
A tabulated summary of all CSP results is provided in SI~Table~\ref{tab:tier1_results}. 

\begin{figure}[htb!]{
\centering 
\includegraphics[trim={0.28cm 0.21cm 0.25cm 0.25cm},clip, width=\textwidth]{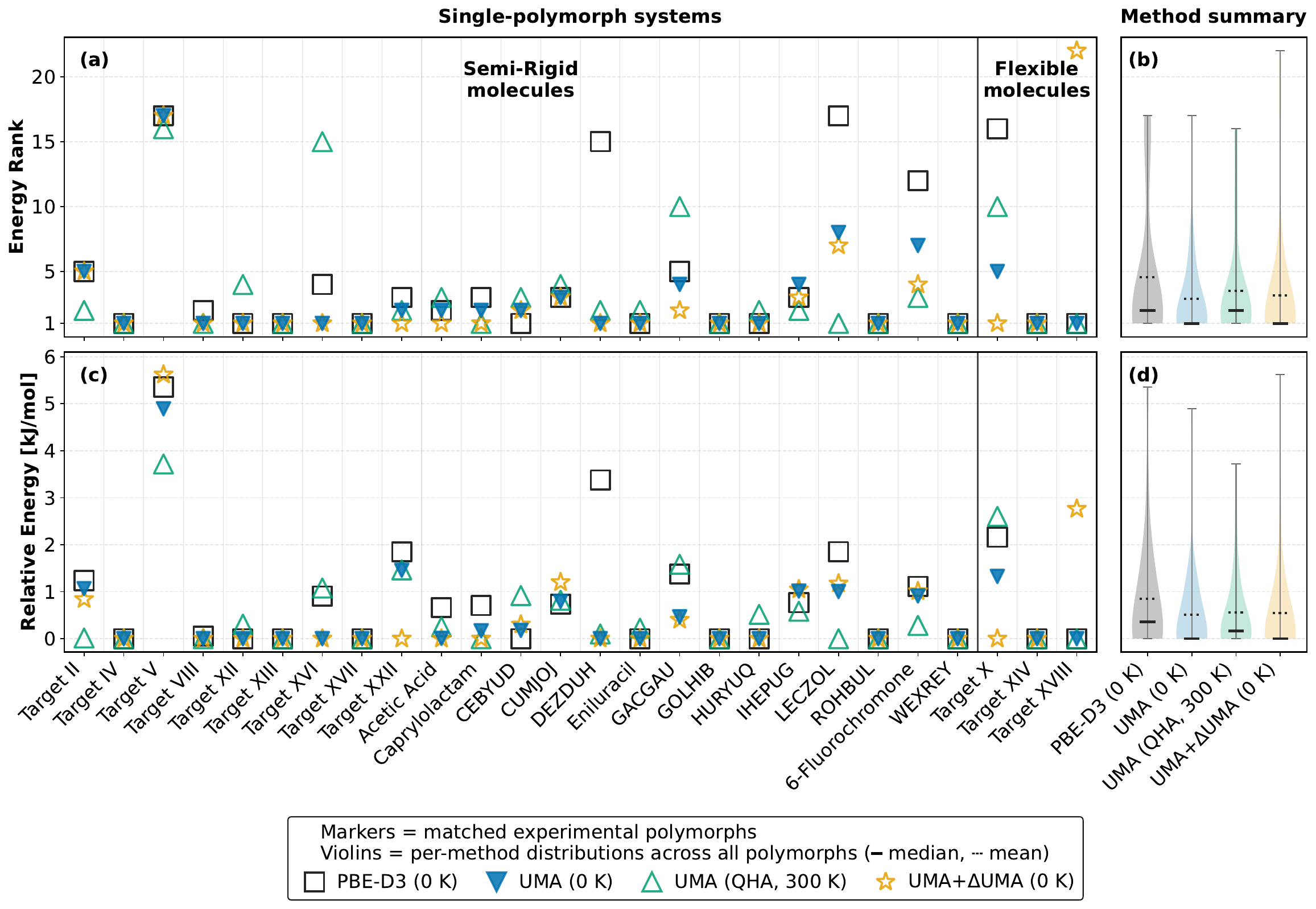} 
\caption{Evaluation of ranking quality for single-polymorph systems. (a) Energy rank and (c) relative energy (in kJ/mol) of experimental polymorph matches. Results obtained using PBE-D3 (at 0 K) are compared to the UMA MLIP method (at 0 K, under the quasi-harmonic approximation [QHA] at 300 K, and with conformer energy corrections). The proximity of experimental forms to the global minimum (rank \#1 and a relative energy of 0 kJ/mol) reflects the ranking quality of the method. Violin plots of (b) energy rank and (d) relative energy distributions indicating the mean (dotted line), median (solid line), minimum, and maximum values.
\label{fig:csp_rank_sp}}
}  \end{figure}

Figure~\ref{fig:csp_rank_sp} shows that for single-polymorph systems, all experimentally observed structures fall within 5~kJ/mol ($\approx$~1.20~kcal/mol) of the lattice energy global minimum, with the vast majority within 2~kJ/mol ($\approx$~0.48~kcal/mol). For 14 of 26 single-polymorph targets, the experimental structure is ranked by UMA as the lattice energy (0~K) global minimum. For 9 additional targets the experimental structure falls within the top~5. The outliers are Target~V (\#17), LECZOL (\#8), and 6-fluorochromone (\#7). In most cases, the UMA ranking is in close agreement with PBE-D3. In some cases UMA even outperforms PBE-D3. For Target VIII, Target XVI, and DEZDUH UMA ranks the experimental structure as the global minimum whereas PBE-D3 does not.
For the outliers Target V, LECZOL, and 6-fluorochromone, as well as for Target XXII, acetic acid, caprylolactam, GACGAU, and Target X UMA ranks the experimental structure better than PBE-D3. For Target~V in particular, it has been shown previously~\cite{zhou2025robust, whittleton2017exchange2, price2023accurate} that lattice energies obtained with various dispersion-inclusive DFT methods misrank the experimental structure (the ranking from~\cite{zhou2025robust} was extracted from an associated repository and is provided in SI~Table~\ref{tab:rank_sc}). For Target X,  it has been shown that the relative stability of the candidate crystal structures depends strongly on the conformational energy of the molecule, which is not described sufficiently accurately by GGA functionals~\cite{whittleton2017exchange2}. Therefore, reliable polymorph ranking requires conformer corrections, as discussed in Section~\ref{sec:energy_corrections}. 

\begin{figure}[htb!]{
\centering 
\includegraphics[trim={0.28cm 0.20cm 0.25cm 0.25cm},clip, width=\textwidth]{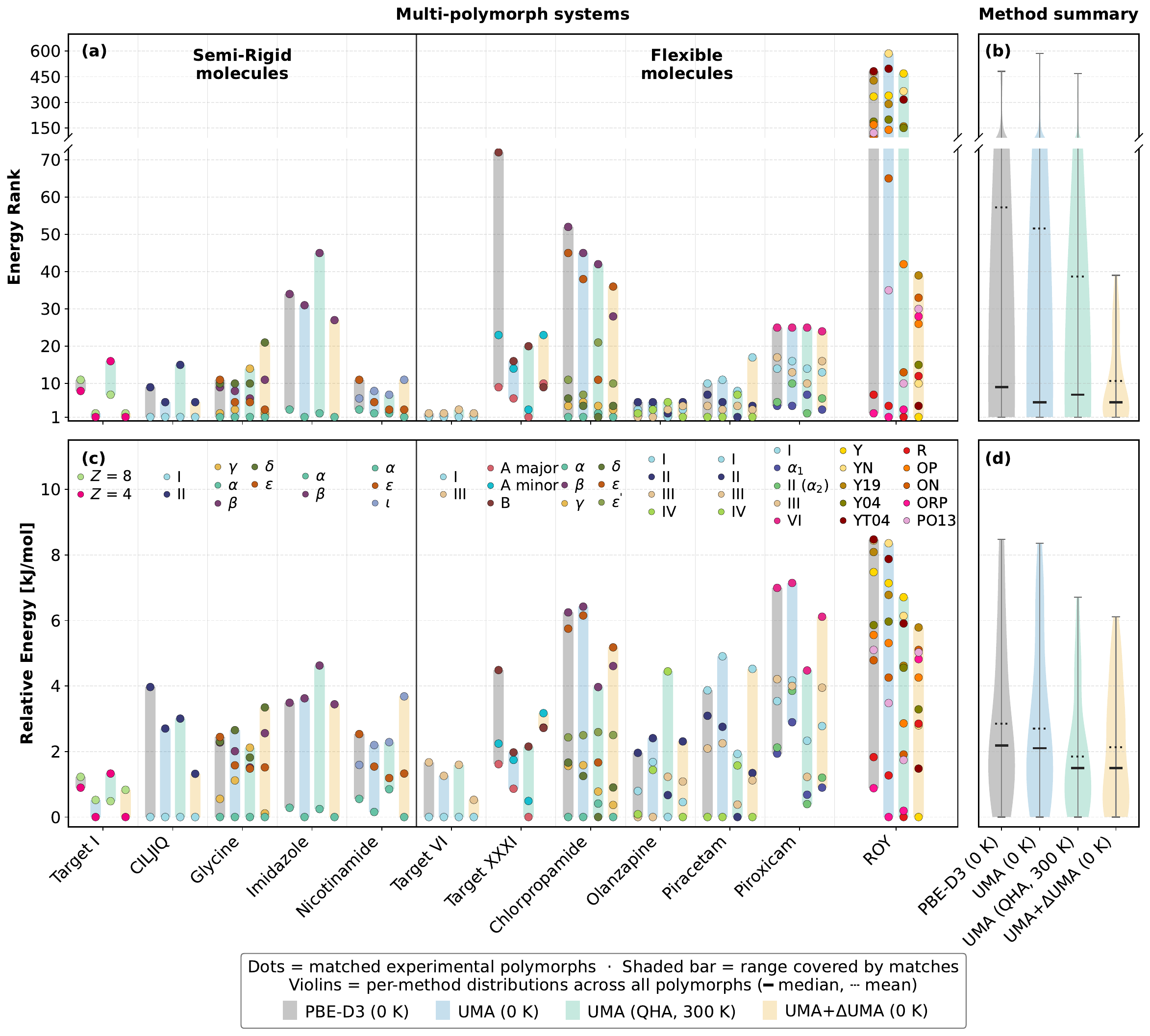} 
\caption{Evaluation of ranking quality for multi-polymorph systems. (a) Energy rank and (c) relative energy (in kJ/mol) of experimental polymorph matches. Results obtained using PBE-D3 (at 0 K) are compared to the UMA MLIP method (at 0 K, under the quasi-harmonic approximation [QHA] at 300 K, and with conformer energy corrections). The proximity of experimental forms to the global minimum (rank \#1 and a relative energy of 0 kJ/mol) reflects the ranking quality of the method. Violin plots of (b) energy rank and (d) relative energy distributions indicating the mean (dotted line), median (solid line), minimum, and maximum values.
\label{fig:csp_rank_mp}}
}
\end{figure}

Figure~\ref{fig:csp_rank_mp} shows that for the multi-polymorph systems, UMA ranks one of the experimentally known polymorphs as the global minimum in 9 out of 12 cases and at least one polymorph is within the top 5 for the remaining 3 cases. All experimentally observed polymorphs fall within 9 kJ/mol of the predicted global minimum. As expected, the matched polymorphs span broader rank and energy ranges for the more flexible systems, most notably Target~XXXI, chlorpropamide, piroxicam, and ROY, reflecting the increased complexity of their energy landscapes. Similar to the single-polymorph systems, the UMA ranking is in close agreement with PBE-D3 in most cases (a detailed discussion of polymorph ordering is presented below). Notably,  UMA outperforms PBE-D3 for Target I and Target XXXI, ranking the experimentally observed polymorphs closer to the global minimum. This demonstrates that UMA provides results on par with dispersion-inclusive DFT at a fraction of the computational cost across diverse molecular systems. 

\subsection{Polymorph Ranking Results}

\begin{figure}[htb!] \centering \includegraphics[width=0.9\textwidth]{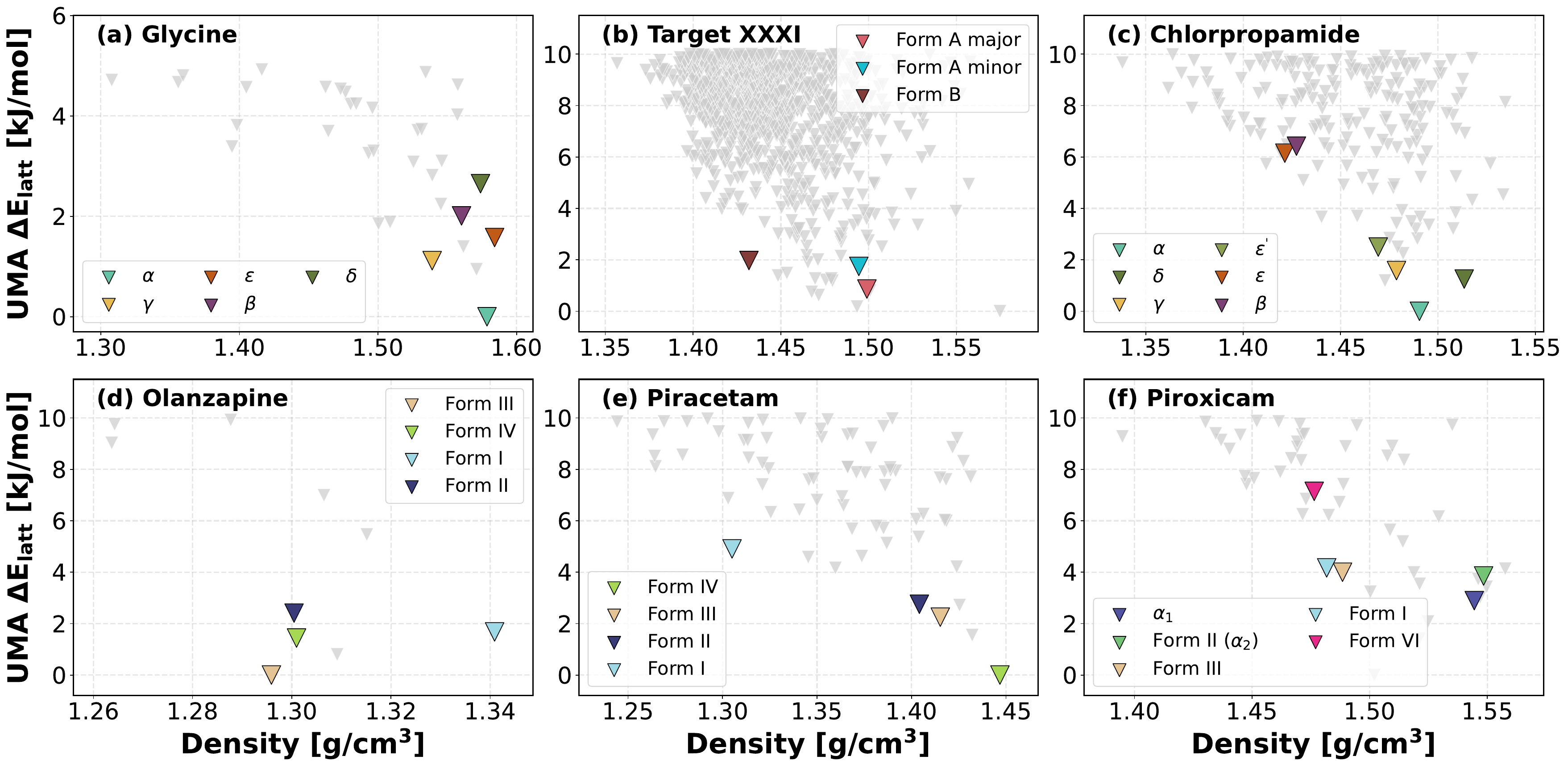} \caption{Final energy landscapes obtained with UMA at 0~K. The relative lattice energy is plotted as a function of density for the polymorphic systems (a) glycine, (b) Target XXXI, (c) chlorpropamide, (d) olanzapine, (e) piracetam, and (f) piroxicam. Colored markers denote matches to experimentally observed polymorphs and gray points show putative structures. The proximity of experimental forms to the global minimum (rank \#1 and a relative energy of 0 kJ/mol) reflects ranking quality.} \label{fig:landscape} \end{figure} 

Figure~\ref{fig:landscape} shows the final energy landscapes obtained with UMA at 0~K for representative polymorphic targets. Results for all targets obtained with UMA at 0~K, UMA with conformer energy corrections at 0~K, UMA at 300~K, and PBE-D3 at 0~K are presented in SI~Figure~\ref{fig:landscape_all}. 
The landscapes of the semi-rigid compounds are relatively sparse compared to the landscapes of the flexible compounds. Typically, the experimentally observed polymorphs are found in the vicinity of the global minimum without many unobserved structures in between. For most of the semi-rigid polymorphic targets, the UMA 0~K ranking correctly reproduces the experimentally observed order of stability. 

\textbf{Target~I} (XULDUD, \ce{C6H6O}, 3,4-cyclobutylfuran) from the first CSP blind test~\cite{lommerse2000test} has a stable $Z=4$ form and a metastable $Z=8$ form. UMA correctly ranks the $Z=4$ form as the global minimum and the $Z=8$ form as \#2, less stable by 0.52~kJ/mol at 0~K, outperforming PBE-D3, which ranks both forms higher than putative structures -- \#7 (0.35 kJ/mol) and \#10  (0.69 kJ/mol), respectively. Previous studies of Target I have reported that some dispersion-inclusive DFT methods misrank its two polymorphs~\cite{gator, t1_stable, zhou2025robust}. 
\textbf{Nicotinamide} (NICOAM, \ce{C6H6N2O}, pyridine-3-carboxamide) is well-known as vitamin $B_3$. It is a highly polymorphic compound with nine known polymorphs~\cite{nico_rank2}. Here, we focus on three polymorphs with $Z' \leq 1$: the most stable $\alpha$ form, the $\iota$ form, and the $\varepsilon$ form. The $\alpha$ form is ranked by UMA as more stable than the $\varepsilon$ and $\iota$ forms by 1.39 and 2.05 kJ/mol, respectively, at 0 K. This is consistent with experimental observations~\cite{nico_rank1, nico_rank2, nico_rank3}, with our PBE-D3 ranking, and with dispersion-inclusive DFT results reported by others~\cite{nico_rank2, zhou2025robust} (the ranking from~\cite{zhou2025robust} was extracted from an associated repository and is provided in SI Table~\ref{tab:rank_sc}). 
UMA also correctly ranks the two polymorphs of \textbf{imidazole} in agreement with experimental observations~\cite{imidazole_rank} and PBE-D3. For \textbf{CILJIQ}, the UMA ranking is consistent with PBE-D3 results. 
The only exception is glycine, whose ranking is discussed in Section \ref{sec:glycine}.

The landscapes of flexible compounds tend to be more dense, and some, such as Target XXXI, chlorpropamide, and ROY (discussed in Section \ref{sec:roy}) contain many structures within an energy window of a few kJ/mol. 
For \textbf{Target VI} (UJIRIO, \ce{C11H11N3O2S}, 6-amino-2-(phenylsulfonylimino)-1,2-dihydropyridine), the structure originally reported  in the second blind test~\cite{motherwell2002crystal} corresponds to Form I, a $Z'=1$ polymorph in the monoclinic space group $P2_1/c$ (No.~14). Two additional polymorphs were subsequently reported: Form II~\cite{vi_ii} with $Z'=2$, and Form III~\cite{vi_iii} with $Z'=1$. 
Form III crystallizes in the orthorhombic space group $Pbca$ (No.~61). Form I has been found to be the most stable at 299 K \cite{vi_iii}. 
Target VI is challenging for conformer generation because it contains two rotatable bonds and can adopt distinct \textit{cis}/\textit{trans} configurations about the S--N=C--N linkage. Forms I and III have different conformations with different orientations of phenyl rings that lead to distinct packing configurations~\cite{nilssonlillMoleculeVISulfonimide2014}. As shown in SI Figure~\ref{fig:landscape_all}, Form I is correctly ranked by UMA as the most stable and Form III as the second most stable by 1.26 kJ/mol. This is consistent with our PBE-D3 results and with previously reported dispersion-inclusive DFT calculations~\cite{chanMoleculeVIBenchmark2011}.

\textbf{Target XXXI} (ZEHFUR, \ce{C12H12F3NO3S}, 3-((difluoro-(2-fluorophenyl)methyl)sulfonyl)-5,5-dimethyl-2l2-iso\-xazolidine), from the most recent blind test~\cite{hunnisett2024seventhgen, hunnisett2024seventhrank}, is an agrochemical containing three rotatable bonds. Target XXXI has three forms labeled as A, B, and C. Form C is excluded from this study because its value of $Z=18$ falls outside the current scope. Forms A and B both crystallize in the monoclinic space group $P2_1/c$ (No.~14) with $Z=4$. In Form A, the \textit{ortho}-fluorophenyl ring is disordered over two sites, denoted Form A$_\mathrm{maj}$ and Form A$_\mathrm{min}$, with an occupancy ratio of 60:40. The experimentally determined order of stability below 55 \textcelsius\ is Form B > Form A~\cite{hunnisett2024seventhrank}. Figure~\ref{fig:landscape}b shows that Form A$_\text{{maj}}$ (ranked \#6) is correctly ranked as more stable than Form A$_\text{{min}}$ (ranked \#13) by 0.88 kJ/mol. However, Form B (ranked \#15) is incorrectly ranked as less stable than Form A. This ranking is consistent with the results of other MLIPs and DFT-based methods reported in the seventh blind test and afterwards~\cite{hunnisett2024seventhrank, mayo2024assessment, nayal_efficient_2025}.

\textbf{Chlorpropamide} (CPA, BEDMIG, \ce{C10H13ClN2O3S}, 1-(4-chlorophenylsulfonyl)-3-propylurea) is a sulfonylurea antidiabetic compound~\cite{cpa}. Nine unique structures have been reported in the CSD. CPA has five stable forms at ambient temperature and pressure, the most stable $\alpha$ form ($P2_12_12_1$, No.~19, $Z=4$)~\cite{cpa, drebushchakTransitionsFivePolymorphs2008}, $\beta$ ($Pbcn$, No.~60, $Z=8$), $\gamma$ ($P2_1$, No.~4, $Z=2$), $\delta$ ($Pbca$, No.~61, $Z=8$), and $\varepsilon$ ($Pna2_1$, No.~33, $Z=4$). The low-temperature $\varepsilon'$ form, which also crystallizes in $Pna2_1$ (No.~33) with $Z=4$, is obtained from the $\varepsilon$ form through a conformational phase transition upon cooling to approximately 200 K~\cite{cpa_epsilon1}. As shown in Figure~\ref{fig:landscape}c, UMA correctly identifies the $\alpha$ form as the global minimum, followed by $\delta$, $\gamma$, $\varepsilon'$, $\varepsilon$, and $\beta$, with relative energies of 1.25, 1.58, 2.50, 6.15, and 6.42 kJ/mol, respectively. The $\beta$ form is ranked as the least stable polymorph, which is consistent with experimental observations~\cite{drebushchakTransitionsFivePolymorphs2008}.  The ordering is in close agreement with our PBE-D3 results. Both methods place $\alpha$ lowest in energy, and the only change is the order of the $\delta$ and $\gamma$ forms: UMA ranks $\delta$ 0.33 kJ/mol below $\gamma$, whereas PBE-D3 reverses this order, placing $\gamma$ 0.10 kJ/mol below $\delta$. Prior work using PBE-D3(BJ) also identified $\alpha$ form as the lowest-energy polymorph, placing $\varepsilon'$ next in energy, while $\gamma$, $\delta$, and $\beta$ were clustered within approximately 5--6 kJ/mol of $\alpha$, and $\varepsilon$ was higher at 9.3 kJ/mol~\cite{cpa}.

\textbf{Olanzapine} (OLZ, UNOGIN, \ce{C17H20N4S}, 2-methyl-4-(4-methylpiperazin-1-yl)-5\textit{H}-thieno[3,2-\textit{c}][1,5]benzo\-di\-azepine) is an antipsychotic drug~\cite{fultonOlanzapine1997}. Although the molecule contains only one rotatable bond, it exhibits a high degree of polymorphism. Over 60 distinct solid-forms have been reported, including four anhydrous polymorphs, 56 crystalline solvates, and an amorphous phase~\cite{reutzel-edensCrystalFormsPharmaceutical2020, bhardwajExploringExperimentalComputed2013}. Here, we focus on the four anhydrous $Z'=1$ polymorphs, Forms I--IV. Forms I--III crystallize in $P2_1/c$ (No.~14) and Form IV crystallizes in $P2_1/n$ (No.~14), with $Z=4$. Form I is the most stable polymorph and Forms II--IV are metastable with respect to Form I~\cite{olz_exp, m.leblancCrystalenergyLandscapesActive2019}. Figure~\ref{fig:landscape}d shows that Form III is incorrectly ranked by UMA as the most stable at 0 K, followed by Form IV 1.44 kJ/mol higher in energy, Form I at 1.68 kJ/mol higher, and Form II at 2.41 kJ/mol higher. This ranking is in agreement with our PBE-D3 results.

\textbf{Piracetam} (BISMEV, \ce{C6H10N2O2}, 2-oxo-1-pyrrolidineacetamide) is a small nootropic, whose crystal polymorphs adopt conformations that differ substantially from the gas-phase minimum with its intramolecular hydrogen bond~\cite{nowellValidationSearchTechnique2005}. Form I is metastable and transforms spontaneously to Form II at room temperature. Forms II and III are stable at lower temperature and are similar in stability near room temperature, with Form II considered as the most stable form. Form IV, obtained under high-pressure conditions, adopts a molecular conformation distinct from those in the previously known polymorphs~\cite{nowellValidationSearchTechnique2005}. Figure~\ref{fig:landscape}e shows that UMA incorrectly ranks Form~IV as the most stable at 0~K, followed by Form~III $>$ Form~II $>$ Form~I, consistent with our PBE-D3 ranking. 

\textbf{Piroxicam} (PXM, BIYSEH, \ce{C15H13N3O4S}, 4-hydroxy-2-methyl-1,1-dioxo-$N$-pyridin-2-yl-1$\lambda^6$,2-benz\-othi\-azine-3-carboxamide) is a nonsteroidal anti-inflammatory drug~\cite{mihalic1986piroxicam}. Over the past four decades, extensive research has been conducted on piroxicam due to its rich polymorphism and nine polymorphs have been reported to date~\cite{yaoPolymorphismPiroxicamNew2020, pxm_viii_ix}. Here, we consider the five known forms with $Z' \leq 1$: Form I ($P2_1/c$, No.~14, $Z=4$), $\alpha_1$ ($Pca2_1$, No.~29, $Z=4$), Form II ($\alpha_2$; $P2_1/c$, No.~14, $Z=4$), Form III ($P\bar{1}$, No.~2, $Z=2$), and Form VI ($P2_1/c$, No.~14, $Z=4$). Experimental melting enthalpies indicate that the thermodynamic stability decreases in the order of Form II $>$ Form I $>$ Form III $>$ Form VI~\cite{yaoPolymorphismPiroxicamNew2020}. The relative stability of $\alpha_1$ was not reported therein. As shown in Figure~\ref{fig:landscape}f, UMA ranks $\alpha_1$ as more stable than Form II ($\alpha_2$) by 0.96 kJ/mol, followed by Form III, Form I, and Form VI at 1.11, 1.28, and 4.25 kJ/mol higher relative to $\alpha_1$, respectively. PBE-D3 also identifies $\alpha_1$ as the lowest-energy form and Form VI as the least stable form, but yields a smaller separation between $\alpha_1$ and Form II ($\alpha_2$) of only 0.18 kJ/mol. The main difference from UMA is the ordering of Form I and Form III: PBE-D3 places Form I 1.60 kJ/mol and Form III 2.27 kJ/mol above $\alpha_1$. In Ref.~\cite{yaoPolymorphismPiroxicamNew2020}, Form II ($\alpha_2$) was ranked as the 0 K minimum, followed by $\alpha_1$, Form I, Form III, and Form VI, using DFT with a different dispersion-corrected GGA functional. 

The ranking of ROY polymorphs is discussed in Section~\ref{sec:roy}.

\subsubsection{Effect of Free Energy Corrections} 
\label{ss:free_energy}

It has been argued that CSP should go beyond 0~K lattice energies and predict the most thermodynamically stable crystal structure under finite temperature and pressure conditions~\cite{hoja2017first, lprice_is_2018, hoja2018first, beran2023frontiers}. 
Free energy corrections have been found to be more significant for larger flexible molecules than for small molecules with limited flexibility~\cite{nyman2015static, nyman2016modelling,yang2020prediction}. Because of the small energy differences between polymorphs, accounting for vibrational free energy contributions can change the relative stability ordering~\cite{nyman2015static, nyman2016modelling, weatherby2022density}. 
However,  the ranking stage of the seventh CSP blind test~\cite{hunnisett2024seventhrank} has demonstrated that free energy corrections are not always beneficial because they improve the ranking for some systems but not for others.

The efficiency of MLIPs enables large-scale free energy calculations for numerous putative structures~\cite{nayal_efficient_2025}. Figure~\ref{fig:csp_rank_mp} shows the change in ranking and relative energy upon switching from lattice energy at 0~K to Gibbs free energy calculated within the quasi-harmonic approximation (QHA) at 300~K. 
In agreement with the conclusions of the seventh blind test, Figure~\ref{fig:recall} shows that for the semi-rigid compounds the free energy corrections make the recall slightly worse than the UMA lattice energy at 0 K. In contrast, for the flexible compounds the free energy corrections lead to an overall improvement in recall, especially with respect to the relative energy.

Of the single-polymorph compounds shown in Figure ~\ref{fig:csp_rank_sp}, the free energy corrections stabilize the experimental structure and improve the ranking of Target II, Target V, caprylolactam, IHEPUG, LECZOL, and 6-fluorochromone, but destabilize the experimental structure and make the ranking worse for Target XIII, Target XVI, acetic acid, CEBYUD, GACGAU, HURYUQ, and Target X. For the remaining single-polymorph compounds, the ranking of the experimental structure is not significantly affected by free energy corrections. 

Of the multi-polymorph compounds, the ranking of CILJIQ, imidazole, nicotinamide, and Target VI is not significantly affected by free energy corrections. For Target I, the free energy correction adversely affects the ordering, destabilizing the $Z=4$ polymorph and making it less stable than the $Z=8$ polymorph and several putative structures. For chlorpropamide, the free energy corrections also adversely affect the stability ordering, misranking the $\delta$ form as the global minimum, followed by the $\alpha$ form, which is supposed to be the most stable.

For Target XXXI, on the one hand, the free energy corrections lead to stabilization of the experimentally observed forms compared to other putative structures, but on the other hand, Form B, which is supposed to be the most stable, is further destabilized with respect to Form A (even above 55~\textcelsius). This contradicts the findings of Ref.~\cite{mayo2024assessment}, in which free energy corrections within the harmonic approximation stabilized Form B. In Ref.~\cite{nayal_efficient_2025}, free energy corrections performed within the QHA using a different MLIP stabilized Form B with respect to Form A$_{\text{min}}$, but it remained higher in energy than Form A$_{\text{maj}}$. 

For piroxicam, the free energy corrections lead to overall stabilization of all the experimentally observed forms compared to other putative structures, similar to Target XXXI. Based on Gibbs free energy, Form II ($\alpha_2$) is ranked as the most stable polymorph in agreement with experiment, followed by the $\alpha_1$ form, whose relative stability is unknown. Form III is now misranked as more stable than Form I, and Form VI is the least stable.
For olanzapine, the free energy corrections at 300 K improve the relative stability of polymorphs. Form I becomes the most stable in agreement with experiment, followed by Forms II ($\alpha_2$), III, and IV at 0.67, 1.23, and 4.44 kJ/mol higher in Gibbs free energy, respectively. 
For piracetam, the free energy corrections also improve the stability ordering of polymorphs. The metastable Form IV is significantly destabilized, restoring the experimentally observed order of Form II > Form III > Form I > Form IV, which is a high-pressure form. 
A detailed analysis of the polymorph ranking of glycine and ROY is presented in Sections~\ref{sec:glycine} and~\ref{sec:roy}, respectively. It is possible that the free energy corrections within the QHA are not universally beneficial because they are not sufficiently accurate, owing to the neglect of anharmonic effects, which may become significant, as the temperature increases~\cite{qha_lim}.

\subsubsection{Effect of Conformer Energy Corrections} 
\label{ss:energy_corrections}

The OMC$\to$OMol ($\Delta$UMA) conformer energy correction (Eq.~\ref{eq:conformer_correction}) rebalances the contributions of the intramolecular conformer energies and the intermolecular interactions to the stability of a molecular crystal. The PBE-D3 level monomer energies from the OMC task are replaced with the more accurate $\omega$B97M-V level energies from the OMol task to mitigate the delocalization error and short-range exchange-correlation inaccuracies in GGA-level functionals~\cite{greenwell2020inaccurate, rana2023correcting, bryentonDelocalizationErrorGreatest2023}.  Figure~\ref{fig:recall} shows that the conformer correction has no appreciable benefit for the semi-rigid compounds. This is expected because the stability of such crystals is primarily driven by intermolecular interactions with relatively minor changes to the molecular conformation. For the flexible compounds, the conformer correction improves recall with respect to both rank and energy thresholds, however, the improvement with respect to the energy threshold is not as significant as that provided by the free energy corrections, and the improvement with respect to the rank is most significant at a relatively high rank, above 20. 
Of the single-polymorph systems, shown in Figure~\ref{fig:csp_rank_sp}, the conformer correction helps stabilize the experimental structures of Target XXII and Target X. 

\textbf{Target XXII} (NACJAF, \ce{C8N4S3}, tricyano-1,4-dithiino[\textit{c}]-isothiazole) from the sixth CSP blind test~\cite{reilly2016report} has unusual intermolecular interactions involving C, S, and N atoms. Although it is considered as semi-rigid, it can bend along the S--S axis of the six-membered ring. It crystallizes in the monoclinic space group $P2_1/n$ (No. 14) with $Z=4$, adopting a bent molecular conformation and a cyclic dimer packing motif~\cite{curtisEffectPackingMotifs2016}. GGA-level DFT functionals misrank the experimental structure of Target XXII because they systematically favor putative structures with a planar molecular conformation and a layered packing motif~\cite{curtisEffectPackingMotifs2016, gator, price2023accurate}. This has been attributed to the self-interaction (delocalization) error in PBE, which favors the more delocalized electron densities of the extended planar conformation and layered packing arrangements. It has been shown that hybrid functionals coupled with a more accurate treatment of dispersion interactions improve the ranking of Target XXII~\cite{curtisEffectPackingMotifs2016, gator, price2023accurate}. The conformer correction stabilizes the bent conformation of the experimental structure and ranks it as the global minimum.

\textbf{Target~X} (HAMTIZ, \ce{C9H9N3O5}, 2-acetamido-4,5-dinitrotoluene) from the third blind test~\cite{day2005third} contains four rotatable bonds and crystallizes in $P2_1/n$ (No. 14) with $Z=4$. No group in the original blind test ranked the experimental structure among their top 3 predictions. It has been shown previously that Target X requires an accurate treatment of both intermolecular and intramolecular energies and that GGA functionals, which tend to favor conformations with more extended conjugations, introduce errors in the relative lattice energies of conformational polymorphs~\cite{whittleton2017exchange2}. 
Indeed, based on UMA-OMC lattice energy the experimental structure is ranked as \#5 and the conformer correction improves it to \#1. The putative structures ranked \#1--4 by UMA-OMC 0~K lattice energy differ from the experimental structure primarily in their intramolecular geometry, specifically in how much the amide group conjugates with the phenyl ring. In the four top-ranked putative structures, the amide lies almost coplanar with the ring (\ce{C-C-N-C} dihedral of $5^\circ$--$15^\circ$), which maximizes intramolecular $\pi$-conjugation and favors $\pi$-stacking or weak \ce{N-H$\cdots$O} interactions as the leading intermolecular contacts. In the experimental form, the amide is rotated to a \ce{C-C-N-C} dihedral of $\sim\!39^\circ$, partially breaking this conjugation. The out-of-plane twist instead exposes the \ce{N-H} and \ce{C=O} groups, enabling strong intermolecular \ce{N-H$\cdots$O} hydrogen bonds. 

Conversely, for \textbf{Target~XVIII} (OBEQET, \ce{C9H7ClN2O3S}) from the fifth blind test~\cite{bardwell2011towards} the conformer correction worsens the ranking of the experimental structure from \#1 to \#22. This result is consistent with the findings of Ref.~\cite{whittleton2017exchange2}. For this molecule, the conformational flexibility is described by three exocyclic torsion angles, and the conjugated \ce{CN2CO} moiety adopts a mostly planar \textit{trans} configuration~\cite{bardwell2011towards}. It therefore does not present the kind of conjugation driven error seen for Target X and Target XXII, where a GGA functional overstabilizes the $\pi$-conjugated conformation of the monomer and the conformer correction is needed to recover the experimental structure~\cite{whittleton2017exchange2, beran2023frontiers}. 

For most of the semi-rigid polymorphic compounds, shown in Figure~\ref{fig:csp_rank_mp}, the conformational corrections do not lead to changes in the relative stability of the observed polymorphs, although they may cause changes in the energy differences between polymorphs. The only exception is glycine, discussed in Section~\ref{sec:glycine}.

The polymorphic flexible compounds are affected by the conformation correction in different ways, depending on their particular chemistry. For Target VI the polymorph ranking does not change, but the energy difference between Form I and Form III decreases. 
For Target XXXI, on the one hand, the conformer correction  recovers the correct order of stability, Form~B $>$ Form~A, but on the other hand, all the experimentally observed polymorphs are destabilized with respect to other putative structures, adversely affecting their ranking. 
For chlorpropamide, the conformer correction correctly maintains the $\alpha$ form as the global minimum, in agreement with experiment. Among the metastable forms, the $\gamma$ form rises from \#4 (1.58~kJ/mol) to \#3 (0.37~kJ/mol), overtaking the $\delta$ form, which drops to \#4, while the correction lowers the $\beta$ form below the higher-energy $\varepsilon$ form (4.61 vs.\ 5.18~kJ/mol).

For olanzapine, the conformer correction misranks Form IV as the global minimum, giving the order Form IV $>$ Form I $>$ Form III $>$ Form II. This is inconsistent with experiment, where Form I is expected to be the most stable. For piracetam, the conformer correction preserves the original order of Form IV $>$ Form~III $>$ Form~II $>$ Form~I and does not help restore the experimentally observed polymorph ordering. This is in contrast to Ref.~\cite{nowellValidationSearchTechnique2005}, where the correct polymorph ranking was recovered by applying a conformer correction on top of a force field. 

For piroxicam, the conformer correction stabilizes all the experimentally observed polymorphs and improves their ranking compared to other putative structures. The stability ordering of Form II ($\alpha_2$) > Form I > Form III > Form VI is in agreement with experiment~\cite{yaoPolymorphismPiroxicamNew2020}. The $\alpha_1$ form (whose relative stability was not reported in~\cite{yaoPolymorphismPiroxicamNew2020}) is persistently ranked as the most stable of the experimentally observed polymorphs.
The conformer correction is most beneficial for ROY, improving the ranking of the experimental ground state (Y form) from rank \#336 to \#1 (Section~\ref{sec:roy}). Altogether, like the free energy corrections, the conformer correction is not universally beneficial.

\subsubsection{Glycine}
\label{sec:glycine}

\begin{figure}[htb!]
    \centering
    \includegraphics[width=0.9\textwidth]{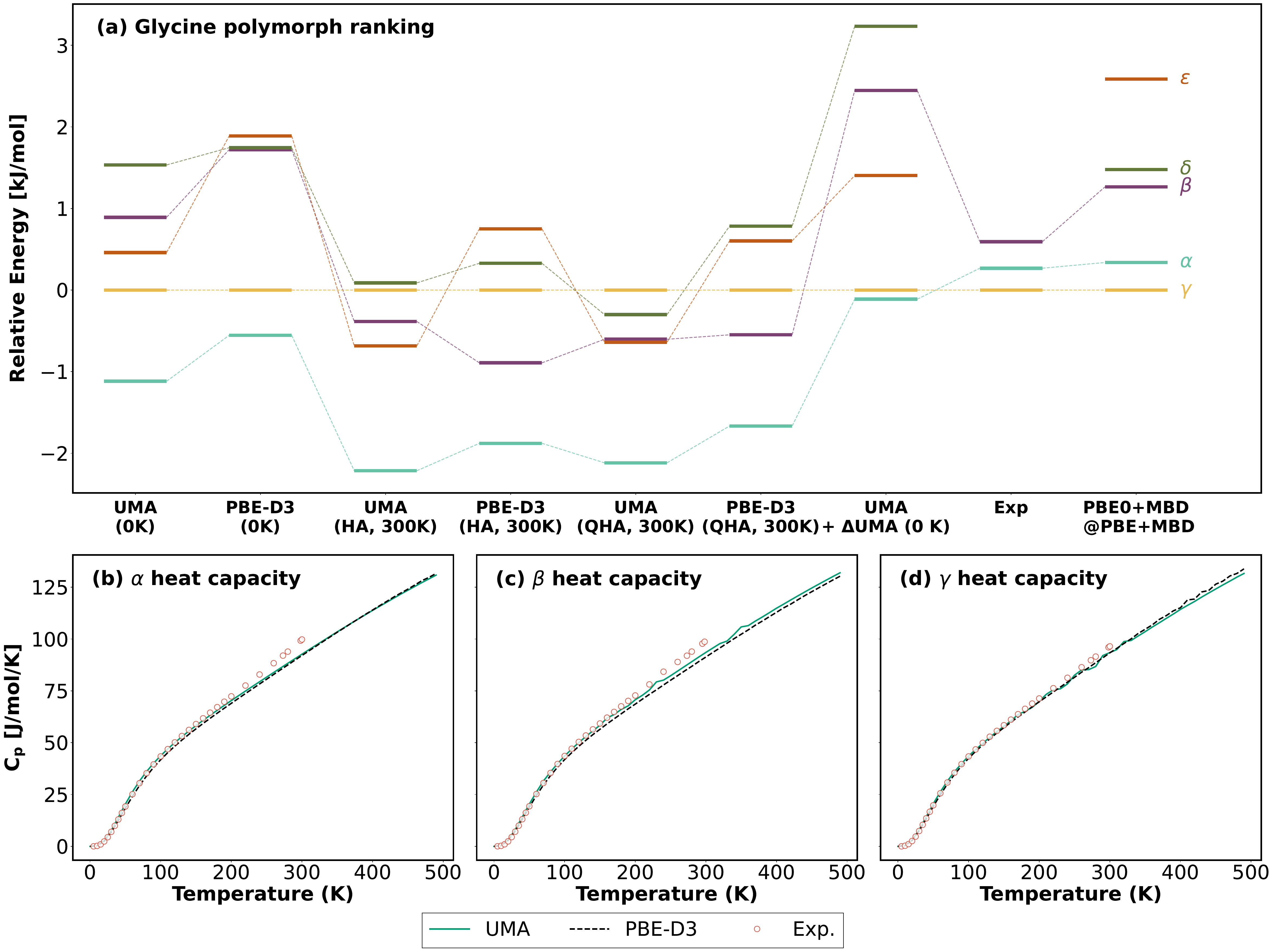}
    \caption{Performance of UMA for the polymorphs of glycine: (a) Relative energy ranking based on UMA and PBE-D3 lattice energy, Helmholtz free energy calculated using the harmonic approximation (HA), Gibbs free energy calculated within the quasi-harmonic approximation (QHA) at 300 K and 1 atm, and conformer energy corrections, also compared against experimental data~\cite{glycine_rank1}. The lattice energy ranking based on single point energy calculations with PBE0+MBD at the geometry optimized with PBE+MBD is also shown. The constant pressure heat capacity ($\textup{C}_\textup{p}$) as a function of temperature for the (b) $\alpha$, (c) $\beta$, and (d) $\gamma$ forms of glycine, calculated using UMA and PBE-D3, compared with experimental data~\cite{glycine_rank3, glycine_exp_data}.}
    \label{fig:glycine}
\end{figure}

Glycine is the simplest amino acid and an important biological building block. In the solid state, glycine adopts a zwitterionic form and crystallizes into several polymorphs stabilized by a close competition among electrostatic, hydrogen-bonding, and dispersion interactions. Under ambient conditions, three anhydrous polymorphs are known: $\alpha$, $\beta$, and $\gamma$, with experimental stability ordering $\gamma > \alpha > \beta$. Two additional high-pressure forms, $\delta$ and $\varepsilon$, are obtained from pressure induced transformations of the $\beta$ and $\gamma$ forms, respectively. It has been shown previously that the experimentally observed stability ranking of $\gamma > \alpha > \beta$ is only recovered when using the more accurate DFT methods, such as hybrid functionals combined with a many-body treatment of dispersion interactions~\cite{glycine_mbd, glycine_vdw, beran2016modeling, glycine_pbe0_free, zhou2025robust}. Therefore, glycine is a challenging case for polymorph ranking.

Figures \ref{fig:landscape}a and \ref{fig:glycine}a present a detailed analysis of the ranking of the glycine polymorphs. All five polymorphs are found within 3 kJ/mol of the global minimum. For the three ambient pressure polymorphs, UMA ranks the $\alpha$ form as the most stable at 0 K, followed by the $\gamma$ form 1.12 kJ/mol higher in energy, and the $\beta$ form at 2.01 kJ/mol higher. This ranking is in agreement with our PBE-D3 results and with previous studies that employed PBE-D3 or similar methods~\cite{glycine_d3_1, glycine_d3_2, glycine_d3_3, weber2025efficient, glycine_ts, glycine_mbd} (the ranking from~\cite{weber2025efficient} was extracted from an associated repository and is provided in SI Table~\ref{tab:rank_sc}). UMA ranks the high-pressure $\varepsilon$ form as more stable than the $\beta$ form, whereas PBE-D3 ranks the $\varepsilon$ form as the least stable. With both UMA and PBE-D3, adding free energy corrections destabilizes the $\gamma$ form with respect to the $\alpha$ form. When adding the phonon contribution via the harmonic approximation (HA) with UMA, the $\gamma$ form becomes less stable than the $\beta$ and $\varepsilon$ forms, whereas with PBE-D3 the $\gamma$ form remains the third most stable after $\alpha$ and $\beta$. The quasi-harmonic approximation (QHA) treats anharmonicity indirectly through the volume dependence of the harmonic frequencies. Applying the QHA with UMA further destabilizes the $\gamma$ form, making it the least stable, whereas with PBE-D3 the $\gamma$ form remains the third most stable.  

Figure~\ref{fig:glycine}b-d shows the temperature dependence of the specific heat of $\alpha$, $\beta$ and $\gamma$-glycine calculated with UMA and PBE-D3 compared to experimental data~\cite{glycine_rank3, glycine_exp_data}. Additional results are provided in SI Figure \ref{fig:glycin_uma_fit_all}. 
The UMA results are in close agreement with PBE-D3 for all three forms of glycine.  The accuracy of calculated vibrational free energies and thermodynamic properties depends on the quality of the phonon spectra. SI Figure~\ref{fig:glycin_uma_dos} compares the phonon band structures and density of states (DOS) of all five glycine polymorphs computed with UMA and PBE-D3. Across all polymorphs, the UMA phonon dispersions closely follow the PBE-D3 reference throughout the Brillouin zone, and the corresponding phonon DOS profiles are nearly identical. In particular, the agreement is excellent in the low-frequency region (0--20 THz), where the modes dominate the vibrational entropy contributions and thus determine the relative free energies~\cite{hoja2018first}. In this range, UMA faithfully reproduces the acoustic branches near the $\Gamma$-point and the positions of the main phonon DOS peaks. This consistency explains the close agreement between the UMA and PBE-D3 specific heats and indicates that UMA accurately captures the soft, low-frequency modes. The absence of imaginary modes in all UMA band structures, especially at the $\Gamma$-point, further confirms that each polymorph is dynamically stable, demonstrating that UMA provides physically reliable phonon spectra suitable for QHA free energy evaluations. This shows that UMA delivers robust performance, equivalent to PBE-D3, in free energy calculations, effectively capturing vibrational contributions and thermal expansion. However, for all three forms of glycine, the UMA and PBE-D3 specific heat start to noticeably deviate from experiment around 200 K. A possible explanation is that the QHA accounts for anharmonicity only to first order and is therefore expected to be less reliable for systems with soft, low-frequency modes and light atoms~\cite{qha_lim}. This may explain why adding free energy corrections within the QHA does not consistently improve the ranking of polymorphs for glycine and across our benchmark set.

Adding the UMA-OMol conformer energy corrections to the lattice energy brings the relative energy $\alpha$ and $\gamma$ forms closest to the experimental ranking. Experimental enthalpies of solution indicate the stability order $\gamma > \alpha > \beta$, with the $\alpha$ and $\beta$ forms lying only 0.268 and 0.593 kJ/mol above $\gamma$, respectively~\cite{glycine_rank1, glycine_mbd}. After the conformer correction, the $\gamma$ form is stabilized and lies only 0.11 kJ/mol above the $\alpha$ form. Thus, although the correction does not fully reverse the order of the two low-energy forms, it reduces their energy differences to the same scale observed experimentally. However the high-pressure $\varepsilon$ form is still ranked as more stable than the $\beta$ form. 

Because the $\Delta$UMA correction replaces the PBE-D3 level description of the intramolecular conformer energies provided by UMA-OMC with the more accurate $\omega$B97M-V/VV10-level description provided by UMA-OMol, it can be helpful when the energy differences between polymorphs stem primarily from conformational changes. However, the intramolecular correction approach assumes that the chosen DFT functional describes the intermolecular interactions well~\cite{greenwell2020inaccurate}. There are notable exceptions, such as systems involving ions or halogen bonds~\cite{otero-de-la-rozaDispersionXDMHybrid2019}, for which hybrid functionals may be required to capture the intermolecular interactions accurately. For glycine, the lattice energies are dominated by the hydrogen-bonded network of zwitterions. The remaining errors are therefore unlikely to arise from the isolated-molecule conformer energy alone. Rather, a more accurate description of intermolecular interactions, including electrostatics and many-body dispersion effects in the crystal environment~\cite{glycine_mbd}, may be required to resolve the very small energy differences between the glycine polymorphs.

\subsubsection{ROY}
\label{sec:roy}

\begin{figure}[htb!]
    \centering
    \includegraphics[width=0.95\textwidth]{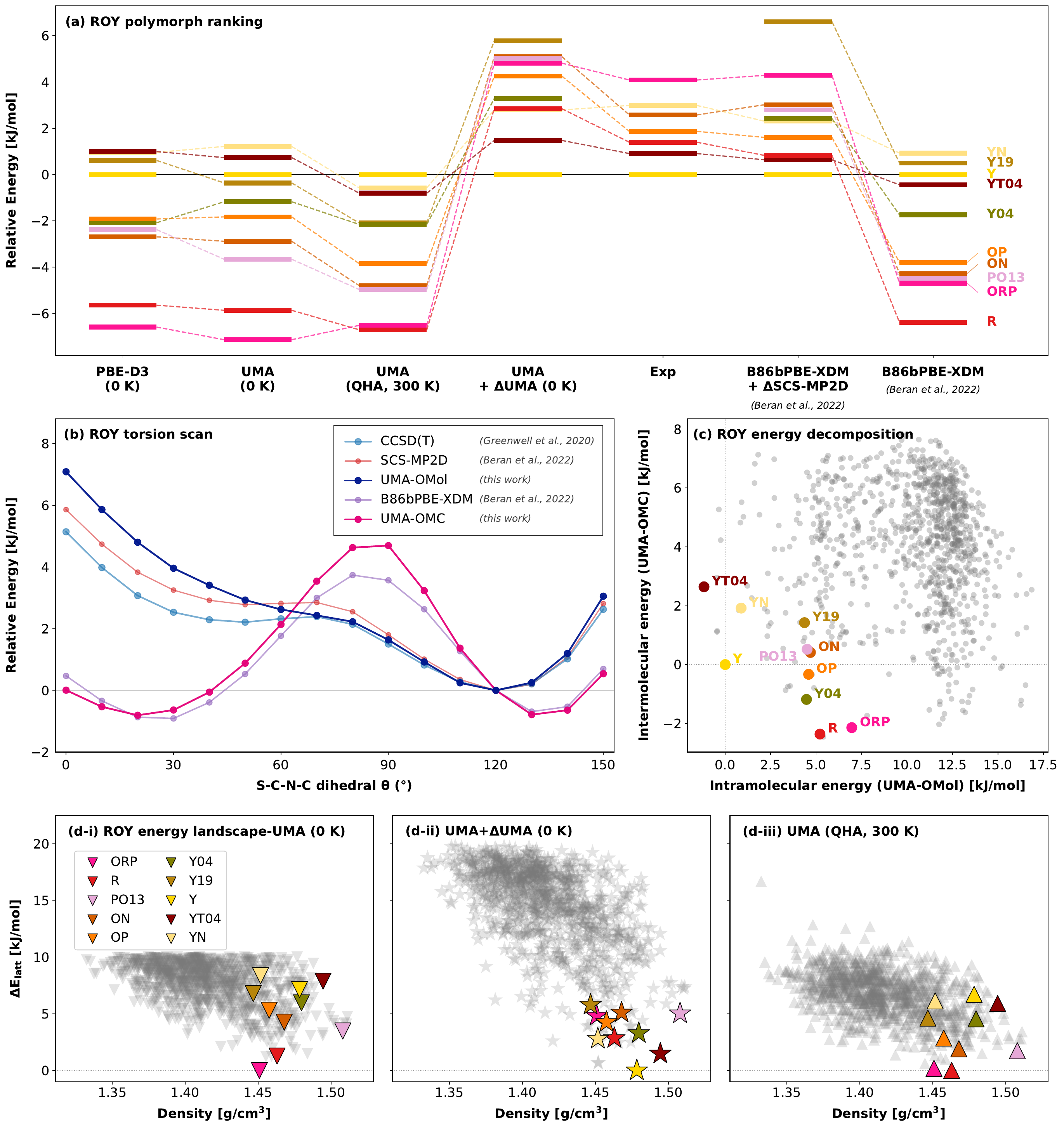}
\caption{
Polymorph energetics, monomer torsional scan, and intramolecular/intermolecular energy decomposition for ROY. (a) Relative energies of the ten experimentally known ROY polymorphs computed using PBE-D3 at 0~K, UMA at 0~K, with free energy corrections within the QHA at 300~K, and with conformer energy corrections, also compared against experimental data~\cite{roy_exp} and literature references~\cite{roy}. All energies are referenced to the experimental ground-state polymorph Y. (b) Gas-phase \ce{S-C-N-C} torsion scan of ROY monomer from
$\theta = 0^{\circ}$ to $150^{\circ}$. Reference curves (CCSD[T], SCS-MP2D, B86bPBE-XDM) are reproduced from Ref.~\cite{greenwellOvercomingDifficultiesPredicting2020, roy}. Each curve is referenced to its own value at the CCSD(T) minimum ($\theta = 120^{\circ}$). 
(c) Intramolecular versus intermolecular energy decomposition for every ROY crystal in our CSP landscape. Both axes are referenced to polymorph Y.
(d) Final energy landscapes obtained with UMA for ROY (i) at 0 K, (ii) with conformer energy corrections, and (iii) quasi-harmonic free energies at 300 K. Matches to the ten experimentally known polymorphs with Z'=1 are highlighted.
}
\label{fig:roy}
\end{figure}

ROY is one of the most extensively studied polymorphic systems and a long-standing benchmark for molecular CSP~\cite{vasileiadisPolymorphsROYApplication2012, tanROYRevisitedAgain2018, nymanAccuracyReproducibilityCrystal2019, roy, galanakisRapidPredictionMolecular2024, weatherstonPolymorphicROYalty14th2025}. Its rich polymorphism arises from the conformational flexibility of the S--C--N--C dihedral angle connecting the thiophene and nitrophenyl rings, which spans a range of approximately $20^{\circ}$ to $150^{\circ}$ across 14 identified ROY forms~\cite{weatherstonPolymorphicROYalty14th2025}. This combination of conformational diversity and a dense polymorph landscape makes ROY an exceptionally challenging target for CSP because the energy ranking requires a simultaneously accurate description of intermolecular interactions and intramolecular conformational energies. In this work, we selected 10 experimentally known $Z'=1$ crystal structures of ROY: Y, ON, R, OP, YN, ORP, YT04, Y04, PO13, and Y19. Based on relative free energies obtained using a differential scanning calorimetry (DSC) method that relies on melting and eutectic melting data~\cite{roy_exp}, the stability decreases in the order Y $>$ YT04 $>$ R $>$ OP $>$ ON $>$ YN $>$ ORP, with Y the most stable form. The relative stabilities of the remaining forms can be inferred from experimental observations. Y04 undergoes spontaneous conversion to YT04 and R, consistent with its metastability relative to both forms, while the specialized crystallization conditions required to isolate Y19 suggest that it is among the least stable forms~\cite{roy_y04, roy}. 

Figure~\ref{fig:roy}a presents the relative lattice energies of the ten $Z'=1$ ROY polymorphs considered in this work. At the UMA (0~K) lattice energy level, the ORP form is ranked as the global minimum, followed by R (\#4, 1.27~kJ/mol), PO13 (\#34, 3.48~kJ/mol), ON (\#63, 4.26~kJ/mol), and OP (\#137, 5.31~kJ/mol). The experimentally most stable Y form is ranked as \#336, 7.14~kJ/mol above ORP, and YT04, which is supposed to be the second most stable form, is ranked as \#494 (7.87~kJ/mol). All ten polymorphs fall within 8.35~kJ/mol window of the UMA global minimum. The ordering obtained with UMA is in agreement with our PBE-D3 results with respect to the ORP form being the most stable followed by the R form, and the Y form is similarly ranked far from the global minimum (\#330, 7.47~kJ/mol), although there is some reshuffling among the other polymorphs. Our results are consistent with PBE-D3 results reported by others~\cite{roy}.
Free energy corrections at 300~K provide a modest improvement for some forms. Under the quasi-harmonic (QHA) correction, ORP remains the top-ranked form with R at \#2, but the energy gap to the higher-ranked forms is reduced. However, the free energy correction alone is insufficient to recover the correct experimental stability ordering.

The systematic misranking of the Y and YT04 forms by both UMA-OMC and PBE-D3 can be attributed to the delocalization error, which is a fundamental limitation of GGA-level exchange-correlation functionals~\cite{bryentonDelocalizationErrorGreatest2023}. Semi-local functionals, such as PBE, overstabilize extended $\pi$-conjugation, artificially lowering the energy of molecular conformations with smaller (more planar) \ce{S-C-N-C} dihedral angles relative to those with larger (more twisted) angles~\cite{greenwell2020inaccurate, rana2023correcting}. The experimentally most stable Y form adopts a large dihedral angle ($\theta \approx 104^{\circ}$), placing it in the twisted, less-conjugated region of the torsional landscape, where GGA functionals impose an energy penalty. Conversely, the R and ORP forms, which are experimentally among the less stable polymorphs, adopt nearly planar conformations ($\theta \approx 21$--$46^{\circ}$) that are artificially stabilized by the delocalization error. This bias propagates to the crystal energetics. Because the incorrectly stabilized planar conformers also tend to pack favorably, the lattice energy ranking becomes doubly biased against the twisted forms. 

Figure~\ref{fig:roy}b illustrates this effect through a gas-phase \ce{S-C-N-C} torsion scan. The UMA-OMC curve closely tracks B86bPBE-XDM~\cite{roy}, a GGA-level density functional that combines B86b exchange with PBE correlation and the exchange-hole dipole moment (XDM) dispersion correction, confirming that UMA faithfully reproduces the PBE-level torsional profile on which it was trained. Both curves overstabilize the planar minimum ($\theta \approx 40^{\circ}$) relative to the complete-basis-set coupled cluster singles, doubles, and perturbative triples (CCSD[T]) benchmark~\cite{greenwellOvercomingDifficultiesPredicting2020} by approximately 4~kJ/mol. In contrast, the UMA-OMol curve, trained on $\omega$B97M-V data from a range-separated hybrid functional, which substantially reduces delocalization error, closely follows the spin-component-scaled dispersion-corrected second-order M\o ller–Plesset perturbation theory (SCS-MP2D)~\cite{mp2d, roy} and correctly places the energy minimum near $\theta = 120^{\circ}$, in agreement with CCSD(T). 

The impact of the conformer energy correction on the polymorph ranking is shown in Figure~\ref{fig:roy}a. Upon applying the OMC$\to$OMol correction, the Y form rises to rank \#1, YT04 to \#4 (1.48~kJ/mol), YN to \#10 (2.80~kJ/mol), and R to \#12 (2.85~kJ/mol). The corrected ranking Y $>$ YT04 $>$ R is in excellent agreement with the experimentally determined stability order Y $>$ YT04 $>$ R $>$ OP $>$ ON $>$ YN $>$ ORP~\cite{roy_exp}. The OP and ON forms are ranked at \#22 and \#29, respectively, representing a significant improvement over their uncorrected positions (\#137 and \#63). ORP, drops from \#1 to \#24 (4.82~kJ/mol), consistent with it being experimentally among the less stable forms. All ten polymorphs are brought to within a 5.79~kJ/mol window relative to Y, compared to 8.35~kJ/mol relative to ORP before the correction, in good agreement with the 6.6~kJ/mol literature estimate for the ROY polymorph energy span~\cite{roy}. The conformer corrected UMA results are in agreement with the B86bPBE-XDM with $\Delta$SCS-MP2D conformer correction from Ref. \cite{roy}.

Figure~\ref{fig:roy}c decomposes the CSP landscape into intramolecular (UMA-OMol) and intermolecular (UMA-OMC lattice energy minus UMA-OMC monomer single-points) energy contributions, both referenced to the Y polymorph. This decomposition reveals a clear trade-off: crystals with low intramolecular energy (twisted conformations) tend to have higher intermolecular energy, whereas those with low intermolecular energy (efficient packing of planar conformations) pay a conformational penalty. The experimentally observed polymorphs span this Pareto front, with Y occupying the position of low intramolecular energy but moderate intermolecular energy, and ORP/R at low intermolecular energy but high conformational cost. The conformer energy correction effectively recalibrates the balance between these two contributions, allowing the ranking to correctly reflect that the conformational penalty of planar forms outweighs their packing advantage. 

These results demonstrate that for conformationally flexible molecules, where the delocalization error significantly distorts the torsional energy surface, the UMA-OMol conformer correction provides an effective and computationally inexpensive remedy. By leveraging the higher-accuracy $\omega$B97M-V-trained UMA-OMol task for monomer energies while retaining the PBE-D3-trained UMA-OMC task for crystal packing, the correction scheme achieves a ranking quality comparable to that reported by Beran et al.~\cite{roy} using B86bPBE-XDM with conformer energy corrections at SCS-MP2D level~\cite{mp2d}, but at a fraction of the computational cost.

\subsection{MLIP evaluation}
\label{sec:mlip_eval}
We assess the performance of UMA by comparing the relaxed geometries and energy ranking of putative structures to PBE-D3. Further validation of UMA for molecular crystals can be found in~\cite{omc25, uma}. The metrics used here reflect the extent to which UMA reproduces the PBE-D3 potential energy surface (PES). If the PBE-D3 results are reproduced reliably, then UMA can completely replace PBE-D3 in CSP workflows with no loss of fidelity. 

\begin{figure}[h!]
    \centering
    \includegraphics[width=0.9\textwidth]{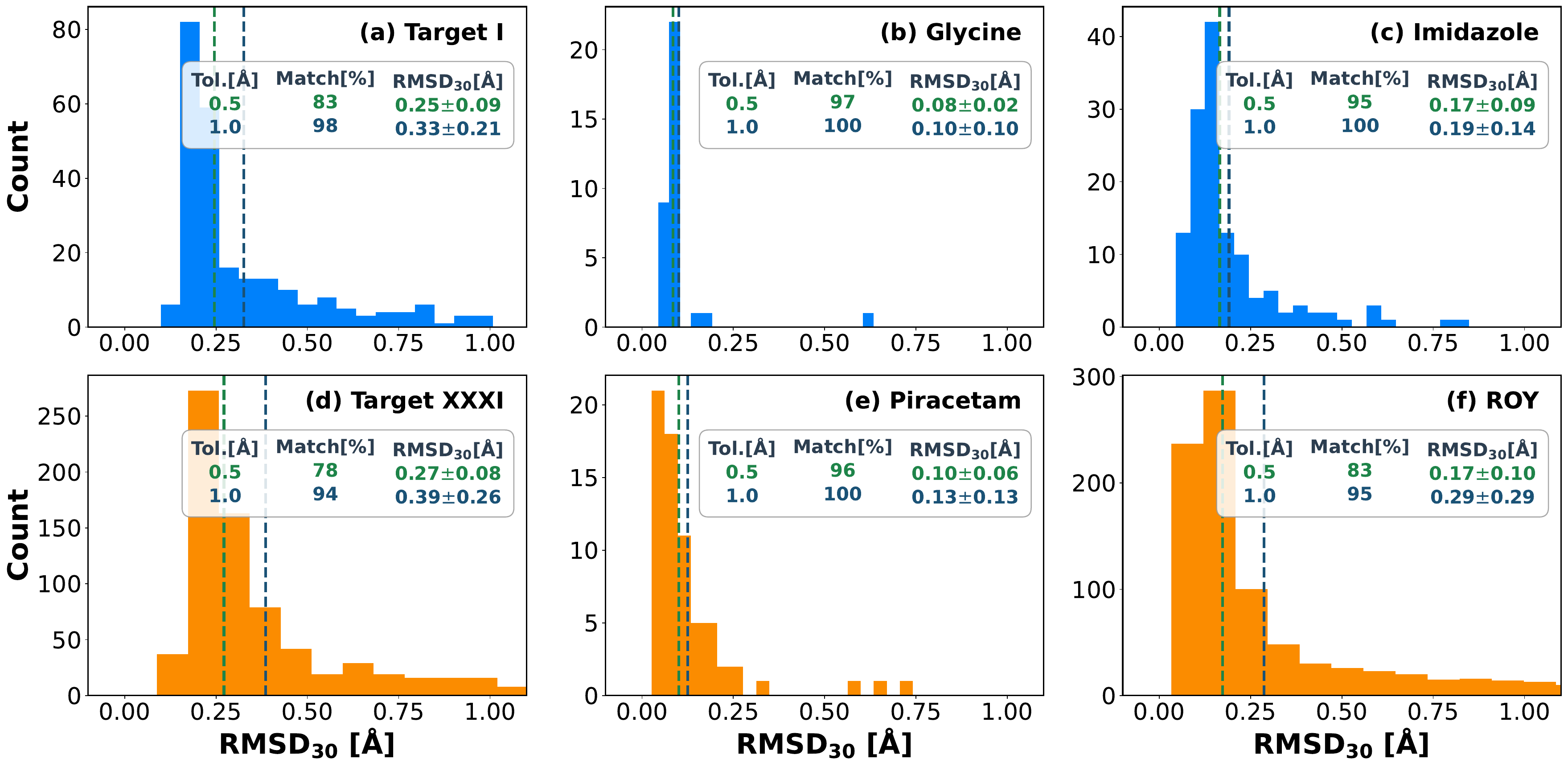}
    \caption{RMSD$_{30}$ histograms comparing the relaxed crystal structures obtained with UMA to those obtained with PBE-D3 for (a) Target I, (b) glycine, (c) imidazole, (d) Target XXXI, (e) piracetam, and (f) ROY. The match rates and mean RMSD$_{30}$ values with standard deviations for two tolerances are reported. The mean RMSD$_{30}$ positions are indicated by vertical dashed lines. These six targets were selected to cover the benchmark, including the two blind-test systems with the largest RMSD$_{30}$ values, two highly polymorphic systems (glycine and ROY), and two representative intermediate cases with near-average RMSD$_{30}$ values (imidazole and piracetam).
    }
    \label{fig:rmsd30}
\end{figure}

To assess how well the structures relaxed with UMA match the corresponding structures relaxed with PBE-D3, we calculated RMSD$_{30}$, using the COMPACK molecular overlay method~\cite{compack}, implemented in the CSD Python API~\cite{csdapi}. Each crystal structure is represented by a cluster of 30 molecules. The root mean squared deviation (RMSD) between the two molecular clusters is calculated based on the molecules that match within 35\% distance and $35^\circ$ angle tolerances, excluding hydrogen atoms. Two structures are considered as matched if all 30 molecules can be overlaid with RMSD$_{30} < 1$ \AA. This is a commonly used matching criterion that was adopted in the seventh CSP blind test~\cite{hunnisett2024seventhgen}. A stricter RMSD$_{30}<0.5$~\AA\ criterion is also reported to provide a more stringent assessment of agreement between UMA and PBE-D3 relaxed structures. 

The match rate between structures relaxed with UMA and PBE-D3 is very high with an average of 99\% across the benchmark set under the RMSD$_{30}<1$~\AA\ criterion, and remaining at 91\% even when the tolerance is a stricter RMSD$_{30}<0.5$~\AA\ criterion. A full account is provided in SI Table~\ref{tab:uma_eval_results} and Figure~\ref{fig:rmsd30_all}. Overall, the structures relaxed with UMA and PBE-D3 are very similar with an average RMSD$_{30}$ of 0.19 \AA\ across the benchmark set. Representative RMSD$_{30}$ histograms are displayed in Figure \ref{fig:rmsd30}. 

For the semi-rigid molecules subset, Target I has the highest mean RMSD$_{30}$ of 0.33~\AA, glycine has the lowest mean RMSD$_{30}$ of 0.10~\AA, and imidazole has a mean RMSD$_{30}$ of 0.19~\AA, close to the overall benchmark average. At the stricter 0.5~\AA\ threshold, the match rates for these three semi-rigid systems are 83\%, 97\%, and 95\%, respectively, increasing to 98\%, 100\%, and 100\% at the 1.0~\AA\ threshold. These results indicate that the agreement is not only robust under the conventional CSP criterion, but also remains high under a stricter comparison criterion. 

Flexible systems generally exhibit broader RMSD$_{30}$ distributions, consistent with their additional conformational degrees of freedom and greater sensitivity to small differences in the relaxation model. Among the flexible systems, Target XXXI has the highest mean RMSD$_{30}$  of 0.39~\AA\ at the 1.0~\AA\ threshold. The  match rates are 78\% and 94\% at the 0.5 and 1.0~\AA\ thresholds, respectively. Piracetam shows close agreement, with a mean RMSD$_{30}$ of 0.13~\AA\ and match rates of 96\% and 100\%. ROY has a mean RMSD$_{30}$ of 0.29~\AA\ and match rates of 83\% and 95\%. Histograms for the remaining targets are provided in SI Figure~\ref{fig:rmsd30_all}. UMA thus demonstrates excellent relaxation performance out-of-the-box for both semi-rigid and flexible molecules with various types of intermolecular interactions and packing motifs, with no system-specific fine-tuning required.

\begin{figure}[htb!]
    \centering
    \includegraphics[width=0.9\textwidth]{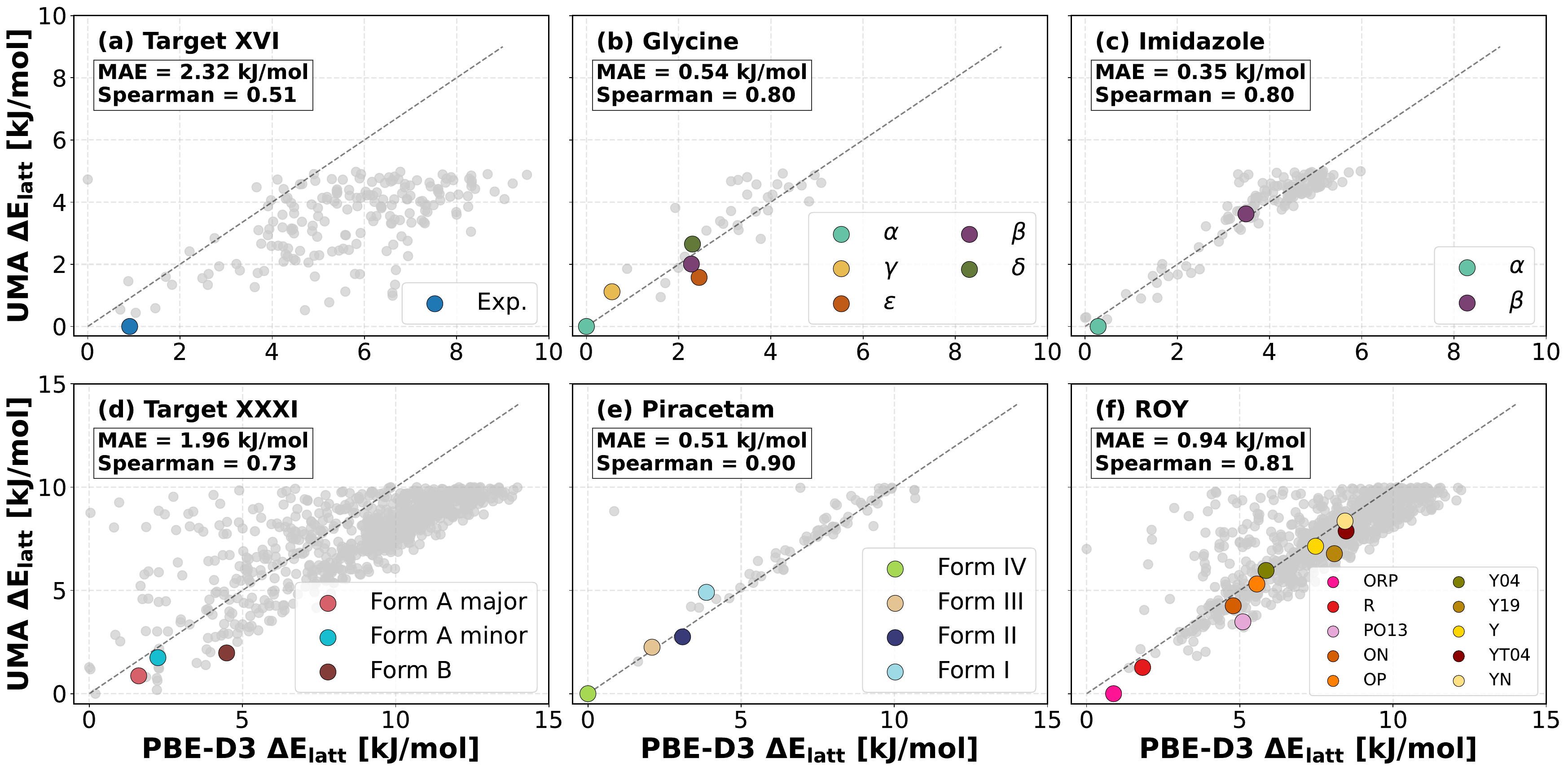}
    \caption{Parity plots comparing the relative lattice energies of structures relaxed with UMA to those relaxed with PBE-D3 for (a) Target XVI, (b) glycine, (c) imidazole, (d) Target XXXI, (e) piracetam, and (f) ROY. The relative energies are referenced to the global minimum of each method. These six targets were selected to cover the benchmark: Target XVI and Target XXXI have the largest MAEs in the benchmark, glycine and ROY serve as highly polymorphic representative systems, and imidazole and piracetam illustrate best-performing, low MAE cases.
    }
    \label{fig:opt}
\end{figure}

Figure~\ref{fig:opt} shows parity plots comparing the relative lattice energies of structures relaxed with UMA and PBE-D3. Plots for the remaining targets, as well as parity plots for single point energies from PBE-D3 obtained at the UMA-relaxed geometries, are provided in SI Figures~\ref{fig:spe_all} and~\ref{fig:opt_all}. Across both the semi-rigid and flexible subsets, UMA produces a mean absolute error (MAE) of 0.84 and 0.90 kJ/mol, respectively, with corresponding Spearman rank correlations of 0.76 and 0.85. 

Among the semi-rigid molecules subset, glycine and imidazole represent successful polymorphic cases, with MAEs of 0.54 and 0.35 kJ/mol, respectively, and Spearman correlations of 0.80 for both. In contrast, Target XVI represents the worst case with an MAE of 2.32 kJ/mol and the lowest Spearman correlation of 0.51. The experimental structure of Target XVI from the fifth blind test includes diazide-carbonyl interactions, which have also proved challenging for classical force fields and other MLIPs~\cite{bardwell2011towards, nickerson2025assessment}. Moreover, molecules with diazo functional groups comprise <0.1\% of the training set of the OMC25 dataset and the specific intermolecular interactions of Target XVI do not appear in any OMC25 training structure~\cite{omc25}. This may explain why UMA is unable to capture the unique intermolecular interactions of Target XVI. Notably, both UMA and other MLIPs~\cite{nickerson2025assessment} perform better for Target XVIII, which also contains a diazo group. This is likely because the other functional groups in Target XVIII, which are better represented in OMC25, have a more dominant contribution to the crystal packing~\cite{nickerson2025assessment}, and this reduces the sensitivity of the energy landscape to the underrepresented diazo environment.

Among the flexible molecules subset, piracetam provides the strongest example of agreement, with an MAE of 0.51 kJ/mol and a Spearman correlation of 0.90. For Target XXXI the performance of UMA is somewhat worse, with an MAE of 1.96 kJ/mol and a Spearman correlation of 0.73. The 3-sulfonyl-4,5-dihydroisoxazoline ring of Target XXXI appears in only one compound in OMC25 and the aryl-\ce{CF2}-\ce{SO2} linker that controls its conformation is entirely absent from the dataset. As with Target XVI, the underrepresented chemistry constitutes the structural core that determines both the intramolecular geometry and the dominant intermolecular contacts, which explains the reduced accuracy. 

Here, the relative energies are referenced to the global minimum of each method, which is consistent with the results presented in Figure \ref{fig:csp_rank_mp}. We note that the comparison of relative energies can be affected by the choice of reference, in particular in cases where UMA and DFT global minimum is significantly different. ROY provides a clear example of this effect. When each method is referenced to its own global minimum, UMA gives a moderate MAE of 0.94 kJ/mol and a Spearman correlation of 0.81. When the lowest-energy PBE-D3 structure is instead used as a common reference, the MAE increases to 7.26 kJ/mol. This increase reflects the different global minima predicted by UMA and PBE-D3 for ROY. Similar reference sensitivity is observed for several other systems, including Target I, Target XVI, CEBYUD, CUMJOJ, DEZDUH, imidazole, nicotinamide, 6-fluorochromone, and Target XXXI (SI~Table~\ref{tab:uma_eval_results}).
Based on the results presented here, UMA delivers excellent out-of-the-box performance in most cases, and even in the isolated cases, whose chemistry is not well-represented in the OMC25 dataset, its performance is still acceptable. Thus, UMA can be used as an effective and considerably more efficient replacement for PBE-D3 in CSP workflows.

\section{Discussion}

We have introduced FastCSP, an open-source, end-to-end workflow for molecular crystal structure prediction powered entirely by a universal machine learning interatomic potential (MLIP), the Universal Model for Atoms (UMA), without any system-specific fine-tuning or DFT calculations. The workflow integrates conformer generation and selection, random crystal structure generation via Genarris, geometry optimization, energy ranking, optional finite-temperature free energy evaluation, and an optional conformer energy correction that treats intramolecular energetics at a higher level of theory. All stages of the workflow are conducted using the same pretrained UMA checkpoint, exploiting the complementarity of its OMC task (trained on periodic molecular crystals at the PBE-D3 level) and OMol task (trained on non-periodic molecular systems and clusters at the $\omega$B97M-V level).  

We have validated FastCSP on a chemically and structurally diverse benchmark of 28 semi-rigid and 10 flexible molecules, covering 74 experimental known polymorphs. The workflow successfully generated all known experimental structures. The only generation failure under the default sampling settings was Forms~I and~IV of olanzapine, which were recovered by increasing the number of structures generated per space group. For single-polymorph systems, the experimentally observed structure is ranked as the lattice energy global minimum in 14 of 26 cases and within the top~5 in 9 additional cases, with all experimental forms falling within 5~kJ/mol of the predicted global minimum. Target~V was the only compound whose experimental structure was ranked somewhat higher by UMA, at \#17 above the global minimum based on lattice energy, however, this ranking was in close agreement with the DFT ranking of \#17. For multi-polymorph systems, one of the experimentally known polymorphs is ranked as the global minimum in 9 out of 12 cases, and at least one polymorph is within the top 5 for the remaining 3 cases. All experimentally observed polymorphs fall within 9 kJ/mol of the predicted global minimum.  

The UMA ranking is in close agreement with PBE-D3 across the benchmark, and in several cases UMA outperforms PBE-D3. Moreover, we have shown that the UMA MLIP reproduces with high fidelity the results of dispersion-inclusive DFT at the PBE-D3 level for relaxation and free energy calculations at a fraction of the computational cost. The only systems for which the relative lattice energies obtained with UMA compared to PBE-D3 are somewhat worse are Target~XVI and Target XXXI, whose chemistry is not well-represented in the OMC25 training dataset. This could be remedied by targeting similar compounds, as well as other underrepresented molecular scaffolds, for data acquisition and retraining models on an even larger dataset.

The computational efficiency of UMA enabled us to conduct large-scale free energy calculations for numerous putative structures. 
Accounting for thermal free energy corrections at 300 K via the quasi-harmonic approximation (QHA) did not systematically improve the stability ranking of experimental structures across the benchmark set. For most of the semi-rigid compounds, the free energy corrections were not beneficial and adversely affected the recall. For the flexible compounds, the free  energy corrections tended to improve the stability of experimentally observed polymorphs compared to other putative structures and improved the recall. However, the free energy corrections did not necessarily improve the stability ranking among polymorphs.  Our findings mirror the variable success of free energy corrections reported in the seventh CSP blind test. Detailed analysis performed for glycine showed that the results of UMA-OMC based phonon and free energy calculations are in agreement with the results of PBE-D3. Therefore, we attribute the non-uniform effect of free energy corrections to the limitations of PBE-D3 and/or the QHA itself.

The OMC$\to$OMol ($\Delta$UMA) conformer energy correction replaces the PBE-D3 level intramolecular energetics with $\omega$B97M-V-level monomer energies to achieve a balanced description of the intermolecular versus intramolecular contributions to the stabilization of molecular crystals. Similar to the free energy corrections, the $\Delta$UMA conformer energy corrections were overall more beneficial for the recall of the flexible molecules than the semi-rigid molecules. The improvement was found to be particularly significant in cases of conformational polymorphism, where the conformer energy is a dominant contribution to the stability ordering. One such example is ROY, a long-standing benchmark case of extreme conformational polymorphism. UMA lattice energies alone ranked the experimentally most stable Y form far from the global minimum, reflecting the well-known delocalization error of GGA-level DFT. UMA+$\Delta$UMA recovered the Y form as the global minimum and reproduced the experimental stability ordering without any system-specific retraining or post-hoc DFT calculation. However, in some cases, such as glycine, although the conformer correction achieved some improvement, an accurate description of the intermolecular interactions, such as that provided by hybrid DFT functionals with non-local dispersion corrections, is still necessary for ranking the $\gamma$ form as the global minimum. Overall, our results demonstrate that a carefully designed correction hierarchy within a single MLIP framework can address some of the fundamental limitations of the underlying DFT training data. 

\section{Conclusion}

Based on the results presented here, the accuracy of UMA obviates the need for classical force fields in early-stage screening and for DFT-based final re-ranking within CSP workflows. We have demonstrated that a single, pretrained MLIP can generalize across chemical space for compounds varying in composition, intermolecular interactions, and crystal packing, effectively replacing the multi-stage hierarchical workflows that have defined CSP practice for decades. Furthermore, the training of system-specific classical force fields or MLIPs would no longer be necessary, unless the compound of interest is substantially different than anything found in the UMA training sets.

The computational efficiency of UMA brings high-throughput CSP within practical reach.
End-to-end results for a single system can be obtained within hours to several days on tens of modern NVIDIA H200 GPUs. This was not achievable previously with DFT-based approaches or system-specific force fields or MLIPs requiring per-molecule training. The FastCSP workflow provides settings for balancing computational cost against polymorph recall (conformer count, space group coverage, deduplication stringency), enabling both exhaustive pharmaceutical-grade searches and rapid high-throughput screening campaigns.
The FastCSP workflow is general: Genarris generates structures in all space groups, and the OMC task of UMA, trained on crystal structures of 50,000 chemically diverse molecules, has excellent transferability. Where the OMC-level description is insufficient for conformational energies, the OMol task trained on 140 million finite molecular systems and clusters provides corrections at a higher level of theory without requiring additional training data for the target molecule. 
The UMA models could be improved  with targeted  data acquisition for compounds that are not well-represented in OMC25, such as metal-organic compounds and/or supplementing OMC25 with more  higher-level DFT data to more accurately capture intermolecular interactions in periodic extended systems.

The open-source release of FastCSP, together with the Genarris~3 structure generator, the UMA MLIP, and the OMC25 and OMol25 datasets, provides the community with the first fully open end-to-end CSP workflow capable of matching the accuracy of dispersion-inclusive DFT. 
We envision this framework being utilized to support the development of products and technologies based on molecular crystals, including pharmaceuticals, agrochemicals, organic electronics, and other domains where polymorph control is critical.

\FloatBarrier

\section*{Acknowledgments}
N.M. acknowledges support from the National Science Foundation (NSF) Designing Materials to Revolutionize and Engineer our Future (DMREF) program via award DMR-2323749. G.J.O.B. acknowledges support from the NSF via grant CHE-1955554. Y.Y. acknowledges support from the Frontera Computational Science Fellowship awarded by the Texas Advanced Computing Center (TACC).

\bibliographystyle{assets/naturemag}
\bibliography{paper}

\clearpage
\newpage
\beginappendix

\setcounter{figure}{0}
\renewcommand{\thefigure}{S\arabic{figure}}
\setcounter{table}{0}
\renewcommand{\thetable}{S\arabic{table}}
\setcounter{lstlisting}{0}
\renewcommand{\thelstlisting}{S\arabic{lstlisting}}

\input{appendix/dataset}
\clearpage
\newpage
\input{appendix/workflow_parameters}
\clearpage
\newpage
\input{appendix/runtime}
\clearpage
\newpage
\input{appendix/conf_eval}
\clearpage
\newpage
\input{appendix/csp_eval}
\clearpage
\newpage
\input{appendix/mlip_eval}

\end{document}

%% file: imports.tex
\usepackage{graphicx}
\usepackage{amsmath}
\usepackage[version=4]{mhchem}
\usepackage{booktabs}
\usepackage{caption}
\usepackage{longtable}
\usepackage[utf8]{inputenc} % allow utf-8 input

\usepackage{hyperref}       % hyperlinks
\usepackage{booktabs}       % professional-quality tables
\usepackage{amsfonts}       % blackboard math symbols
\usepackage[T1]{fontenc}    % use 8-bit T1 fonts
\usepackage{url}            % simple URL typesetting
\usepackage{nicefrac}       % compact symbols for 1/2, etc.
\usepackage{microtype}      % microtypography
\usepackage{xifthen}
\usepackage{multirow}
\usepackage{mciteplus}
\usepackage{epstopdf}
\usepackage{shellesc}
\usepackage{subcaption}

\usepackage{siunitx}
\usepackage{mhchem}
\usepackage{threeparttable}
\usepackage{cleveref}

\newcommand{\kindatiny}{\fontsize{6pt}{7.2pt}\selectfont}

\newlength\savewidth
\newcommand{\tablestyle}[2]{%
    \fontfamily{ptm}\selectfont%
    \let\itold\it%
    \def\it{\itold \fontfamily{ptm}\selectfont}%
    \setlength{\tabcolsep}{#1}\renewcommand{\arraystretch}{#2}\centering\kindatiny%
    \let\citeold\cite%
    \renewcommand{\cite}[1]{\normalfont\fontfamily{ptm}\selectfont\tiny\citeold{##1}}%
}

\newcolumntype{x}[1]{>{\centering\arraybackslash}p{#1pt}}
\newcolumntype{y}[1]{>{\raggedright\arraybackslash}p{#1pt}}
\newcolumntype{z}[1]{>{\raggedleft\arraybackslash}p{#1pt}}
\newcolumntype{a}{>{\centering\arraybackslash}p{16pt}}

\definecolor{c0-title-bkg}{HTML}{ffffff}
\definecolor{c0-title-text}{HTML}{000000}
\definecolor{c0-item-bkg}{HTML}{ffffff}
\definecolor{c0-item-text}{HTML}{818589}

\definecolor{c1-title-bkg}{HTML}{d1e2dd}
\definecolor{c1-title-text}{HTML}{005953}
\definecolor{c1-item-bkg}{HTML}{e6efec}
\definecolor{c1-item-text}{HTML}{2d7b6d}

\definecolor{c2-title-bkg}{HTML}{cfe1e1}
\definecolor{c2-title-text}{HTML}{005760}
\definecolor{c2-item-bkg}{HTML}{e4eeed}
\definecolor{c2-item-text}{HTML}{24797b}

\definecolor{c3-title-bkg}{HTML}{cddfe5}
\definecolor{c3-title-text}{HTML}{330704}
\definecolor{c3-item-bkg}{HTML}{facac5}
\definecolor{c3-item-text}{HTML}{330704}

\definecolor{c4-title-bkg}{HTML}{cedce8}
\definecolor{c4-title-text}{HTML}{191b1c}
\definecolor{c4-item-bkg}{HTML}{e2edf4}
\definecolor{c4-item-text}{HTML}{191b1c}

\definecolor{c5-title-bkg}{HTML}{d0d9eb}
\definecolor{c5-title-text}{HTML}{0e1834}
\definecolor{c5-item-bkg}{HTML}{d0e6f9}
\definecolor{c5-item-text}{HTML}{0e1834}

\definecolor{c6-title-bkg}{HTML}{d3d5ed}
\definecolor{c6-title-text}{HTML}{220b1b}
\definecolor{c6-item-bkg}{HTML}{f5d8f1}
\definecolor{c6-item-text}{HTML}{220b1b}

\definecolor{c7-title-bkg}{HTML}{dad1ed}
\definecolor{c7-title-text}{HTML}{1c0b14}
\definecolor{c7-item-bkg}{HTML}{d58cb5}
\definecolor{c7-item-text}{HTML}{1c0b14}

\definecolor{c8-title-bkg}{HTML}{ded1ec}
\definecolor{c8-title-text}{HTML}{633273}
\definecolor{c8-item-bkg}{HTML}{D8BFD8}
\definecolor{c8-item-text}{HTML}{7c5997}

\definecolor{c9-title-bkg}{HTML}{e5d1eb}
\definecolor{c9-title-text}{HTML}{6c2f6b}
\definecolor{c9-item-bkg}{HTML}{f0e0f6}
\definecolor{c9-item-text}{HTML}{885591}

\definecolor{c10-title-bkg}{HTML}{ebd1e7}
\definecolor{c10-title-text}{HTML}{722e5f}
\definecolor{c10-item-bkg}{HTML}{f5e2f3}
\definecolor{c10-item-text}{HTML}{915487}

\definecolor{avg-title-bkg}{HTML}{f3f3f3}
\definecolor{avg-title-text}{HTML}{000000}
\definecolor{avg-item-bkg}{HTML}{f3f3f3}
\definecolor{avg-item-text}{HTML}{000000}

\ExplSyntaxOn

\cs_generate_variant:Nn \coffin_typeset:Nnnnn {Nffff}

\NewDocumentCommand\rotbox{ O{l,H} D<>{0pt,0pt} m m}{
    \hcoffin_set:Nn \l_tmpa_coffin {#4}
    \coffin_rotate:Nn \l_tmpa_coffin {#3}
    \coffin_typeset:Nffff \l_tmpa_coffin 
        {\clist_item:nn{#1}{1}}
        {\clist_item:nn{#1}{2}}
        {\clist_item:nn{#2}{1}}
        {\clist_item:nn{#2}{2}}
}
\ExplSyntaxOff

\newlength{\ccustomlen}
\newcommand{\ccustom}[3][c0]{%
    \cellcolor{#1-item-bkg}{%
        \rotbox[l,t]{90}{%
            \parbox[t]{\ccustomlen}{%
                \ifthenelse{\isempty{#3}}{%
                    \mbox{%
                        \kindatiny\textcolor{#1-title-text}{#2}%
                    }%
                }{%
                    \kindatiny\textcolor{#1-title-text}{#2} \\%
                    \tiny{\textcolor{#1-item-text}{\it #3}}%
                }%
            }%
        }%
    }%
}

\usepackage{xspace}

\makeatother

%% file: appendix/dataset.tex
\section{Compounds considered for FastCSP benchmark}
\label{app:tier1}

\begin{table}[h!tbp]
    \small
    \centering
    \caption{Summary of the Semi-Rigid molecules subset, listing compound names, corresponding selected CSD reference codes, molecular formula, number of molecules per unit cell ($Z$), space groups, and polymorph details. All structures are measured under ambient pressure except for the two high-pressure glycine polymorphs and $\beta$ form of imidazole.}
    \label{tab:all_molecules}
    \begin{adjustbox}{max width=0.8\textwidth, max height=0.75\textheight, center}
    \begin{tabular}{@{}clllcll@{}}
    \toprule
    \textbf{No.} 
    & \textbf{Compound} 
    & \textbf{CSD Refcode} 
    & \textbf{Formula} 
    & $\boldsymbol{Z}$ 
    & \textbf{Space Group} & \textbf{Polymorph} \\
    %& \textbf{$T_{\text{exp}}$}~\cite{csd} \\
     \midrule
        % z=4 > z=8
         \multirow{2}{*}{1} & \multirow{2}{*}{Target I} & XULDUD01 &\multirow{2}{*}{\ce{C6 H6 O}} & 4 & 14 ($P2_1/c$) & Monoclinic  \\%& 293 K  \\
         & & XULDUD &  & 8 & 61 ($Pbca$) & Orthorhombic \\%& 293 K  \\
        \midrule
        2 & Target II & GUFJOG & \ce{C5 H3 N O S} & 4 & 14 ($P2_1/n$) & - \\%& 295 K  \\
        \midrule
        3 & Target IV & BOQQUT & \ce{C8 H11 N O2} & 4 & 14 ($P2_1/a$) & - \\%& 297 K \\
        \midrule
        4 & Target V & BOQWIN & \ce{C10 H14 Br N O2 S} & 4 & 19 ($P2_12_12_1$) & - \\%& 296 K \\
        \midrule
        5 & Target VIII & PAHYON01 & \ce{C3 H4 N2 O2} & 8 & 15 ($C2/c$) & - \\%& 190 K \\
        \midrule
        6 & Target XII & AXOSOW01 & \ce{C3 H4 O} & 8 & 61 ($Pbca$) & - \\%& 150 K \\
        \midrule
        7 & Target XIII & SOXLEX01 & \ce{C6 H2 Br2 Cl F} & 4 & 14 ($P2_1/c$) & - \\%& 173 K \\
        \midrule
        8 & Target XVI & OBEQUJ & \ce{C6 H4 N2 O} & 8 & 61 ($Pbca$) & - \\%& 174 K \\
        \midrule
        9 & Target XVII & OBEQOD & \ce{C6 H2 Cl2 N2 O4} & 4 & 14 ($P2_1/c$) & - \\%& 174 K \\
        \midrule
        10 & Target XXII & NACJAF & \ce{C8 N4 S3} & 4 & 14 ($P2_1/n$) & - \\%& 150 K \\
        \midrule
        11 & Acetic Acid & ACETAC03 & \ce{C2 H4 O2} & 4 & 33 ($Pna2_1$) & Orthorhombic \\%& 133 K  \\
        \midrule
        12 & Caprylolactam & CAPRYL & \ce{C8 H15 N O} & 4 & 9 ($Cc$) & - \\%& 295 K  \\
        \midrule
        13 & CEBYUD & CEBYUD & \ce{C4 N4 S} & 3 & 145 ($P3_2$) & - \\%& 295 K  \\
        \midrule
        \multirow{2}{*}{14} & \multirow{2}{*}{CILJIQ} & CILJIQ & \multirow{2}{*}{\ce{C7 H5 Br O2}} & 4 & 19 ($P2_12_12_1$) & Form I \\%& 295 K  \\
         & & CILJIQ01 & & 4 & 14 ($P2_1/c$) & Form II \\%& 293 K  \\
        \midrule
        15 & CUMJOJ & CUMJOJ & \ce{C8 H12 N2 O2} & 8 & 114 ($P\overline{4}2_1c$) & - \\%& 138 K \\
        \midrule
        16 & DEZDUH & DEZDUH & \ce{C8 H6 F6} & 4 & 33 ($Pna2_1$) & - \\%& 173 K  \\
        \midrule
        17 & Eniluracil & DOHFEM$^*$ & \ce{C6 H4 N2 O2} & 4 & 14 ($P2_1/n$) & - \\%& 150 K  \\
        \midrule
        18 & GACGAU & GACGAU & \ce{C8 H8 O3} & 8 & 61 ($Pbca$) & - \\%& 100 K  \\
        \midrule
        \multirow{5}{*}{19} & \multirow{5}{*}{Glycine} & GLYCIN16 & \multirow{5}{*}{C2 H5 N O2} & 3 & 145 ($P3_2$) & $\gamma$ \\%& 83 K \\
         & & GLYCIN20& & 4 & 14 ($P2_1/n$) & $\alpha$ \\%& 301 K  \\
         & & GLYCIN32 & & 2 & 4 ($P2_1$) & $\beta$ \\%& 150 K  \\
         & & GLYCIN67 & & 4 & 14 ($P2_1/a$) & $\delta$ [1.9 GPa]~\cite{glycine_p2} \\%& 293 K  \\
         & & GLYCIN68 & & 2 & 7 ($Pn$) & $\varepsilon$ [4.3 GPa]~\cite{glycine_p2} \\%& 293 K \\
        \midrule
        20 & GOLHIB & GOLHIB & \ce{C11 H10 N2 O} & 4 & 9 ($Cc$) & - \\%& 273 K \\
        \midrule
        21 & HURYUQ & HURYUQ & \ce{C11 H14 O3} & 1 & 1 ($P1$) & - \\%& 293 K \\
        \midrule
        22 & IHEPUG & IHEPUG & \ce{C8 H11 N O2} & 8 & 92 ($P4_12_12$) & - \\%& 293 K \\
        \midrule
        \multirow{2}{*}{23} & \multirow{2}{*}{Imidazole} & IMAZOL06 & \multirow{2}{*}{\ce{C3 H4 N2}} & 4 & 14 ($P2_1/c$) & $\alpha$ \\%& 103 K \\
         & & IMAZOL25 & & 8 & 41 ($Aba2$) & $\beta$ [1.2 GPa] ~\cite{imidazole_rank} \\%& 293 K \\
        \midrule
        24 & LECZOL & LECZOL & \ce{C12 H6 O3} & 6 & 170 ($P6_5$) & - \\%& 100 K \\
        \midrule
        % melting enthalpy, alpha > epsilon
        \multirow{3}{*}{25} & \multirow{3}{*}{Nicotinamide} &  NICOAM03 & \multirow{3}{*}{\ce{C6 H6 N2 O}} & 4 & 14 ($P2_1/c$) & $\alpha$ \\%& 295 K \\
         & & NICOAM07 & & 2 & 4 ($P2_1$) & $\varepsilon$ \\
         & & NICOAM17 & & 4 & 14 ($P2_1/c$) & $\iota$ \\%& 100 K \\
        \midrule
        26 & ROHBUL &  ROHBUL & \ce{C5 H8 N4 O2} & 8 & 61 ($Pbca$) & - \\%& 295 K \\
        \midrule
        27 & 6-Fluorochromone & UMIMIO & \ce{C9 H5 F O2} & 1 & 1 ($P1$) & - \\%& 100 K \\
        \midrule
        28 & WEXREY & WEXREY & \ce{C6 H5 N3} & 4 & 33 ($Pna2_1$) & - \\%& 295 K \\
        \midrule
    \end{tabular}
    \end{adjustbox}{}
    \begin{tablenotes}
  \item \small{{$^*$} DOHFEM is disordered over two sites in a ratio 0.74:0.26. Here we chose the major component.}
\end{tablenotes}
\end{table}

\begin{table}[h!tbp]
    \ContinuedFloat
    \small
    \centering
    \caption{\textbf{(continued)} Summary of the Flexible molecules subset, listing compound names, corresponding selected CSD reference codes, molecular formula, number of molecules per unit cell ($Z$), space groups, and polymorph details.}
    \begin{adjustbox}{max width=0.8\textwidth, max height=0.75\textheight, center}
    \begin{tabular}{@{}clllcll@{}}
    \toprule
    \textbf{No.} 
    & \textbf{Compound} 
    & \textbf{CSD Refcode} 
    & \textbf{Formula} 
    & $\boldsymbol{Z}$ 
    & \textbf{Space Group} & \textbf{Polymorph} \\
\midrule
% Form I > Form III
\multirow{2}{*}{29} & \multirow{2}{*}{Target VI} & UJIRIO & \multirow{2}{*}{\ce{C11H11N3O2S}} & 4 & 14 ($P2_1/c$) & Form I \\

 & &UJIRIO05 & & 8 & 61 ($Pbca$) & Form III \\
\midrule
30 & Target X & HAMTIZ01 & \ce{C9H9N3O5} & 4 & 14 ($P2_1/n$) & - \\
\midrule
31 & Target XIV & WIDBAO & \ce{C10H10N2S3} & 4 & 14 ($P2_1/c$) & - \\
\midrule
32 & Target XVIII & OBEQET & \ce{C9H7Cl1N2O3S} & 8 & 61 ($Pbca$) & - \\
\midrule
% Form B > Form A
\multirow{3}{*}{33} & \multirow{3}{*}{Target XXXI} & ZEHFUR & \multirow{3}{*}{\ce{C12H12F3NO3S}} & 4 & 14 ($P2_1/c$) & Form B \\
 & & ZEHFUR02$^\dagger$ &  & 4 & 14 ($P2_1/c$) & Form A maj \\
 & & ZEHFUR02$^\dagger$ & & 4 & 14 ($P2_1/c$) & Form A min \\
 
\midrule
% most stable: alpha
\multirow{6}{*}{34} & \multirow{6}{*}{Chlorpropamide} & BEDMIG08 & \multirow{7}{*}{\ce{C10H13Cl1N2O3S}} & 4 & 19 ($P2_12_12_1$) & $\alpha$ \\
 & & BEDMIG01 & & 8 & 60 ($Pbcn$) & $\beta$ \\
 & & BEDMIG19 & & 2 & 4 ($P2_1$) & $\gamma$ \\
 & & BEDMIG03 & & 8 & 61 ($Pbca$) & $\delta$ \\
 & & BEDMIG06 & & 4 & 33 ($Pna2_1$) & $\varepsilon$ \\
 & & BEDMIG05 & & 4 & 33 ($Pna2_1$) & $\varepsilon^{'}$ \\
 % & & BEDMIG20 & & 4 & 14 ($P2_1/c$) & $\zeta$ \\
\midrule
% most stable: Form I
\multirow{4}{*}{35} & \multirow{4}{*}{Olanzapine} & UNOGIN03 & \multirow{4}{*}{\ce{C17H20N4S}} & 4 & 14 ($P2_1/c$) & Form I \\
 & & UNOGIN04 & & 4 & 14 ($P2_1/c$) & Form II \\
 & & UNOGIN06 & & 4 & 14 ($P2_1/c$) & Form III \\
 & & UNOGIN05 & & 4 & 14 ($P2_1/n$) & Form IV \\
\midrule
% Form II >= Form III > Form I > Form IV
\multirow{4}{*}{36} & \multirow{4}{*}{Piracetam} & BISMEV11 &\multirow{4}{*}{\ce{C6H10N2O2}}  & 2 & 2 (\textit{P$\overline{1}$}) & Form II \\
 & & BISMEV02 & & 4 & 14 ($P2_1/n$) & Form III \\
 & & BISMEV05$^\dagger$ &  & 4 & 14 ($P2_1/n$) & Form I maj \\
 & & BISMEV04 & & 4 & 14 ($P2_1/c$) & Form IV \\
\midrule
%  Form II > Form I > Form III > Form VI,
\multirow{5}{*}{37} & \multirow{5}{*}{Piroxicam} & BIYSEH05 & \multirow{5}{*}{\ce{C15H13N3O4S}} & 4 & 14 ($P2_1/c$) & Form II ($\alpha_2$) \\
 & & BIYSEH03 & & 4 & 14 ($P2_1/c$) & Form I \\
 & & BIYSEH07 & & 2 & 2 ($P\overline{1}$) & Form III \\
 & & BIYSEH17 & & 4 & 14 ($P2_1/c$) & Form VI \\
 & & BIYSEH08 & & 4 & 29 ($Pca2_1$) & $\alpha_1$ \\
 
\midrule
% Y > YT04 > R > OP > ON > YN > ORP
\multirow{10}{*}{38} & \multirow{10}{*}{ROY} & QAXMEH01 & \multirow{10}{*}{\ce{C12H9N3O2S}} & 4 & 14 ($P2_1/n$) & Y \\
 & & QAXMEH12 & & 4 & 14 ($P2_1/n$) & YT04 \\
 & & QAXMEH81 & & 2 & 2 ($P\overline{1}$) & R \\
 & & QAXMEH03 & & 4 & 14 ($P2_1/n$) & OP \\
 & & QAXMEH32 &  & 4 & 14 ($P2_1/c$) & ON \\
 & & QAXMEH04 & & 2 & 2 ($P\overline{1}$) & YN \\
 & & QAXMEH05 & & 8 & 61 ($Pbca$) & ORP \\
 & & QAXMEH59 & & 4 & 14 ($P2_1/c$) & PO13 \\
 & & QAXMEH53 & & 2 & 2 ($P\overline{1}$) & Y04 \\
 & & QAXMEH60 & & 4 & 14 ($P2_1/c$) & Y19 \\
\midrule
\end{tabular}
\end{adjustbox}{}
\begin{tablenotes}
\item \small{{ {$^\dagger$} "maj" and "min" refer to the major and minor components of disorder, respectively.}}
\end{tablenotes}
\end{table}

%% file: appendix/workflow_parameters.tex
\section{Workflow Setup and Parameters}
\label{app:hyperparams}

\textbf{Conformer Generation.} 
Conformers were generated by an approach inspired by the multi-sampler RDKit pipeline of~\cite{zhou_deep_2023}, with two modifications: the energy (force-field-optimized) sampler is split into two embedding-initialization variants, and the final clustering selection is replaced by Butina clustering~\cite{butina1999unsupervised} with an energy gate (described below). All passes use \textsc{RDKit}~\cite{rdkit} with ETKDGv3 distance-geometry embedding~\cite{etkdg,etkdgv3} (chirality enforced and small-ring torsion preferences enabled). For a target pool size $N=500$, we request $\lceil N/2.5 \rceil$ conformers via ETKDGv3 embedding followed by MMFF force-field optimization~\cite{mmff94}, $\lceil N/2.5 \rceil$ via the same embedding initialized from random Cartesian coordinates followed by MMFF optimization, $\lceil N/10 \rceil$ via ETKDGv3 embedding without force-field optimization, and $\lceil N/10 \rceil$ by uniformly sampling each rotatable torsion in half-open $[0, 2\pi]$ range from a single ETKDGv3-embedded reference. Across all four passes, conformers with any pair of atoms closer than 0.6~\AA, and conformers whose RDKit-inferred covalent connectivity (including hydrogens) differs from that of the input SMILES, are discarded. The remaining conformers are clustered with the Butina algorithm using all-atom RMSD at a 0.25~\AA\ threshold. The cluster representative is the member with the most neighbors within the threshold (i.e.\ the conformer which is maximally similar to other conformers in that region of conformational space). At this stage the clustering is purely geometric.

Each retained representative is then relaxed with the OMol task of UMA-Small v1.1~\cite{uma,omol} using the formal charge of the input SMILES and spin multiplicity 1 (closed shell), via ASE's~\cite{ase} BFGS optimizer with a force-convergence threshold of 0.05~eV/\AA\ and a hard cap of 100 steps. Conformers whose covalent connectivity changes during relaxation are discarded. Conformers whose stereochemistry inverts relative to the input SMILES are tagged in the output. In the next clustering step, these stereo-flipped conformers are deprioritized: within each Butina cluster the first stereo-correct member is chosen as the representative, and the original Butina centroid is retained only if every member of the cluster is stereo-flipped. The relaxed pool is then re-clustered by Butina with the same 0.25~\AA\ RMSD threshold combined with a within-cluster energy gate of 1.5~kJ/mol (pairs separated by $\geq$ 1.5~kJ/mol are not merged), and finally truncated to a 40~kJ/mol window above the lowest-energy survivor (higher than the thresholds discussed in~\cite{thompson2014conformations, chattopadhyay2025lattice}).

This results in an energy-sorted list of optimized conformers (in general much fewer than the original $N$ requested). From the relaxed energy-sorted pool we forward the lowest-energy conformers per target to structure generation: one for the Semi-Rigid subset, twenty for the Flexible subset, and five for nicotinamide because of its rotatable amide group. If at least one relaxed conformer survives all filters, the unrelaxed pool is never used to top up the relaxed selection, even when fewer than the requested number remain. The unrelaxed pool is used as a fallback only when the relaxed pool is completely empty. Glycine is the one molecule in our benchmark where this fallback path is triggered: starting from its zwitterionic SMILES, every UMA relaxation proton-transfers to neutral glycine (the expected gas-phase form~\cite{glycine_exp_data}), fails the connectivity check, and is dropped. With an empty relaxed pool, the workflow automatically substitutes the unrelaxed RDKit-generated geometries (ranked by their UMA-OMol single-point energy) for the single lowest-energy template handed to structure generation.

\textbf{Structure Generation.} 
For structure generation, we used the Genarris 3 package~\cite{genarrisv3}. Genarris 3 automatically identifies all space groups compatible with the molecular point group symmetry of each selected conformer and the requested number of molecular formula units per unit cell ($Z \in \{1, 2, 3, 4, 6, 8\}$), including configurations in which molecules occupy special Wyckoff positions ($Z' \leq 1$). The Semi-Rigid subset included all compatible space groups, generating 500 structures per (conformer, $Z$, space group) combination. To balance computational costs, the Flexible molecules subset targeted only the top 10 most prevalent organic space groups in the CSD~\cite{csd} (representing over 90\% of organic crystals), yielding 1,500 structures per combination. This top-10 restriction was expanded for piroxicam to include $Pca2_{1}$ (No. 29, the 11th most frequent space group in CSD) and for chlorpropamide to include $Pbcn$ (No. 60, the 13th most frequent space group). This addition was necessary because no top-10 space group produced a match for one of the experimental polymorphs of these systems. Consequently, a strict top-10 limit would have automatically resulted in two omitted structures. The larger pool for the Flexible subset reflects both the larger accessible packing space per conformer and the empirical observation that the more pronounced packing ambiguities in flexible molecules require more sampling to recover the experimental motifs. Note that when \texttt{spg\_distribution\_type = "standard"} is used for the Semi-Rigid molecules, if the molecular point group admits a special Wyckoff position within a given space group, the number of generated structures is reduced by the ratio of the special-position multiplicity to the requested $Z$, so the per-space-group count is no longer exactly the requested 500. When several Wyckoff positions are simultaneously compatible, Genarris distributes the structures evenly across them. The target unit cell volume was determined using the Genarris 3 default setting~\cite{bier2020machine}, scaling the estimated volume by a factor of 1.5 to ensure sufficient placement space. Structure generation proceeded until the desired count per space group was reached. The generated structures were then compressed under the symmetry of their parent space group using the Rigid Press feature of Genarris 3, with a BFGS optimizer, a convergence tolerance of 0.01, and a contact-distance gate of 0.85. Listing~\ref{lst:genarris_config} shows the full Genarris configuration file used for our experiments.\\

\begin{adjustwidth}{0.5in}{0.5in}
\begin{lstlisting}[language=ini, caption={The base Genarris 3 configuration file used for random molecular crystal generation that is run for each conformer and $Z$ value.}, label={lst:genarris_config}, stringstyle=\color{black}]
(*\textcolor{blue}{[master]}*) 
name = <target name>
molecule_path = [<path to conformer geometry file>]
z = <selected Z value>
log_level = info

(*\textcolor{blue}{[workflow]}*) 
tasks = ['generation', 'symm_rigid_press']

(*\textcolor{blue}{[generation]}*) 
num_structures_per_spg = <500 for Semi-Rigid molecules and 1500 for Flexible molecules>
spg_distribution_type = <"standard" for Semi-Rigid and Top-10 organic CSD + experimental space groups for Flexible molecules>
sr = 0.95
max_attempts_per_spg = 100000000
tol = 0.01
unit_cell_volume_mean = predict
volume_mult = 1.5
max_attempts_per_volume = 10000000
generation_type = crystal
natural_cutoff_mult = 1.2

(*\textcolor{blue}{[symm\_rigid\_press]}*) 
sr = 0.85
method = BFGS
tol = 0.01
natural_cutoff_mult = 1.2
debug_flag = False
\end{lstlisting}
\end{adjustwidth}

To reduce redundancy before relaxation, we deduplicated the generated pool using \textsc{Pymatgen}'s \textsc{StructureMatcher}~\cite{ong2013python} with relatively loose tolerances: a fractional length tolerance (\texttt{ltol}) of $0.3$, a site tolerance (\texttt{stol}) of $0.4$, an angle tolerance (\texttt{angle\_tol}) of $5^{\circ}$, and excluding hydrogen atoms. Structures are first bucketed by the triple (conformer, $Z$, generated space group), so that two structures coming from different conformers, different $Z$ values, or different parent space groups are never compared at this stage. Within each bucket, we apply a greedy clustering algorithm: each structure is tested in turn against existing cluster representatives, and any structure that matches an existing representative is added to that cluster. Note that this clustering is order-dependent and not strictly transitive: if $A \leftrightarrow B$ and $A \leftrightarrow C$ but not $B \leftrightarrow C$, the three structures are nonetheless grouped together as long as $A$ is encountered first. To further manage computational cost we skip \textsc{StructureMatcher} entirely for any pair of structures whose densities differ by more than $0.05$~g/cm$^{3}$, or whose sorted Niggli-reduced lattice constants differ by more than \texttt{ltol} or \texttt{angle\_tol}. Within each cluster the structure whose density is closest to the cluster median is retained as the representative. On average this stage removes about 67.0$\pm$11.0\% of the generated structures for the Semi-Rigid molecules and 42.6$\pm$5.3\% for the Flexible molecules, substantially reducing the relaxation cost downstream.

\textbf{MLIP Relaxations.} 
All geometry relaxations were performed using the OMC task of UMA-Small version 1.1~\cite{uma}. Relaxations were carried out with the BFGS optimizer implemented in \textsc{ASE}~\cite{ase} and a force convergence threshold of $0.01$~eV/\AA. Both atomic positions and lattice parameters were optimized using \textsc{ASE}'s \textsc{FrechetCellFilter} class without enforcing any symmetry constraints. Each optimization was capped at 1,000 steps, and structures that did not converge (7.8$\pm$3.1\% for the Semi-Rigid molecules and 3.5$\pm$1.4\% for the Flexible molecules) or whose intramolecular covalent connectivity changed during relaxation (2.0$\pm$4.8\% for the Semi-Rigid molecules and 0.1$\pm$0.1\% for the Flexible molecules) were discarded.

For downstream analysis we retained, for every target, all unique polymorphs within $10$~kJ/mol of the global minimum, with the exception of the free energy calculations and the DFT re-ranking validation for the Semi-Rigid molecules subset, for which we used a tighter window of $5$~kJ/mol. We also constrained the density to the range $0.5$--$3$~g/cm$^{3}$. Following this energy and density filtering, we applied a second round of structure deduplication with tighter matching criteria (\texttt{ltol} $= 0.2$, \texttt{stol} $= 0.3$, \texttt{angle\_tol} $= 5^{\circ}$, and excluding hydrogen atoms). At this point, to find all unique structures regardless of provenance, we pool all structures of a given target together --- no per-(conformer, $Z$, space group) buckets --- and apply the same greedy matching as above. To manage computational cost we restrict the structural fit to pairs that differ by at most $0.1$~g/cm$^{3}$ in density and $2$~kJ/mol per molecule in relaxed energy. Dropping these prefilters and performing an all-to-all comparison removes only fewer than 2\% additional duplicates at substantially higher cost. Within each cluster the lowest-energy structure is retained as the representative. We further verified that the structures surviving second-round deduplication on the $5$~kJ/mol pool for the Semi-Rigid subset are not strictly a subset of those surviving on the $10$~kJ/mol pool, owing to the non-determinism of the greedy matching, with the differences, however, being minuscule ($<$ 3\%).

\textbf{MLIP Free Energy Calculations.}
Harmonic free energy calculations were performed using the finite displacement supercell method implemented in \textsc{Phonopy}~\cite{phonopy-phono3py-JPCM, phonopy-phono3py-JPSJ}. Force constants were calculated using supercells of relaxed structures with lattice vectors of at least 15~\AA\ and atom displacements of 0.01~\AA. $\Gamma$-Centered $q$ meshes, with the number of points $n$ along each reciprocal-lattice direction chosen so that $n \times a \geq 100$~\AA, where $a$ is the lattice parameter, were used to compute the phonon density of states and thermal properties, including Helmholtz free energy, entropy, and heat capacity at constant volume. Calculations were performed at temperature intervals from 0 to 500 K, in increments of 10 K.

Quasi-harmonic free energy calculations were carried out by performing harmonic free energy calculations over seven fixed volumes, ranging between $\pm$6\% of the 0~K equilibrium volume. For each volume, volume-constrained geometry relaxations of the atomic positions and cell shape were conducted with a convergence criterion of 0.01~eV/\AA, allowing up to 1,000 relaxation steps. The Helmholtz free energies $F(T, V)$ computed from the seven harmonic calculations at different volumes $V$ were used to fit a Vinet equation of state~\cite{vinet1989universal} to determine the equilibrium volume at each temperature. Then the Gibbs free energy $G(T, P)$ was approximated by Legendre transformation,
\begin{equation}
    G(T, P) = \min_{V} \left\{F(T, V) + PV\right\}\ ,
\end{equation}
where $F(T, V)$ is the Helmholtz free energy, $V$ is the cell volume, $T$ is temperature, and $P$ is pressure.

Finally, the heat capacity at constant pressure ($C_P$) was estimated using the following thermodynamic relation,
\begin{equation}
    C_P = C_V + \alpha^2B_TVT\ ,
\end{equation}
where $C_V$ is the heat capacity at constant volume, $\alpha$ is the thermal expansion coefficient, and $B_T$ is the isothermal bulk modulus~\cite{callen1985}.

\textbf{Conformer Energy Corrections.} 
For re-ranking we apply a single-point correction that retains the OMC task for the intermolecular contribution to the lattice energy while substituting the OMol task for the intramolecular conformer contribution. We first identify the $Z$ molecular fragments within each unit cell, unwrapped under periodic boundary conditions. Every fragment was validated against the expected molecular formula and atom count, and all were valid in our runs. Single-point energies of each fragment were then evaluated with both the OMC and OMol tasks at the same checkpoint (\texttt{charge}~$=0$ and \texttt{spin}~$=1$ for the OMol task). The corrected crystal energy is then
\begin{equation}
    E^{\text{UMA+$\Delta$UMA}}_{\text{crystal}}
   \;=\; E^{\text{UMA-OMC}}_{\text{crystal}}
        \;-\; \sum_{i=1}^{Z} \left(E^{\text{UMA-OMC}}_{\text{conformer}_i}
        \;-\; E^{\text{UMA-OMol}}_{\text{conformer}_i}\right)\ ,
\end{equation}
i.e.\ the OMC molecular self-energies are replaced by the corresponding
OMol single-molecule energies, leaving the OMC intermolecular contribution intact.

\textbf{Structure Matching Criteria.}
To assess the matching criteria between the predicted and experimental crystal structures, we applied the COMPACK crystal packing similarity method~\cite{compack}, implemented in the CSD Python API~\cite{csdapi}. Following the conventions of the seventh CSP blind test~\cite{hunnisett2024seventhgen, hunnisett2024seventhrank}, the similarity measure, RMSD$_{30}$, is the root mean squared deviation calculated from the atomic positions of matched molecules between two clusters of 30 molecules. Matching molecules were identified using 35\% distance and 35$^\circ$ angle tolerances, excluding hydrogen atoms and not allowing molecular differences. Two crystal structures were considered matching if all 30 molecules could be overlaid with RMSD$_{30}$ < 1.0~\AA. Where multiple predicted structures matched the same experimentally observed polymorph, we report only the one with the lowest relative (0~K) energy according to UMA-S-1.1 (OMC). Conversely, where a single predicted structure matched multiple experimental polymorphs, we report the match with the experimental polymorph with the lowest RMSD$_{30}$.

\textbf{DFT Reranking for Validation.}
We reranked the UMA-relaxed structures with dispersion-inclusive DFT to benchmark the accuracy of the ML-based energy and rank predictions. All DFT calculations used the same computational settings as those employed in the generation of the OMC25 dataset~\cite{omc25}, and we refer the reader to the OMC25 paper for full methodological details. Briefly, we used VASP version 6.3~\cite{vasp1,vasp2,vasp3} with the PBE exchange-correlation functional~\cite{pbe} combined with Grimme's D3 dispersion correction (zero-damping)~\cite{dftd3} and the PBE pseudopotentials distributed with VASP 5.4. Input sets were created using \textsc{RelaxSetGenerator} class from \textsc{Atomate2}~\cite{ganose2025atomate2}. The atomic positions and lattice vectors were relaxed until the maximum per-atom residual force fell below 0.001~eV/\AA, or the relaxation exceeded 1,500 steps. The total-energy convergence tolerance was set to 10$^{-6}$~eV, the plane-wave energy cut-off was fixed at 520~eV with an augmentation cut-off of 1360~eV, and $k$-point meshes were generated automatically with a reciprocal grid density of 64 $k$-points per \AA$^{-3}$ using the $\Gamma$-centered scheme.

\textbf{DFT Validation for Glycine.} We performed DFT free energy calculations on five matched polymorphs of glycine using FHI-aims (version 240920\_2)~\cite{fhiaims1, fhiaims2, fhiaims3} using the Tier 1 basis sets and \textit{light} species defaults of FHI-aims. The number of k-points ($n$) was determined by $n\times a \geq 25$\ \AA, where $a$ is the lattice parameter in a given direction. Geometry relaxations were performed using the trust radius enhanced variant of the BFGS algorithm as implemented in FHI-aims, with a force convergence criterion of 0.01 eV/\AA. Harmonic free energy calculations were performed at PBE-D3 level using the finite displacement supercell method implemented in \textsc{Phonopy}~\cite{phonopy-phono3py-JPCM, phonopy-phono3py-JPSJ}. Force constants were calculated using $2\times2\times2$ supercells of relaxed structures with atom displacements of 0.01 \AA. $\Gamma$-Centered $q$-meshes were used, with the number of points along each direction chosen so that $n \times a \geq 100$ \AA. Calculations were performed at temperature intervals from 0 to 500 K, in increments of 10 K.

Quasi-harmonic free energy calculations were carried out by performing harmonic free energy calculations over six fixed volumes, ranging from -6\% to +10\% of the 0~K equilibrium volume. For each volume, volume-constrained geometry relaxations were conducted with a convergence criterion of 0.01 eV/\AA. The Helmholtz free energies $F(T, V)$ computed from the six harmonic calculations at different volumes $V$ were used to fit a Vinet equation of state following the procedure outlined above for the MLIP free energy calculations.

The heat capacity at constant pressure ($C_P$) can be calculated by
\begin{equation}
    C_P(T, P) = -T\frac{\partial^2 G(T, P)}{\partial T^2}
\end{equation}

The five predicted polymorphs of glycine were further re-optimized with PBE, combined with the many-body dispersion (MBD) correction method~\cite{mbd1, mbd2, libmbd} using the Tier 2 basis sets and \textit{tight} species defaults. These structures were then re-ranked based on single-point energy evaluations at the PBE-based hybrid functional, PBE0~\cite{pbe0}, paired with the MBD method using the Tier 2 basis sets and \textit{intermediate} species defaults of FHI-aims.

%% file: appendix/runtime.tex
\section{Computational Cost and Scalability}
\label{app:si_runtime}

Perhaps the most striking aspect of FastCSP is not any single timing number, but a qualitative shift in what is now computationally accessible. Accurate molecular CSP has historically required either dispersion-inclusive DFT itself---prohibitively expensive for the thousands of candidate structures generated during a search---or the training of system-specific MLIPs for each molecule of interest, which demands considerable expertise, DFT data generation, and per-target computational investment~\cite{hunnisett2024seventhgen}. In practice, the field has relied on multi-stage hierarchical workflows that combine classical force fields for initial screening with DFT for final re-ranking. As a result, CSP campaigns have remained the domain of specialized groups with access to supercomputing resources and bespoke in-house workflows.

The use of a single pretrained MLIP for all stages of the workflow yields substantial computational savings compared to traditional DFT-based or system-specific MLIP approaches. Here we break down the cost of each stage, provide detailed per-structure timing benchmarks, and contextualize these against the equivalent DFT cost.

\textbf{Stage-by-stage costs.} Conformer generation and selection (RDKit sampling, UMA-OMol relaxation, and clustering) is a negligible fraction of the total budget, completing in minutes to tens of minutes per target using a single GPU. Crystal structure generation with Genarris~3 and Rigid Press consumes on the order of 100 CPU-core$\cdot$h per conformer in our study, depending on the number of compatible space groups and $Z$ values sampled.
The dominant cost is MLIP relaxation, which accounts for most of the GPU-hour budget. Among the optional stages, conformer energy corrections are inexpensive, requiring only single-point UMA-OMC and UMA-OMol evaluations on each of the $Z$ constituent molecules extracted from each relaxed crystal.
Free energy calculations within the quasi-harmonic approximation are considerably more expensive, as they require phonon calculations on supercells at multiple volumes for each retained candidate structure. Detailed per-system runtime breakdowns are provided in SI~Table~\ref{tab:runtime}.

\begin{table}[h!b]{
\centering
\caption{Runtime analysis of the FastCSP workflow across all evaluated systems. For the Semi-Rigid molecules (No. 1–28), a single conformer was used, with the exception of nicotinamide which used 5 conformers. For Flexible molecules (No. 29–38), 20 conformers were used, except for Target XIV which used 13. All computations were executed on a cluster equipped with NVIDIA V100 (16 GB and 32 GB) GPUs.}
\label{tab:runtime}
\begin{adjustbox}{max width=\textwidth, max height=\textheight, center}
\begin{tabular}{@{}cl cccccccccc @{}}
\toprule
\multirow{2}{*}{\textbf{No.}} & \multirow{2}{*}{\textbf{Compound}} & \multicolumn{2}{c}{\textbf{Conformer generation}} & \multicolumn{2}{c}{\textbf{Structure generation}} & \multicolumn{2}{c}{\textbf{Structure relaxations}} & \multicolumn{2}{c}{\textbf{Conformer energy corrections}} & \multicolumn{2}{c}{\textbf{Free energy evaluations}} \\
\cmidrule(lr){3-4} \cmidrule(lr){5-6} \cmidrule(lr){7-8} \cmidrule(lr){9-10} \cmidrule(l){11-12}
 &  & \textbf{GPU$\cdot$min} & \textbf{\# conformers} & \textbf{CPU-core$\cdot$hour} & \textbf{\# structures} & \textbf{\# structures} & \textbf{GPU$\cdot$hour} & \textbf{\# structures} & \textbf{GPU$\cdot$min} & \textbf{\# structures} & \textbf{GPU$\cdot$hour} \\
\midrule
1 & Target I & 1 & 1 & 203 & 86500 & 13171 & 309 & 1929 & 43 & 251 & 84 \\ \midrule
2 & Target II & 1 & 2 & 156 & 56007 & 16617 & 626 & 860 & 20 & 100 & 22 \\ \midrule
3 & Target IV & 3 & 3 & 208 & 53860 & 22083 & 922 & 202 & 11 & 30 & 28 \\ \midrule
4 & Target V & 8 & 1 & 348 & 52540 & 29166 & 1351 & 117 & 6 & 17 & 12 \\ \midrule
5 & Target VIII & 1 & 1 & 146 & 56025 & 15868 & 593 & 292 & 7 & 46 & 18 \\ \midrule
6 & Target XII & 1 & 3 & 104 & 57000 & 10176 & 330 & 2326 & 62 & 436 & 88 \\ \midrule
7 & Target XIII & 1 & 1 & 173 & 95215 & 29621 & 952 & 939 & 19 & 80 & 16 \\ \midrule
8 & Target XVI & 1 & 1 & 213 & 63666 & 18530 & 764 & 1343 & 33 & 217 & 67 \\ \midrule
9 & Target XVII & 1 & 4 & 168 & 56001 & 22222 & 735 & 693 & 15 & 62 & 31 \\ \midrule
10 & Target XXII & 1 & 3 & 204 & 53815 & 20531 & 722 & 189 & 3 & 8 & 2 \\ \midrule
11 & Acetic Acid & 1 & 4 & 91 & 57001 & 10115 & 335 & 909 & 26 & 331 & 78 \\ \midrule
12 & Caprylolactam & 9 & 28 & 270 & 55843 & 25188 & 1030 & 392 & 18 & 49 & 57 \\ \midrule
13 & CEBYUD & 1 & 1 & 212 & 66791 & 15704 & 595 & 845 & 22 & 63 & 10 \\ \midrule
14 & CILJIQ & 3 & 4 & 153 & 56510 & 21538 & 745 & 250 & 5 & 29 & 8 \\ \midrule
15 & CUMJOJ & 4 & 8 & 313 & 54018 & 26419 & 1189 & 214 & 9 & 44 & 41 \\ \midrule
16 & DEZDUH & 1 & 1 & 200 & 54949 & 20448 & 557 & 1512 & 36 & 164 & 64 \\ \midrule
17 & Eniluracil & 1 & 1 & 313 & 63584 & 18822 & 802 & 121 & 3 & 27 & 12 \\ \midrule
18 & GACGAU & 1 & 2 & 218 & 55104 & 18619 & 642 & 292 & 7 & 22 & 15 \\ \midrule
19 & Glycine & 2 & 9 & 124 & 52029 & 11803 & 373 & 289 & 6 & 34 & 9 \\ \midrule
20 & GOLHIB & 2 & 1 & 336 & 53198 & 25918 & 1110 & 201 & 10 & 17 & 8 \\ \midrule
21 & HURYUQ & 2 & 2 & 383 & 53489 & 28182 & 1193 & 215 & 8 & 25 & 12 \\ \midrule
22 & IHEPUG & 1 & 1 & 246 & 53225 & 19524 & 840 & 337 & 17 & 46 & 28 \\ \midrule
23 & Imidazole & 1 & 1 & 196 & 67773 & 10056 & 352 & 1115 & 30 & 134 & 36 \\ \midrule
24 & LECZOL & 1 & 1 & 339 & 68456 & 15202 & 530 & 790 & 41 & 196 & 88 \\ \midrule
25 & Nicotinamide & 4 & 5 & 867 & 280035 & 102125 & 4105 & 732 & 17 & 47 & 22 \\ \midrule
26 & ROHBUL & 1 & 1 & 220 & 54489 & 15881 & 664 & 16 & 0 & 13 & 6 \\ \midrule
27 & 6-Fluorochromone & 1 & 1 & 323 & 64803 & 21397 & 829 & 636 & 16 & 126 & 33 \\ \midrule
28 & WEXREY & 1 & 1 & 294 & 66726 & 19463 & 816 & 568 & 17 & 63 & 23 \\ \midrule
29 & Target VI & 10 & 46 & 1950& 300000& 163368& 7758& 31 & 2 & 31 & 30 \\ \midrule
30 & Target X & 20 & 45 & 1551& 300000& 176172& 7558& 438 & 16 & 438 & 232 \\ \midrule
31 & Target XIV & 10 & 13 & 937 & 195000& 108545& 4088& 26 & 1 & 26 & 16 \\ \midrule
32 & Target XVIII & 18 & 52 & 1292& 300000& 155487& 6489& 403 & 17 & 403 & 184 \\ \midrule
33 & Target XXXI & 22 & 91 & 2333& 300000& 188767& 11848& 893 & 31 & 893 & 668 \\ \midrule 
34 & Chlorpropamide & 17 & 59 & 2374& 330000& 200913& 10299& 188 & 8 & 188 & 141 \\ \midrule
35 & Olanzapine & 26 & 48 & 3691& 300000& 189722& 10788& 8 & 0.4 & 8 & 12 \\ \midrule
36 & Piracetam & 23 & 45 & 973 & 300000& 143033& 5516& 70 & 1 & 70 & 35 \\ \midrule
37 & Piroxicam & 23 & 28 & 3720& 330000& 208956& 9106& 51 & 2 & 51 & 43 \\ \midrule
38 & ROY & 16 & 59 & 1825& 300000& 167322& 7758& 1037& 41 & 1037& 677 \\ \midrule
\end{tabular}
\end{adjustbox}
}
\end{table}

\textbf{MLIP relaxation cost.} All structure relaxations in this work were performed using UMA-Small v1.1 with the OMC task in default inference mode on NVIDIA V100 (16~GB and 32~GB) GPUs. 
The inference time per BFGS step scales roughly linearly with the number of atoms in the unit cell, ranging from 50--150~ms per step (7--19 steps/s) for small cells ($Z\!=\!1$--$2$) to 300--650~ms per step (1.5--3 steps/s) for the largest cells in our set ($Z = 8$ of the 10 flexible molecules, up to ${\sim}$340 atoms). 
The averaged inference cost is ${\sim}$370~ms per step (${\sim}$2.7 steps/s), reflecting that most steps in our campaign came from large-$Z$ flexible-molecule structures. 
Equivalently, the throughput averages ${\sim}$280 atoms/s.
The mean relaxation time is 138~s per structure (${\approx}$~2.3~min) for the semi-rigid molecules and 172~s per structure (${\approx}$~2.9~min) for the flexible molecules.

\textbf{Comparison to DFT.} To place the MLIP cost in context, we computed the ratio of DFT compute cost (CPU-core$\cdot$h, using VASP~6.3 on 16--32 CPU cores) to MLIP compute cost (GPU$\cdot$h on a single V100) for the same atomic configurations. Across our DFT relaxations, a single full-relaxation-averaged DFT ionic step is on average ${\sim}$5,600$\times$ more expensive than the corresponding MLIP step (CPU-core$\cdot$h per GPU$\cdot$h; median: 3,000$\times$; 10th--90th percentile: 1,500--13,000$\times$). The initial DFT single-point calculation at the ML-relaxed geometry (VASP's first ionic iteration, starting from scratch and converging the electronic self-consistent field (SCF) to \texttt{EDIFF}~$= 10^{-6}$) is on average ${\sim}$22,000$\times$ more expensive (median: 16,000$\times$; 10th--90th percentile: 7,000--42,000$\times$).

Extrapolating from these per-step measurements to the full CSP campaign---2.3$\times$10$^{6}$ structure relaxations and ${\sim}$10$^{9}$ total BFGS steps---the MLIP relaxations consumed ${\sim}$10$^{5}$~GPU$\cdot$h, split roughly 1:3 between the two molecule subsets: 2.4$\times$10$^{4}$~GPU$\cdot$h for the 28 semi-rigid molecules and 8$\times$10$^{4}$~GPU$\cdot$h for the 10 flexible molecules, which required denser sampling (roughly 3$\times$ the aggregate cost of the semi-rigid subset and ${\sim}$10$\times$ per molecule).
Reproducing the same trajectories with DFT would require 1.9$\times$10$^{8}$~CPU-core$\cdot$h (22,000 CPU-core$\cdot$years) for the 28 semi-rigid molecules and 1.5$\times$10$^{9}$~CPU-core$\cdot$h (170,000~CPU-core$\cdot$years) for the 10 flexible molecules, for a combined 1.6$\times$10$^{9}$~CPU-core$\cdot$h (192,000~CPU-core$\cdot$years)---dominated (88\%) by the flexible molecules. This places a DFT-based CSP campaign of comparable scope out of reach for essentially all academic and industrial compute budgets.

\textbf{Hardware scaling and throughput optimization.} The runtimes reported here were obtained on NVIDIA V100 GPUs using the default inference settings. It is important to distinguish between two types of speedup.

\textit{Per-structure runtime} (wall-clock time for a single sequential BFGS relaxation). Because BFGS is inherently sequential, the runtime of a single relaxation can only be reduced by decreasing the time per inference step. Transitioning from a V100 to a current-generation NVIDIA H200 GPU increases inference speed under identical inference settings.

\textit{Effective throughput} (amortized GPU-time per structure when processing many structures). Parallelized batch inference---evaluating multiple independent structures simultaneously on a single GPU---does not reduce the runtime of any individual relaxation, but dramatically increases the number of structures processed per GPU$\cdot$h by better utilizing GPU compute capacity. 
The GPU is not fully utilized during sequential inference, especially for the smallest unit cells in our pool (Z = 1--2 of the semi-rigid molecules), and batching recovers this idle capacity.

Combined with the per-step hardware speedup, the overall throughput improvement is approximately 6$\times$ on an NVIDIA H200 GPU with parallelized batch inference relative to V100 sequential processing (estimated from typical GPU utilization patterns).
Additional gains are accessible through model compilation and mixed-precision inference~\cite{uma}. These engineering optimizations do not alter the model weights or accuracy but they only improve throughput.

\textbf{Cost control knobs.} The primary knobs for controlling computational cost are the number of conformers forwarded to structure generation, the number of space groups and $Z$ values sampled (for instance, excluding $Z = 8$ eliminates roughly half the generated structures and, because these are also the most expensive to relax, an even larger share of the total relaxation cost), the number of structures generated per (conformer, $Z$, space group) combination, and the deduplication stringency prior to relaxation. As shown in SI Section~\ref{app:conf_eval}, 20 conformers per flexible target provided ample coverage, but fewer than 10 would have sufficed for most systems. Deduplication strategy also plays a significant role: aggressive deduplication substantially reduces the number of structures passed to relaxation (67\% and 43\% reduction for the semi-rigid and flexible subsets, respectively), but overly strict thresholds risk discarding structures that would have relaxed to a match with an experimental polymorph. Several structure comparison methods with varying stringency are available for this purpose~\cite{ong2013python,martirossyan2026all,hunnisett2024seventhrank,hunnisett2024seventhgen}.

For high-throughput screening campaigns where exhaustive polymorph recovery is less critical, a lean search is sufficient. For pharmaceutical development where missing a single polymorph carries regulatory risk, the full search with generous conformer pools and relaxed deduplication thresholds remains advisable. The efficiency of UMA makes both regimes accessible within the same framework, bringing high-throughput molecular crystal structure prediction within practical reach on modest academic GPU clusters.

%% file: appendix/conf_eval.tex
\section{Conformer Generation Quality}
\label{app:conf_eval}

The conformer generation stage provides the molecular geometries that seed all downstream crystal structure generation. We evaluate how faithfully our conformer generation and selection reproduces the experimental molecular geometries extracted from the crystal structures, reporting RMSD between each generated gas-phase conformer and its experimental crystallographic counterpart. Conformers are ranked by ascending UMA-OMol gas-phase energy, previously shown to result in mean absolute energy difference of 0.5 kJ/mol, well within chemical accuracy versus a high performance DFT functional ($\omega$B97M-V/def2-TZVPD) for conformer ranking~\cite{omol}. Conformer~\#1 is the lowest-energy generated geometry and higher ranks correspond to progressively less stable gas-phase conformations that may nonetheless be stabilized by crystal packing. We denote the gas-phase energy of a conformer relative to conformer~\#1 as $\Delta E_\text{conf}$. 

For the molecules where a single conformer was used for structure generation (all semi-rigid molecules except nicotinamide, SI~Table~\ref{tab:tier1_conf}), the workflow reliably recovered the experimental gas-phase geometry in all cases. Under the appropriate symmetry interpretation, all heavy-atom RMSDs fall below 0.16~\AA{} and all-atom RMSDs fall below 0.17~\AA, except all forms of glycine attaining all-atom RMSDs up to 0.46~\AA. The largest values occur for the glycine polymorphs, which adopt a zwitterionic conformation in the crystal. Here, the unrelaxed RDKit-generated zwitterionic geometry was used as noted in SI~Section~\ref{app:hyperparams}. 

For the systems where multiple conformers were passed to crystal generation (the entire flexible molecules subset and nicotinamide from the semi-rigid subset, SI Table~\ref{tab:tier2_conf}), up to 20 conformers per target were forwarded to structure generation (5 for nicotinamide), and the pool captured all distinct molecular conformations observed across the known polymorphs. For piroxicam, all five experimental conformations are already represented by conformer~\#1 (the gas-phase global minimum), reflecting the fact that the conformational differences between its polymorphs are subtle. At the other extreme, chlorpropamide, whose six ambient polymorphs each adopt a distinct molecular conformation, requires conformers up to rank~\#16 ($\Delta E_\mathrm{conf} = 14.0$~kJ/mol) to cover all experimental forms. Target~VI similarly requires high-energy conformers (ranks~\#12 and~\#15, $\Delta E_\mathrm{conf} = 16.2$ and 17.7~kJ/mol), consistent with its flexible sulfonamide linkage adopting energetically unfavorable geometries under crystal packing forces. ROY, despite having multiple known polymorphs spanning a wide range of S--C--N--C dihedral angles, requires only 8 conformers ($\Delta E_\mathrm{conf} \leq 2.2$~kJ/mol) to cover all experimental forms. For most other flexible targets, the first experimentally matching conformer appears within the top~10 and within $\Delta E_\mathrm{conf} < 7$~kJ/mol of the gas-phase minimum. Across the entire flexible subset, the highest-ranked conformer required to recover any experimental polymorph is rank~\#17 (Form~III of piracetam with $\Delta E_\mathrm{conf} = 8.9$~kJ/mol), indicating that a pool of $\sim$20 conformers provides comfortable coverage of the experimentally relevant conformational space. For most individual targets, fewer than 10 conformers would have sufficed, suggesting that adaptive conformer selection could substantially reduce the number of downstream generated crystal structures without sacrificing polymorph recovery. The best heavy-atom RMSD across the conformer pool is below 0.38~\AA, except for the $\epsilon$ form of chlorpropamide with an RMSD of 0.57~\AA. While these results confirm the generated conformers provide geometries of sufficient quality to seed accurate crystal structure prediction, they do not guarantee the success of downstream prediction efforts.

\begin{table}[h!tbp]
\centering
\caption{Conformer evaluation for systems where a single conformer was passed to downstream structure generation. $N_{\text{rot}}$: rotatable bonds, $N_{\text{gen}}$: generated conformers, $N_{\text{used}}$: conformers passed to structure generation (here always~1). Each RMSD cell reports the symmetry-appropriate conformer-match score (\AA): for non-Sohncke space groups (which contain improper symmetry operations that generate both hands of the molecule in the unit cell), the value shown is minimum of direct and inverse RMSDs, where ``inverse'' denotes point-inversion of the probe, and for Sohncke (chiral) space groups, only the direct RMSD is reported. }
\label{tab:tier1_conf}
\begin{adjustbox}{max width=0.75\textwidth, max height=0.5\textheight, center}
\begin{tabular}{@{}cl ccc l c cc@{}} \toprule \multirow{2}{*}{\textbf{No.}} & \multirow{2}{*}{\textbf{Compound}} & \multirow{2}{*}{$\boldsymbol{N_\mathrm{rot}}$} & \multirow{2}{*}{$\boldsymbol{N_\mathrm{gen}}$} & \multirow{2}{*}{$\boldsymbol{N_\mathrm{used}}$} & \multirow{2}{*}{\textbf{Polymorph}} & {\textbf{Space Group}} & \multirow{2}{*}{\textbf{RMSD\textsubscript{heavy} [\AA]}} & \multirow{2}{*}{\textbf{RMSD\textsubscript{all} [\AA]}} \\ & & & & & & {\textbf{Non-Sohncke}}& & \\ \midrule
    \multirow{2}{*}{1} & \multirow{2}{*}{Target I} & \multirow{2}{*}{0} & \multirow{2}{*}{1} & \multirow{2}{*}{1} & $Z=8$ & Yes & 0.01& 0.08\\
     &  &  &  &  & $Z=4$ & Yes & 0.01& 0.07\\
    \midrule
    2 & Target II & 0 & 2 & 1 & Exp. & Yes & 0.02& 0.09\\
    \midrule
    3 & Target IV & 0 & 3 & 1 & Exp. & Yes & 0.03& 0.09\\
    \midrule
    4 & Target V & 0 & 1 & 1 & Exp. & No & 0.12& 0.17\\
    \midrule
    5 & Target VIII & 0 & 1 & 1 & Exp. & Yes & 0.03& 0.13\\
    \midrule
    6 & Target XII & 1 & 3 & 1 & Exp. & Yes & 0.03& 0.10\\
    \midrule
    7 & Target XIII & 0 & 1 & 1 & Exp. & Yes & 0.02& 0.05\\
    \midrule
    8 & Target XVI & 0 & 1 & 1 & Exp. & Yes & 0.04& 0.08\\
    \midrule
    9 & Target XVII & 2 & 4 & 1 & Exp. & Yes & 0.09& 0.09\\
    \midrule
    10 & Target XXII & 0 & 3 & 1 & Exp. & Yes & 0.04& 0.04\\
    \midrule
    11 & Acetic Acid & 0 & 4 & 1 & Exp. & Yes& 0.01& 0.15\\
    \midrule
    12 & Caprylolactam & 0 & 28 & 1 & Exp. & Yes& 0.03& 0.11\\
    \midrule
    13 & CEBYUD & 0 & 1 & 1 & Exp. & No & 0.03& 0.03\\
    \midrule
    \multirow{2}{*}{14} & \multirow{2}{*}{CILJIQ} & \multirow{2}{*}{1} & \multirow{2}{*}{4} & \multirow{2}{*}{1} & Form I & No & 0.05& 0.12\\
     &  &  &  &  & Form II & Yes & 0.04& 0.10\\
    \midrule
    15 & CUMJOJ & 0 & 8 & 1 & Exp. & Yes& 0.07& 0.12\\
    \midrule
    16 & DEZDUH & 0 & 1 & 1 & Exp. & Yes& 0.03& 0.11\\
    \midrule
    17 & Eniluracil & 0 & 1 & 1 & Exp.$^*$ & Yes & 0.04& 0.09\\
    \midrule
    18 & GACGAU & 0 & 2 & 1 & Exp. & Yes & 0.01& 0.03\\
    \midrule
    \multirow{5}{*}{19} & \multirow{5}{*}{Glycine} & \multirow{5}{*}{1} & \multirow{5}{*}{9} & \multirow{5}{*}{1} & $\gamma$ & No & 0.10& 0.37\\
     &  &  &  &  & $\alpha$ & Yes & 0.12& 0.38\\
     &  &  &  &  & $\beta$ & No & 0.14& 0.46\\
     &  &  &  &  & $\delta$ & Yes & 0.16& 0.37\\
     &  &  &  &  & $\varepsilon$ & Yes& 0.05& 0.35\\
    \midrule
    20 & GOLHIB & 0 & 1 & 1 & Exp. & Yes& 0.06& 0.12\\
    \midrule
    21 & HURYUQ & 0 & 2 & 1 & Exp. & No & 0.07& 0.14\\
    \midrule
    22 & IHEPUG & 0 & 1 & 1 & Exp. & No & 0.03& 0.10\\
    \midrule
    \multirow{2}{*}{23} & \multirow{2}{*}{Imidazole} & \multirow{2}{*}{0} & \multirow{2}{*}{1} & \multirow{2}{*}{1} & $\alpha$ & Yes & 0.01& 0.02\\
     &  &  &  &  & $\beta$ & Yes& 0.02& 0.11\\
    \midrule
    24 & LECZOL & 0 & 1 & 1 & Exp. & No & 0.01& 0.09\\
    \midrule
    26 & ROHBUL & 0 & 1 & 1 & Exp. & Yes & 0.06& 0.09\\
    \midrule
    27 & 6-Fluorochromone & 0 & 1 & 1 & Exp. & No & 0.01& 0.06\\
    \midrule
    28 & WEXREY & 0 & 1 & 1 & Exp. & Yes& 0.02& 0.09\\
    \bottomrule
    \end{tabular}
\end{adjustbox}
    \begin{tablenotes}
    \item \small{$^*$ DOHFEM is disordered over two sites in a ratio 0.74:0.26. Here we chose the major component.}
    \end{tablenotes}
\end{table}

\begin{table}[h!tbp]
\centering
\caption{Conformer evaluation for systems where multiple conformers were passed to downstream structure generation. Conformers are ranked $1, 2, \dots, N_\text{gen}$ by ascending gas-phase relaxed energy. $N_{\text{rot}}$: rotatable bonds, $N_{\text{gen}}$: generated conformers, $N_{\text{used}}$: conformers passed to structure generation. $\text{Rank}_\text{1st\,match}$: rank of the lowest ranked conformer that recovered an experimental polymorph, $\Delta E_\text{1st\,match}$: its gas-phase energy relative to conformer~\#1. Each RMSD cell reports the symmetry-appropriate conformer-match score (\AA): for non-Sohncke space groups (which contain improper symmetry operations that generate both hands of the molecule in the unit cell), the value shown is minimum of direct and inverse RMSDs, where ``inverse'' denotes point-inversion of the probe, and for Sohncke (chiral) space groups, only the direct RMSD is reported. ``First'' uses conformer~\#1, ``Best'' is the minimum across the $N_\text{used}$ conformer pool with the conformer rank shown as a subscript, and ``1st Match'' uses conformer $\text{Rank}_\text{1st\,match}$.}
\label{tab:tier2_conf}
\begin{adjustbox}{max width=\textwidth, max height=0.95\textheight, center}
\begin{tabular}{@{}cl ccc cc l c ccc ccc@{}} \toprule \multirow{2}{*}{\textbf{No.}} & \multirow{2}{*}{\textbf{Compound}} & \multirow{2}{*}{$\boldsymbol{N_\textbf{rot}}$} & \multirow{2}{*}{$\boldsymbol{N_\textbf{gen}}$} & \multirow{2}{*}{$\boldsymbol{N_\textbf{used}}$} & \multirow{2}{*}{${\textbf{Rank}_\textbf{1st\,match}}$} & \multirow{2}{*}{\shortstack{$\boldsymbol{\Delta E_\textbf{1st\,match}}$\\[2pt]\textbf{[kJ/mol]}}} & \multirow{2}{*}{\textbf{Polymorph}} & \multirow{2}{*}{\shortstack{\textbf{Space Group}\\\textbf{Non-Sohncke}}} & \multicolumn{3}{c}{\textbf{RMSD\textsubscript{heavy} [\AA]}} & \multicolumn{3}{c}{\textbf{RMSD\textsubscript{all} [\AA]}} \\ \cmidrule(lr){10-12}\cmidrule(lr){13-15} & & & & & & & & & \textbf{First} & \textbf{Best}$_{\boldsymbol{\,\textbf{rank}}}$ & \textbf{1st Match} & \textbf{First} & \textbf{Best}$_{\boldsymbol{\,\textbf{rank}}}$ & \textbf{1st Match} \\ \midrule
    \multirow{3}{*}{25} & \multirow{3}{*}{Nicotinamide} & \multirow{3}{*}{1} & \multirow{3}{*}{5} & \multirow{3}{*}{5} & 1 & 0.00 & $\alpha$ & Yes & 0.03 & $0.03_{1}$ & 0.03 & 0.04 & $0.04_{1}$ & 0.04 \\
     &  &  &  &  & 3 & 3.90 & $\epsilon$ & No & 0.91 & $0.04_{3}$ & 0.04 & 1.08 & $0.11_{3}$ & 0.11 \\
     &  &  &  &  & 1 & 0.00 & $\iota$ & Yes & 0.06 & $0.06_{1}$ & 0.06 & 0.19 & $0.19_{1}$ & 0.19 \\
    \midrule
    \multirow{2}{*}{29} & \multirow{2}{*}{Target VI} & \multirow{2}{*}{2} & \multirow{2}{*}{46} & \multirow{2}{*}{20} & 12 & 16.21 & Form I & Yes & 1.06 & $0.29_{15}$ & 0.45 & 1.20& $0.45_{15}$ & 0.66 \\
     &  &  &  &  & 15 & 17.69 & Form III & Yes & 1.16 & $0.34_{15}$ & 0.34 & 1.35 & $0.52_{15}$ & 0.52 \\
    \midrule
    30 & Target X & 3 & 45 & 20 & 4 & 0.80 & Exp. & Yes & 0.43 & $0.26_{4}$ & 0.26 & 0.51 & $0.40_{4}$ & 0.40 \\
    \midrule
    31 & Target XIV & 0 & 13 & 13 & 1 & 0.00 & Exp. & Yes & 0.11 & $0.09_{5}$ & 0.11 & 0.26 & $0.14_{4}$ & 0.26 \\
    \midrule
    32 & Target XVIII & 2 & 52 & 20 & 15 & 7.42 & Exp. & Yes & 0.86 & $0.23_{16}$ & 0.25 & 1.19 & $0.36_{15}$ & 0.36 \\
    \midrule
    \multirow{3}{*}{33} & \multirow{3}{*}{Target XXXI} & \multirow{3}{*}{2} & \multirow{3}{*}{91} & \multirow{3}{*}{20} & 10 & 6.07 & Form A major & Yes & 2.07 & $0.22_{17}$ & 0.44 & 2.23 & $0.32_{17}$ & 0.61 \\
     &  &  &  &  & 11 & 6.31 & Form A minor & Yes & 1.73 & $0.27_{18}$ & 0.51 & 1.81 & $0.42_{8}$ & 0.59 \\
     &  &  &  &  & 10 & 6.07 & Form B & Yes & 2.14 & $0.20_{14}$ & 0.35 & 2.26 & $0.28_{14}$ & 0.55 \\
    \midrule
    \multirow{6}{*}{34} & \multirow{6}{*}{Chlorpropamide} & \multirow{6}{*}{4} & \multirow{6}{*}{59} & \multirow{6}{*}{20} & 4 & 12.42 & $\alpha$ & No & 1.83 & $0.32_{7}$ & 1.02 & 2.31 & $0.45_{12}$ & 1.46 \\
     &  &  &  &  & 8 & 12.85 & $\delta$ & Yes & 1.81 & $0.38_{5}$ & 0.51 & 2.50 & $0.78_{13}$ & 0.79 \\
     &  &  &  &  & 16 & 13.98 & $\gamma$ & No & 1.38 & $0.20_{16}$ & 0.20 & 1.67 & $0.30_{16}$ & 0.30 \\
     &  &  &  &  & 4 & 12.42 & $\epsilon'$ & Yes & 1.74 & $0.32_{7}$ & 0.32 & 2.29 & $0.47_{12}$ & 0.50 \\
     &  &  &  &  & 16 & 13.98 & $\epsilon$ & Yes & 1.03 & $0.56_{16}$ & 0.56 & 1.46 & $0.74_{16}$ & 0.74 \\
     &  &  &  &  & 16 & 13.98 & $\beta$ & Yes & 1.18 & $0.15_{16}$ & 0.15 & 1.56 & $0.22_{16}$ & 0.22 \\
    \midrule
    \multirow{4}{*}{35} & \multirow{4}{*}{Olanzapine} & \multirow{4}{*}{0} & \multirow{4}{*}{48} & \multirow{4}{*}{20} & $1^{\bigtriangleup}$ & $0.00^{\bigtriangleup}$ & Form I & Yes & 0.16 & $0.09_{4}$ & $0.16^{\bigtriangleup}$ & 0.25 & $0.21_{4}$ & $0.25^{\bigtriangleup}$ \\
     &  &  &  &  & 1 & 0.00 & Form II & Yes & 0.26 & $0.14_{4}$ & 0.26 & 0.36 & $0.21_{4}$ & 0.36 \\
     &  &  &  &  & 1 & 0.00 & Form III & Yes & 0.26 & $0.16_{4}$ & 0.26 & 0.39 & $0.30_{4}$ & 0.39 \\
     &  &  &  &  & $4^{\bigtriangleup}$ & $1.06^{\bigtriangleup}$ & Form IV & Yes & 0.12 & $0.12_{1}$ & $0.14^{\bigtriangleup}$ & 0.31 & $0.31_{1}$ & $0.33^{\bigtriangleup}$ \\
    \midrule
    \multirow{4}{*}{36} & \multirow{4}{*}{Piracetam} & \multirow{4}{*}{2} & \multirow{4}{*}{45} & \multirow{4}{*}{20} & 1 & 0.00 & Form IV & Yes & 0.38 & $0.17_{15}$ & 0.38 & 0.49 & $0.24_{16}$ & 0.49 \\
     &  &  &  &  & 17 & 8.89 & Form III & Yes & 0.97 & $0.23_{19}$ & 0.77 & 1.12 & $0.41_{19}$ & 0.99 \\
     &  &  &  &  & 11 & 5.99 & Form II & Yes & 0.99 & $0.26_{19}$ & 0.84 & 1.14 & $0.39_{19}$ & 1.10 \\
     &  &  &  &  & 1 & 0.00 & Form I maj & Yes & 0.87 & $0.07_{19}$ & 0.87 & 1.01 & $0.15_{19}$ & 1.01 \\
    \midrule
    \multirow{5}{*}{37} & \multirow{5}{*}{Piroxicam} & \multirow{5}{*}{2} & \multirow{5}{*}{28} & \multirow{5}{*}{20} & 1 & 0.00 & $\alpha_1$ & No & 0.06 & $0.06_{1}$ & 0.06 & 0.12 & $0.12_{1}$ & 0.12 \\
     &  &  &  &  & 1 & 0.00 & Form II ($\alpha_2$) & Yes & 0.05 & $0.05_{1}$ & 0.05 & 0.30 & $0.29_{2}$ & 0.30 \\
     &  &  &  &  & 1 & 0.00 & Form III & Yes & 0.13 & $0.11_{2}$ & 0.13 & 0.18 & $0.17_{2}$ & 0.18 \\
     &  &  &  &  & 1 & 0.00 & Form I & Yes & 0.14 & $0.11_{2}$ & 0.14 & 0.23 & $0.20_{2}$ & 0.23 \\
     &  &  &  &  & 1 & 0.00 & Form IV & Yes & 0.11 & $0.11_{1}$ & 0.11 & 0.17 & $0.17_{1}$ & 0.17 \\
    \midrule
    \multirow{10}{*}{38} & \multirow{10}{*}{ROY} & \multirow{10}{*}{3} & \multirow{10}{*}{59} & \multirow{10}{*}{20} & 4 & 1.23 & ORP & Yes & 1.10 & $0.10_{17}$ & 0.99 & 1.13 & $0.15_{17}$ & 1.03 \\
     &  &  &  &  & 7 & 1.86 & R & Yes & 1.18 & $0.10_{16}$ & 0.73 & 1.27 & $0.14_{16}$ & 0.85 \\
     &  &  &  &  & 1 & 0.00 & PO13 & Yes & 0.22 & $0.15_{9}$ & 0.22 & 0.30 & $0.27_{2}$ & 0.30 \\
     &  &  &  &  & 8 & 2.17 & ON & Yes & 0.87 & $0.13_{8}$ & 0.13 & 0.92 & $0.17_{8}$ & 0.17 \\
     &  &  &  &  & 4 & 1.23 & OP & Yes & 1.00 & $0.18_{17}$ & 0.93 & 1.05 & $0.23_{17}$ & 0.98 \\
     &  &  &  &  & 1 & 0.00 & Y04 & Yes & 0.57 & $0.16_{11}$ & 0.57 & 0.61 & $0.25_{7}$ & 0.61 \\
     &  &  &  &  & 6 & 1.73 & Y19 & Yes & 0.79 & $0.08_{8}$ & 0.58 & 0.85 & $0.12_{8}$ & 0.68 \\
     &  &  &  &  & 8 & 2.17 & Y & Yes & 0.23 & $0.17_{6}$ & 0.61 & 0.30 & $0.24_{6}$ & 0.71 \\
     &  &  &  &  & 1 & 0.00 & YT04 & Yes & 0.20 & $0.10_{6}$ & 0.20 & 0.22 & $0.21_{4}$ & 0.22 \\
     &  &  &  &  & 1 & 0.00 & YN & Yes & 0.21 & $0.20_{5}$ & 0.21 & 0.35 & $0.27_{6}$ & 0.35 \\
    \bottomrule
    \end{tabular}
\end{adjustbox}
    \begin{tablenotes}
    \item \small{$^\bigtriangleup$ Obtained from a larger initial pool generated by 5 lowest-energy conformers and 3,000 structures per space group.}
    \end{tablenotes}
\end{table}

%% file: appendix/csp_eval.tex
\section{Crystal Structure Prediction Results}
\label{app:csp_matches}

\begin{table}[h!tbp]
    \small
    \centering
    \caption{Summary of the CSP results for the Semi-Rigid molecules subset, listing compound names, corresponding CSD reference codes, and performance metrics for experimental matches with different methods.}
    \label{tab:tier1_results}
    \begin{adjustbox}{max width=\textwidth, max height=\textheight, center}
    \begin{tabular}{@{}cllcccccccccccc@{}}
    \toprule
    \multirow{2}{*}{\textbf{No.}} 
    & \multirow{2}{*}{\textbf{Compound}}
    & \multirow{2}{*}{\textbf{Polymorph}}
    & \multicolumn{3}{c}{\textbf{UMA (0 K)}} & \multicolumn{2}{c}{\textbf{UMA + \boldmath{$\Delta$}UMA (0 K)}} & \multicolumn{3}{c}{\textbf{PBE-D3 (0 K)}} & \multicolumn{4}{c}{\textbf{UMA (QHA, 300 K)}} \\
    \cmidrule(lr){4-6} \cmidrule(lr){7-8} \cmidrule(lr){9-11} \cmidrule(l){12-15}
    & & & \textbf{RMSD$_\mathbf{30}$ [\AA]} & \boldmath{$\Delta E$} \textbf{[kJ/mol]} & \textbf{Rank} & \boldmath{$\Delta E$} \textbf{[kJ/mol]} & \textbf{Rank} & \textbf{RMSD$_\mathbf{30}$ [\AA]} & \boldmath{$\Delta E$} \textbf{[kJ/mol]} & \textbf{Rank} & \boldmath{$\Delta F_{\textbf{\textup{HA}}}$\ \textbf{[kJ/mol]}} & \textbf{Rank} & \boldmath{$\Delta G_{\textbf{\textup{QHA}}}$\ \textbf{[kJ/mol]}} & \textbf{Rank} \\
    \midrule

    \multirow{2}{*}{1} & \multirow{2}{*}{Target I} & $Z=8$  & 0.18 & 0.52 & 2 & 0.00 & 1 & 0.09 & 0.69 & 10 & 0.76 & 9 & 0.48 & 7 \\
    & & $Z=4$  & 0.13 & 0.00 & 1 & 0.83 & 2 &  0.06 & 0.35 & 7 & 1.35 & 14 & 1.33 & 16 \\
    \midrule
    2 & Target II & Exp.  & 0.61 & 1.07 & 5 & 0.84 & 5 &  0.62 & 1.24 & 5 & 0.15 & 3 & 0.02 & 2 \\
    \midrule
    3 & Target IV & Exp.  & 0.14 & 0.00 & 1 & 0.00 & 1 &  0.20 & 0.00 & 1 & 0.00 & 1 & 0.00 & 1 \\
    \midrule
    4 & Target V & Exp.  & 0.23 & 4.89 & 17 & 5.62 & 17 &  0.20 & 5.36 & 17 & 3.69 & 14 & 3.72 & 16 \\
    \midrule
    5 & Target VIII & Exp.  & 0.18 & 0.00 & 1 & 0.00 & 1 &  0.18 & 0.00 & 1 & 0.00 & 1 & 0.00 & 1 \\
    \midrule
    6 & Target XII & Exp.  & 0.24 & 0.00 & 1 & 0.00 & 1 &  0.11 & 0.00 & 1 & 0.42 & 6 & 0.31 & 4 \\
    \midrule
    7 & Target XIII & Exp.  & 0.09 & 0.00 & 1 & 0.00 & 1 &  0.07 & 0.00 & 1 & 0.00 & 1 & 0.00 & 1 \\
    \midrule
    8 & Target XVI & Exp.  & 0.16 & 0.00 & 1 & 0.00 & 1 &  0.19 & 0.91 & 4 & 0.88 & 7 & 1.08 & 15 \\
    \midrule
    9 & Target XVII & Exp.  & 0.13 & 0.00 & 1 & 0.00 & 1 &  0.05 & 0.00 & 1 & 0.00 & 1 & 0.00 & 1 \\
    \midrule
    10 & Target XXII & Exp.  & 0.19 & 1.45 & 2 & 0.00 & 1 &  0.11 & 1.85 & 3 & 1.76 & 3 & 1.46 & 2 \\
    \midrule
    11 & Acetic Acid & Exp.  & 0.07 & 0.01 & 2 & 0.00 & 1 &  0.10 & 0.66 & 2 & 0.00 & 1 & 0.13 & 2 \\
    \midrule
    12 & Caprylolactam & Exp.  & 0.16 & 0.17 & 2 & 0.00 & 1 & 0.17  & 0.71 & 2 & 0.00 & 1 & 0.00 & 1 \\
    \midrule
    13 & CEBYUD & Exp.  & 0.10 & 0.18 & 2 & 0.30 & 2 &  0.04 & 0.00 & 1 & 0.18 & 2 & 0.92 & 3 \\
    \midrule
    \multirow{2}{*}{14} & \multirow{2}{*}{CILJIQ} & Form I & 0.28 & 0.00 & 1 & 0.00 & 1 &  0.23 & 0.00 & 1 & 0.00 & 1 & 0.00 & 1 \\
    & & Form II &  0.12 & 2.70 & 5 & 1.32 & 5  &  0.13 & 3.97 & 9 & 1.17 & 5 & 3.00 & 15 \\
    \midrule
    15 & CUMJOJ & Exp. & 0.16 & 0.79 & 3 & 1.20 & 3 &  0.05 & 0.74 & 3 & 1.38 & 4 & 0.82 & 4  \\
    \midrule
    16 & DEZDUH & Exp. & 0.17 & 0.00 & 1 & 0.00 & 1 &  0.11 & 3.38 & 15 & 0.00 & 1 & 0.10 & 2 \\
    \midrule
    17 & Eniluracil & Exp.$^*$ & 0.13 & 0.00 & 1 & 0.00 & 1 &  0.10 & 0.00 & 1 & 0.00 & 1 &  0.22 & 2 \\
    \midrule
    18 & GACGAU & Exp. & 0.22 & 0.46 & 4 & 0.40 & 2 & 0.14 & 1.37 & 5 & 1.76 & 8 & 1.58 & 10 \\
    \midrule
    \multirow{5}{*}{19} & \multirow{5}{*}{Glycine} & $\gamma$ & 0.14 & 1.12 & 3 & 0.11 & 2 &  0.06 & 0.55 & 2 & 2.21 & 7 & 2.12 & 14 \\
    & & $\alpha$ & 0.09 & 0.00 & 1 & 0.00 & 1 &  0.08 & 0.00 & 1 & 0.00 & 1 & 0.00 & 1 \\
    & & $\beta$ & 0.11 & 2.01 & 8 & 2.56 & 11 &  0.06 & 2.28 & 9 & 1.83 & 4 & 1.51 & 6 \\
    & & $\delta$ & 0.25 & 2.65 & 10 & 3.35 & 21 &  0.20 & 2.30 & 10 & 2.30 & 8 & 1.82 & 10 \\
    & & $\varepsilon$ & 0.37 & 1.58 & 5 & 1.52 & 3 &  0.31 & 2.44 & 11 & 1.53 & 3 & 1.48 & 5 \\
    \midrule
    20 & GOLHIB & Exp. & 0.06 & 0.00 & 1 & 0.00 & 1 &  0.06 & 0.00 & 1 & 0.00 & 1 & 0.00 & 1 \\
    \midrule
    21 & HURYUQ & Exp. & 0.10 & 0.00 & 1 & 0.00 & 1 &  0.15 & 0.00 & 1 & 1.05 & 3 & 0.52 & 2 \\
    \midrule
    22 & IHEPUG & Exp. & 0.14 & 1.01 & 4 & 1.05 & 3 &  0.17 & 0.77 & 3 & 1.28 & 3 & 0.59 & 2 \\
    \midrule
    \multirow{2}{*}{23} & \multirow{2}{*}{Imidazole} & $\alpha$ & 0.19 & 0.00 & 1 & 0.00 & 1 &  0.11 & 0.28 & 3 & 0.00 & 1 & 0.24 & 2 \\
    & & $\beta$ & 0.16 & 3.62 & 31 & 3.44 & 27 & 0.16 & 3.49 & 34 & 4.72 & 75 & 4.62 & 45 \\
    \midrule
    24 & LECZOL & Exp. & 0.22 & 1.00 & 8 & 1.17 & 7 &  0.14 & 1.85 & 17 & 0.90 & 4 & 0.00 & 1 \\
    \midrule
    \multirow{3}{*}{25} & \multirow{3}{*}{Nicotinamide} & $\alpha$ & 0.22 & 0.15 & 2 & 0.00 & 1 & 0.21 & 0.55 & 3 & 0.78 & 5 & 0.85 & 2 \\
    & & $\varepsilon$ & 0.08 & 1.54 & 5 & 1.33 & 3 & 0.06 & 2.53 & 11 & 0.68 & 4 & 1.19 & 3 \\
    & & $\iota$ & 0.37 & 2.20 & 8 & 3.68 & 11 &  0.27 & 1.59 & 6 & 1.31 & 8 & 2.29 & 7 \\
    \midrule
    26 & ROHBUL & Exp. & 0.04 & 0.00 & 1 & 0.00 & 1 &  0.09 & 0.00 & 1 & 0.00 & 1 & 0.00 & 1 \\
    \midrule
    27 & 6-Fluorochromone & Exp. & 0.15 & 0.91 & 7 & 1.01 & 4 & 0.10  & 1.11 & 9 & 0.00 & 1 & 0.27 & 2 \\
    \midrule
    28 & WEXREY & Exp. & 0.29 & 0.00 & 1 & 0.00 & 1 &  0.14 & 0.00 & 1 & 0.00 & 1 & 0.00 & 1 \\

    \midrule
    \end{tabular}
    \end{adjustbox}
    \begin{tablenotes}
    \item \small{$^*$ DOHFEM is disordered over two sites in a ratio 0.74:0.26. Here we chose the major component.}
    \end{tablenotes}
\end{table}

\begin{table}[h!tbp]
    \ContinuedFloat
    \small
    \centering
    \caption{\textbf{(continued)} Summary of the CSP results for the Flexible molecules subset, listing compound names, corresponding CSD reference codes, and performance metrics for experimental matches with different methods.}
    \begin{adjustbox}{max width=\textwidth, max height=\textheight, center}
   \begin{tabular}{@{}cllcccccccccccc@{}}
    \toprule
    \multirow{2}{*}{\textbf{No.}} 
    & \multirow{2}{*}{\textbf{Compound}}
    & \multirow{2}{*}{\textbf{Polymorph}}
    & \multicolumn{3}{c}{\textbf{UMA (0 K)}} & \multicolumn{2}{c}{\textbf{UMA + \boldmath{$\Delta$}UMA (0 K)}} & \multicolumn{3}{c}{\textbf{PBE-D3 (0 K)}} & \multicolumn{4}{c}{\textbf{UMA (QHA, 300 K)}} \\
    \cmidrule(lr){4-6} \cmidrule(lr){7-8} \cmidrule(lr){9-11} \cmidrule(l){12-15}
    & & & \textbf{RMSD$_\mathbf{30}$ [\AA]} & \boldmath{$\Delta E$} \textbf{[kJ/mol]} & \textbf{Rank} & \boldmath{$\Delta E$} \textbf{[kJ/mol]} & \textbf{Rank} & \textbf{RMSD$_\mathbf{30}$ [\AA]} & \boldmath{$\Delta E$} \textbf{[kJ/mol]} & \textbf{Rank} & \boldmath{$\Delta F_{\textbf{\textup{HA}}}$\ \textbf{[kJ/mol]}} & \textbf{Rank} & \boldmath{$\Delta G_{\textbf{\textup{QHA}}}$\ \textbf{[kJ/mol]}} & \textbf{Rank} \\
    \midrule

    \multirow{2}{*}{29} & \multirow{2}{*}{Target VI} & Form I & 0.17 & 0.00 & 1 & 0.00 & 1 & 0.22 &  0.00 & 1 & 0.00 & 1 & 0.00 & 1 \\
    & & Form III & 0.11 & 1.26 & 2 & 0.53 & 2 & 0.05 &  1.67 & 2  & 2.33 & 3 & 1.59 & 3 \\
    \midrule
    30 & Target X & Exp. & 0.09 & 1.32 & 5 & 0.00 & 1 & 0.13 &  2.16 & 16 &  1.28 & 7 & 2.60 & 10 \\
    \midrule
    31 & Target XIV & Exp. & 0.04 & 0.00 & 1 & 0.00 & 1 & 0.07 & 0.00 & 1  &  0.00 & 1 & 0.00 & 1 \\
    \midrule
    32 & Target XVIII & Exp. & 0.22 & 0.00 & 1 & 2.76 & 22 & 0.12 &  0.00 & 1 &  0.00 & 1 & 0.00 & 1 \\
    \midrule
    \multirow{3}{*}{33} & \multirow{3}{*}{Target XXXI} & Form A major & 0.32 & 0.86 & 6 & 2.73 & 10 & 0.15 & 1.61 & 9 &  0.26 & 3 & 0.00 & 1 \\
    & & Form A minor & 0.29 & 1.74 & 13 & 3.17 & 23 & 0.27 & 2.24 & 22 &  0.67 & 7 & 0.49 & 2 \\
    & & Form B & 0.31 & 1.97 & 15 & 2.72 & 9 & 0.26 & 4.48 & 71 &  0.68 & 8 & 2.15 & 19 \\
    \midrule
    \multirow{6}{*}{34} & \multirow{6}{*}{Chlorpropamide} & $\alpha$ & 0.09 & 0.00 & 1 & 0.00 & 1 & 0.08 &  0.00 & 1 &  0.22 & 2 & 0.41 & 2 \\
    & & $\delta$ & 0.17 & 1.25 & 3 & 0.90 & 4 & 0.21 & 1.66 & 5 &  0.00 & 1 & 0.00 & 1 \\
    & & $\gamma$ & 0.19 & 1.58 & 4 & 0.37 & 3 & 0.24 & 1.56 & 3 &  1.34 & 4 & 0.78 & 4 \\
    & & $\varepsilon^{'}$ & 0.11 & 2.50 & 6 & 2.50 & 9 & 0.08 & 2.43 & 10 &  2.67 & 18 & 2.59 & 20 \\
    & & $\varepsilon$ & 0.30 & 6.15 & 37 & 5.18 & 35 & 0.40 &  5.75 & 44 &  2.40 & 14 & 1.66 & 10 \\
    & & $\beta$ & 0.24 & 6.42 & 44 & 4.61 & 27 & 0.25 &  6.24 & 51 &  4.30 & 34 & 3.97 & 41 \\
    \midrule
    \multirow{6}{*}{35} & \multirow{6}{*}{Olanzapine} & Form III & 0.21 & 0.00 & 1 & 0.67 & 2 & 0.27 & 0.00 & 1 &  1.25 & 2 & 0.56 & 2 \\
    & & Form II & 0.17 & 2.41 & 3 & 1.89 & 3 & 0.14 & 1.96 & 3 &  0.00 & 1 & 0.00 & 1 \\
    \cmidrule(lr){3-15}
    & & Form III & 0.21 & 0.00 & 1 & 1.08 & 4 & 0.27 & 0.00 & 1 &  1.94 & 3 & 1.23 & 3 \\
    & & Form II & 0.17 & 2.41 & 5 & 2.31 & 5 & 0.14 & 1.96 & 5 &  0.69 & 2 & 0.67 & 2 \\
    & & Form IV$^\bigtriangleup$ & 0.22 & 1.44 & 3 & 0.00 & 1 & 0.32 & 0.09 & 2 & 5.88 & 5 & 4.44 & 5\\
    & & Form I$^\bigtriangleup$ & 0.09 & 1.68 & 4 & 0.46 & 3 & 0.11 & 0.79 & 3 &  0.00 & 1 & 0.00 & 1\\

    \midrule
    \multirow{4}{*}{36} & \multirow{4}{*}{Piracetam} & Form IV & 0.13 & 0.00 & 1 & 0.00 & 1 & 0.11 & 0.00 & 1 &  0.66 & 4 & 1.57 & 6 \\
    & & Form III & 0.10 & 2.25 & 3 & 1.12 & 3 & 0.12 & 2.09 & 4 &  0.02 & 2 & 0.38 & 3 \\
    & & Form II & 0.09 & 2.75 & 5 & 1.34 & 4 & 0.09 & 3.09 & 6 &  0.00 & 1 & 0.00 & 1 \\
    & & Form I maj & 0.26 & 4.91 & 10 & 4.52 & 16 & 0.18 & 3.87 & 9 & 2.48 & 9 & 1.93 & 7 \\
    \midrule
    \multirow{5}{*}{37} & \multirow{5}{*}{Piroxicam} & $\alpha_1$ & 0.06 & 2.89 & 4 & 0.90 & 3 & 0.10 & 1.94 & 4 &  0.94 & 7 & 0.68 & 6 \\
    & & Form II ($\alpha_2$) & 0.08 & 3.85 & 10 & 1.20 & 6 & 0.13 & 2.12 & 5 &  0.10 & 2 & 0.39 & 2 \\
    & & Form III & 0.19 & 4.00 & 12 & 3.95 & 14 & 0.63 & 4.21 & 15 &  1.47 & 8 & 1.22 & 8 \\
    & & Form I & 0.23 & 4.17 & 14 & 2.78 & 12 & 0.25 & 3.54 & 12 &  1.73 & 9 & 2.33 & 12 \\
    & & Form VI & 0.22 & 7.14 & 22 & 6.11 & 21 & 0.27 & 6.99 & 22 &  5.32 & 22 & 4.47 & 22 \\
    \midrule
    \multirow{10}{*}{38} & \multirow{10}{*}{ROY} & ORP & 0.32 & 0.00 & 1 & 4.82 & 24 & 0.36 & 0.88 & 2 &  0.00 & 1 & 0.19 & 2 \\
    & & R & 0.14 & 1.27 & 4 & 2.85 & 12 & 0.11 & 1.83 & 7 &  0.14 & 2 & 0.00 & 1 \\
    & & PO13 & 0.10 & 3.48 & 34 & 5.03 & 26 & 0.11 & 5.10 & 119 &  1.66 & 5 & 1.74 & 9 \\
    & & ON & 0.22 & 4.26 & 63 & 5.10 & 29 & 0.25 & 4.78 & 98  &  1.88 & 8 & 1.91 & 11 \\
    & & OP & 0.14 & 5.31 & 137 & 4.26 & 22 & 0.18 & 5.56 & 165 &  2.52 & 15 & 2.86 & 40 \\
    & & Y04 & 0.29 & 5.96 & 197 & 3.29 & 14 & 0.23 & 5.86 & 183 &  4.84 & 135 & 4.56 & 148 \\
    & & Y19 & 0.16 & 6.78 & 287 & 5.79 & 34 & 0.13 & 8.09 & 424 &  4.59 & 117 & 4.62 & 157 \\
    & & Y & 0.22 & 7.14 & 336 & 0.00 & 1 & 0.26 & 7.47 & 330 &  6.36 & 336 & 6.71 & 465 \\
    & & YT04 & 0.22 & 7.87 & 494 & 1.48 & 4 & 0.23 & 8.47 & 478 &  5.50 & 218 & 5.91 & 313 \\
    & & YN & 0.27 & 8.35 & 583 & 2.80 & 10 & 0.25 & 8.43 & 472 &  7.07 & 479 & 6.14 & 362 \\

    \bottomrule
    \end{tabular}
    \end{adjustbox}
    \begin{tablenotes}
    \item \small{$^\bigtriangleup$ Obtained from a larger initial pool generated by 5 lowest-energy conformers and 3,000 structures per space group.}
    \end{tablenotes}
\end{table}

\begin{figure}[h!tbp]
    \centering
    \includegraphics[width=\textwidth]{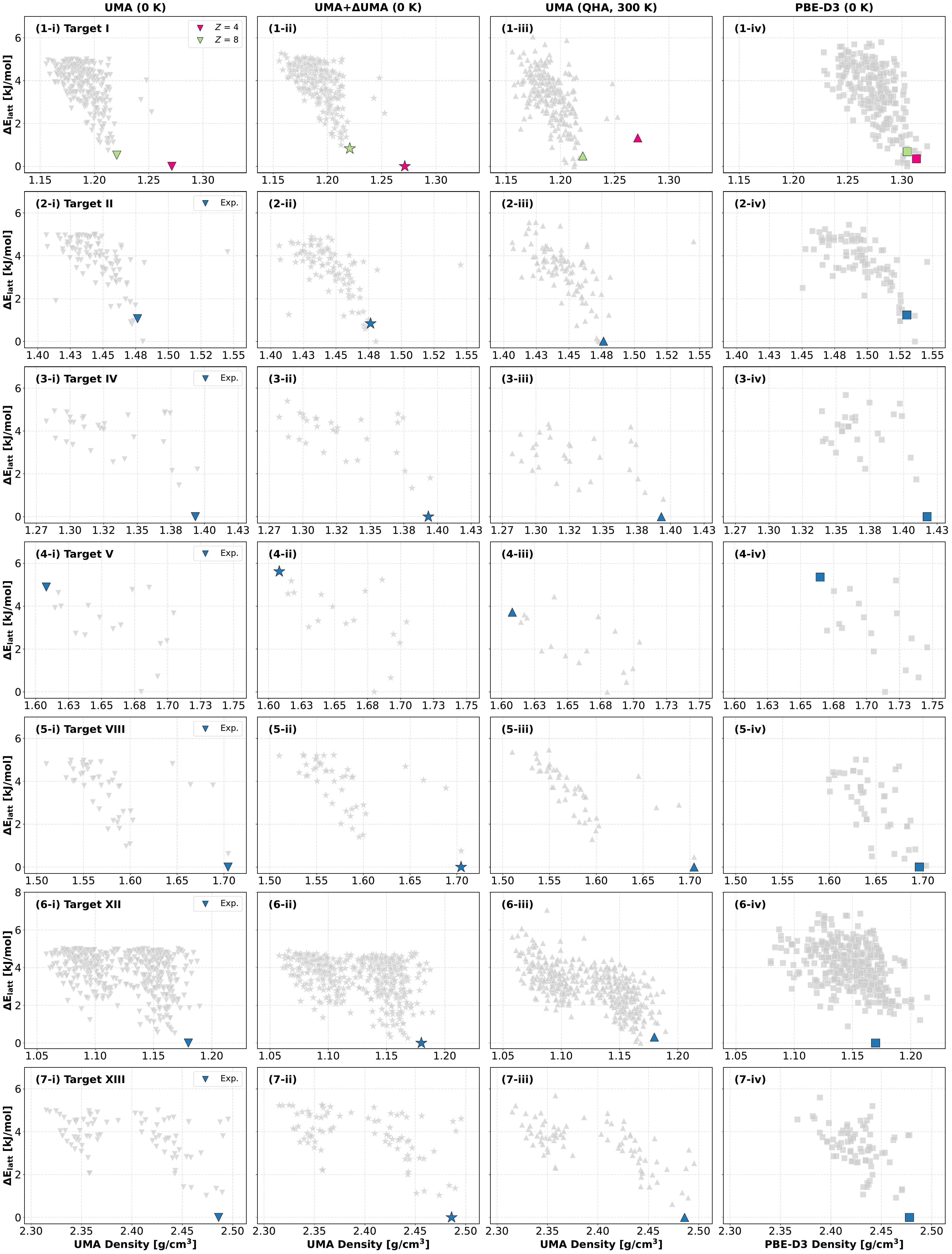}
    \caption{Energy landscapes for all 28 compounds in the Semi-Rigid molecules subset. For each compound, the relative energy landscape is shown as a function of density. Each row corresponds to one compound, and obtained using (i) UMA at 0 K, (ii) UMA+$\Delta$UMA at 0 K, (iii) UMA Gibbs free energies at 300 K, and (iv) PBE-D3 at 0 K. Experimentally observed polymorphs are colored.}
    \label{fig:landscape_all}
\end{figure}

\begin{figure}[h!tbp]
    \ContinuedFloat
    \centering
    \captionsetup{list=false}
    \includegraphics[width=\textwidth]{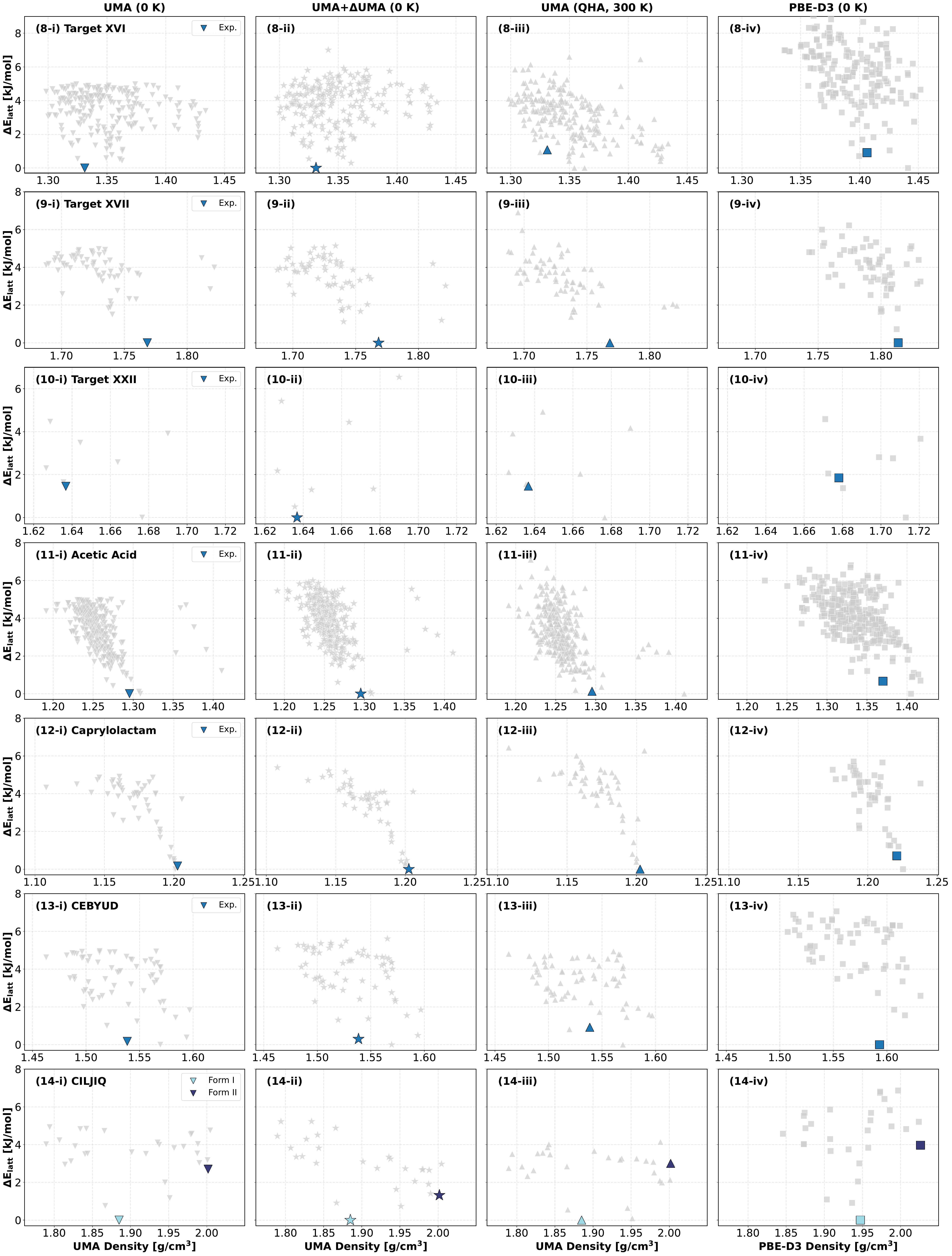}
    \caption{\textbf{(continued)} Energy landscapes for all 28 compounds in the Semi-Rigid molecules subset. For each compound, the relative energy landscape is shown as a function of density. Each row corresponds to one compound, and obtained using (i) UMA at 0 K, (ii) UMA+$\Delta$UMA at 0 K, (iii) UMA Gibbs free energies at 300 K, and (iv) PBE-D3 at 0 K. Experimentally observed polymorphs are colored.}
\end{figure}

\begin{figure}[h!tbp]
    \ContinuedFloat
    \centering
    \captionsetup{list=false}
    \includegraphics[width=\textwidth]{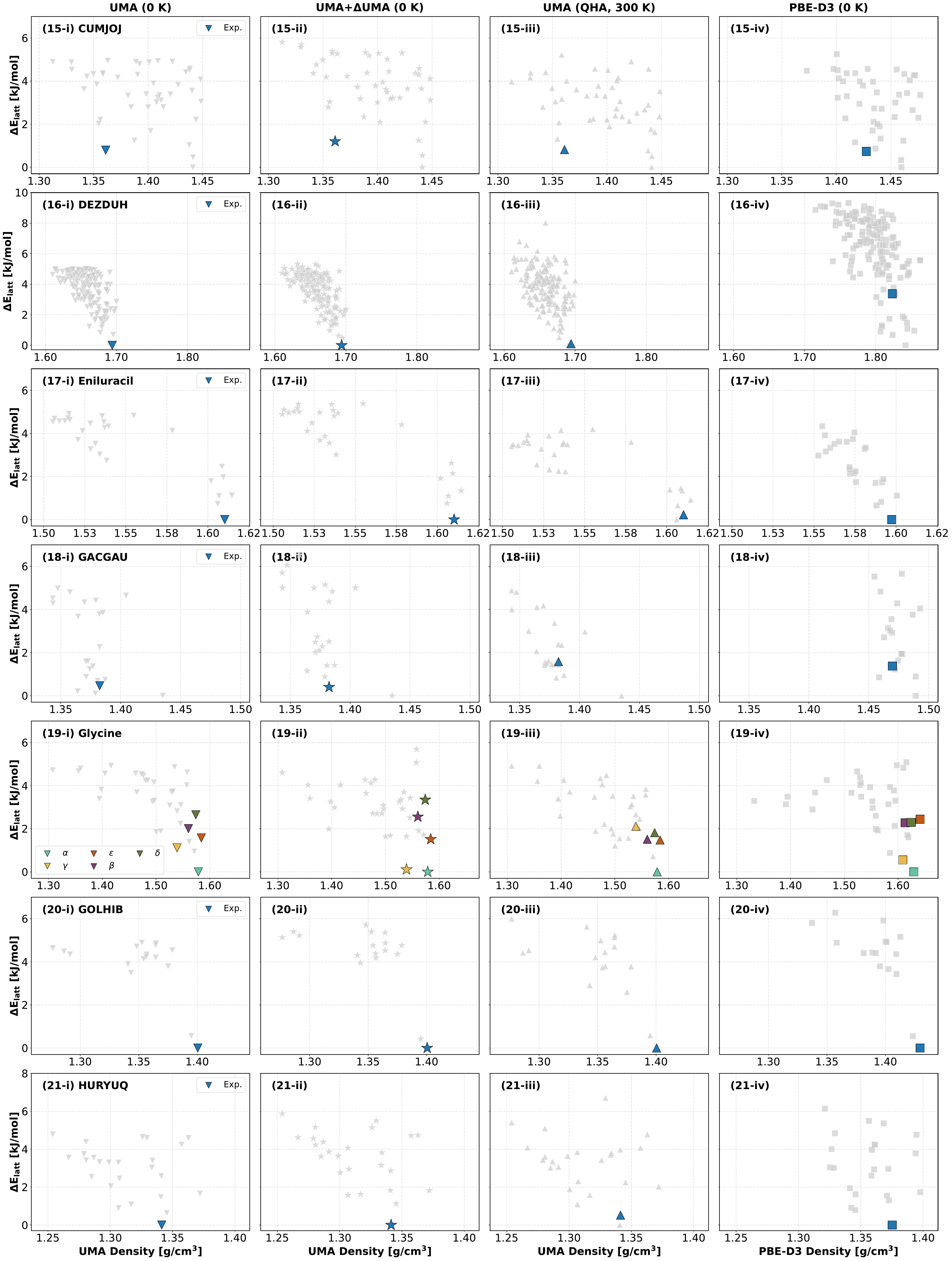}
    \caption{\textbf{(continued)} Energy landscapes for all 28 compounds in the Semi-Rigid molecules subset. For each compound, the relative energy landscape is shown as a function of density. Each row corresponds to one compound, and obtained using (i) UMA at 0 K, (ii) UMA+$\Delta$UMA at 0 K, (iii) UMA Gibbs free energies at 300 K, and (iv) PBE-D3 at 0 K. Experimentally observed polymorphs are colored.}
\end{figure}

\begin{figure}[h!tbp]
    \ContinuedFloat
    \centering
    \captionsetup{list=false}
    \includegraphics[width=\textwidth]{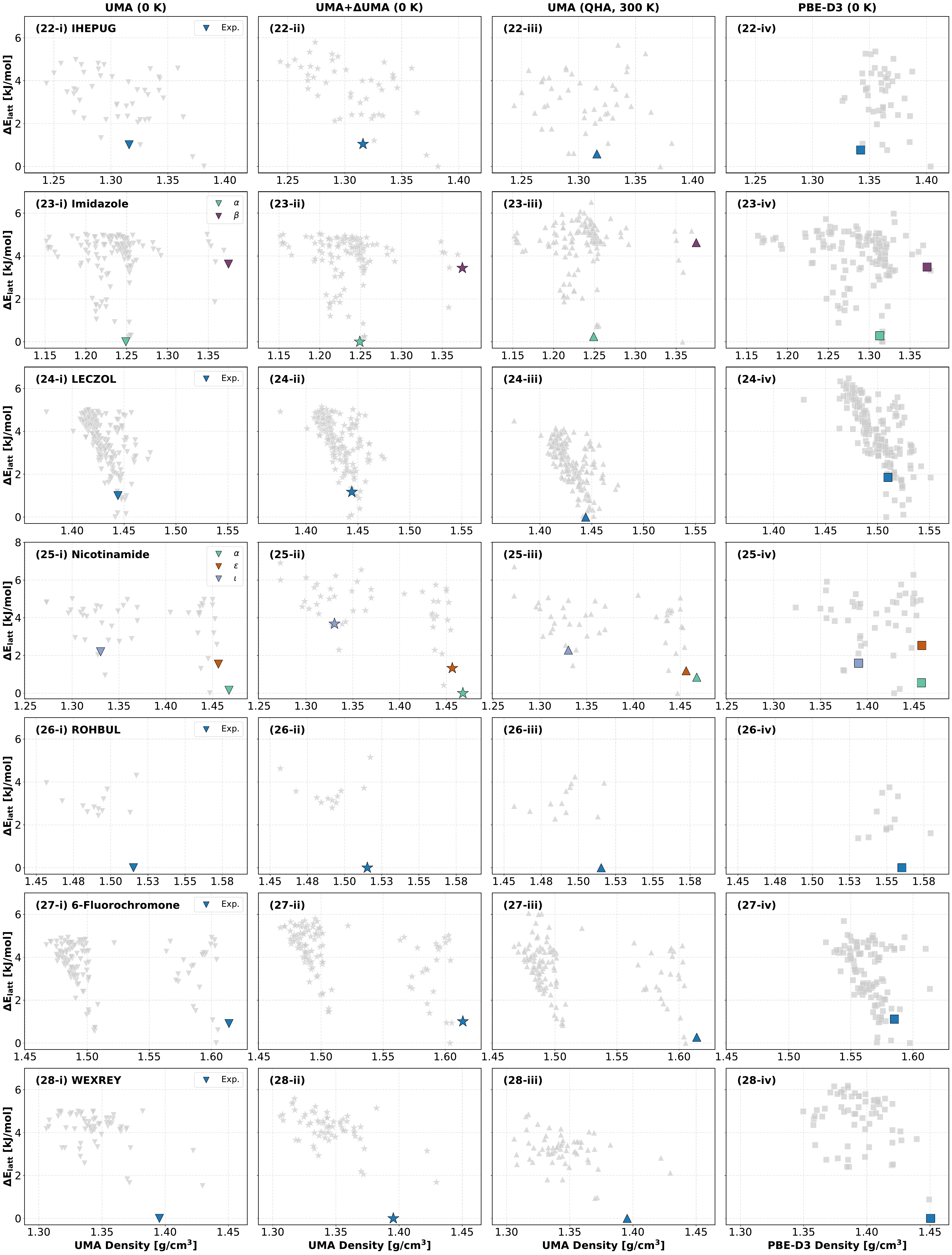}
    \caption{\textbf{(continued)} Energy landscapes for all 28 compounds in the Semi-Rigid molecules subset. For each compound, the relative energy landscape is shown as a function of density. Each row corresponds to one compound, and obtained using (i) UMA at 0 K, (ii) UMA+$\Delta$UMA at 0 K, (iii) UMA Gibbs free energies at 300 K, and (iv) PBE-D3 at 0 K. Experimentally observed polymorphs are colored.}
\end{figure}

\begin{figure}[h!tbp]
    \ContinuedFloat
    \centering
    \captionsetup{list=false}
    \includegraphics[width=\textwidth]{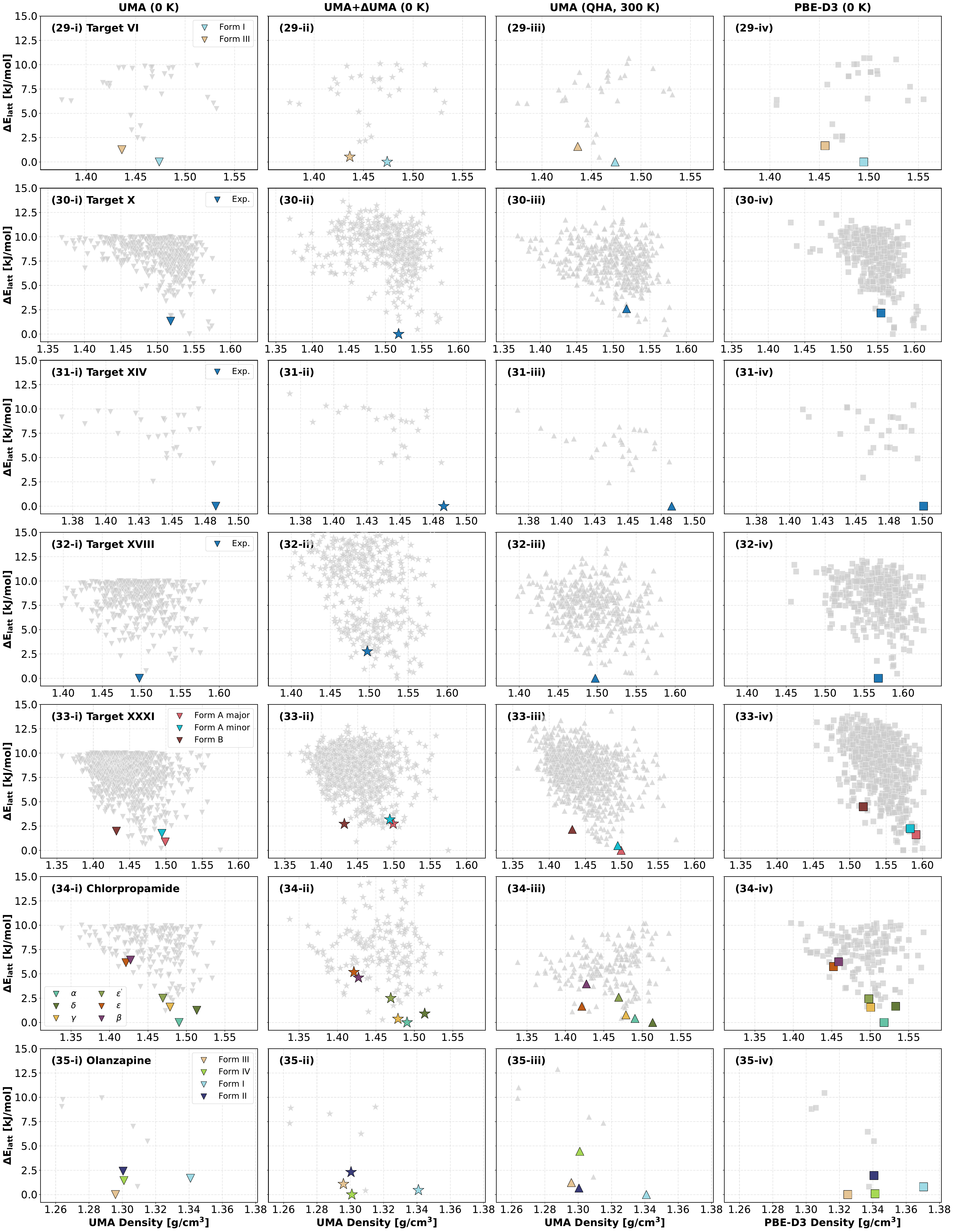}
    \caption{\textbf{(continued)} Energy landscapes for all 10 compounds in the Flexible molecules subset. For each compound, the relative energy landscape is shown as a function of density. Each row corresponds to one compound, and obtained using (i) UMA at 0 K, (ii) UMA+$\Delta$UMA at 0 K, (iii) UMA Gibbs free energies at 300 K, and (iv) PBE-D3 at 0 K. Experimentally observed polymorphs are colored.}
\end{figure}

\begin{figure}[h!tbp]
    \ContinuedFloat
    \centering
    \captionsetup{list=false}
    \includegraphics[width=\textwidth]{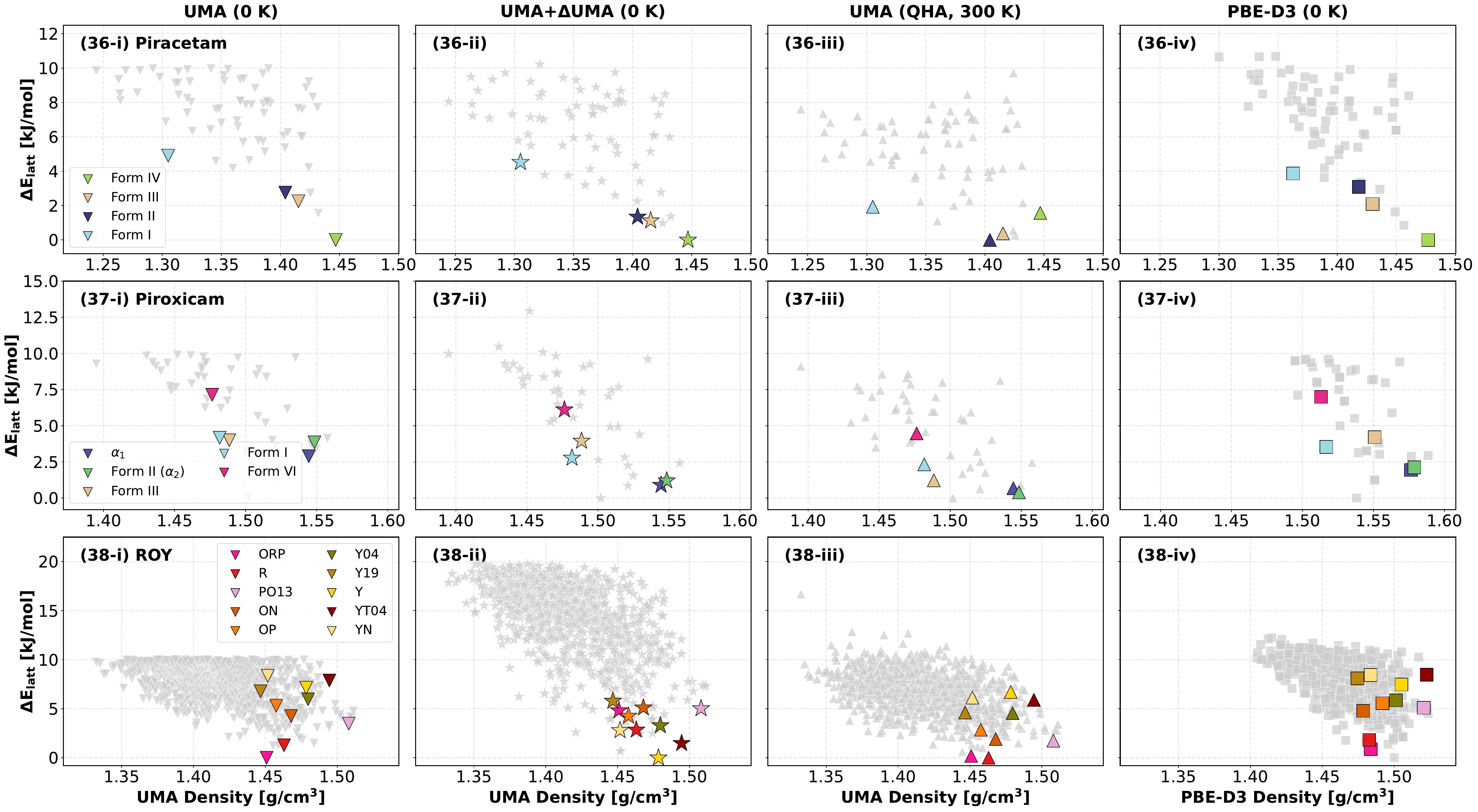}
    \caption{\textbf{(continued)} Energy landscapes for all 10 compounds in the Flexible molecules subset. For each compound, the relative energy landscape is shown as a function of density. Each row corresponds to one compound, and obtained using (i) UMA at 0 K, (ii) UMA+$\Delta$UMA at 0 K, (iii) UMA Gibbs free energies at 300 K, and (iv) PBE-D3 at 0 K. Experimentally observed polymorphs are colored.}
\end{figure}

\clearpage
\begin{figure}[htb!]
    \centering
    \includegraphics[width=0.87\textwidth]{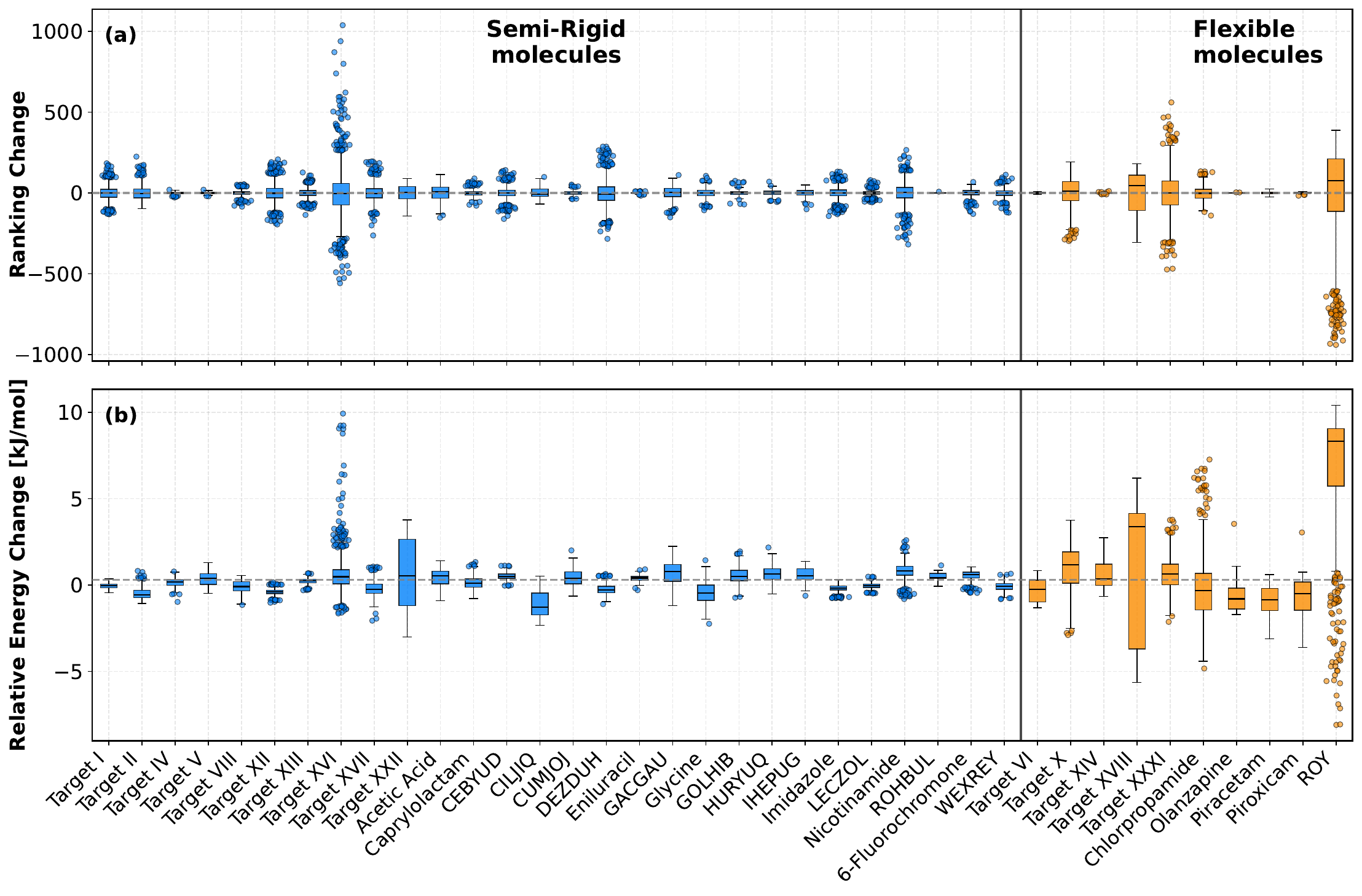}
    \caption{The distribution of (a) ranking changes and (b) relative energy changes between the lattice energy at 0 K and the free energy at 300 K, obtained using UMA. The gray dashed horizontal lines show the median values.}
    \label{fig:free_change}
\end{figure}

\begin{figure}[htb!]
    \centering
    \includegraphics[width=0.87\textwidth]{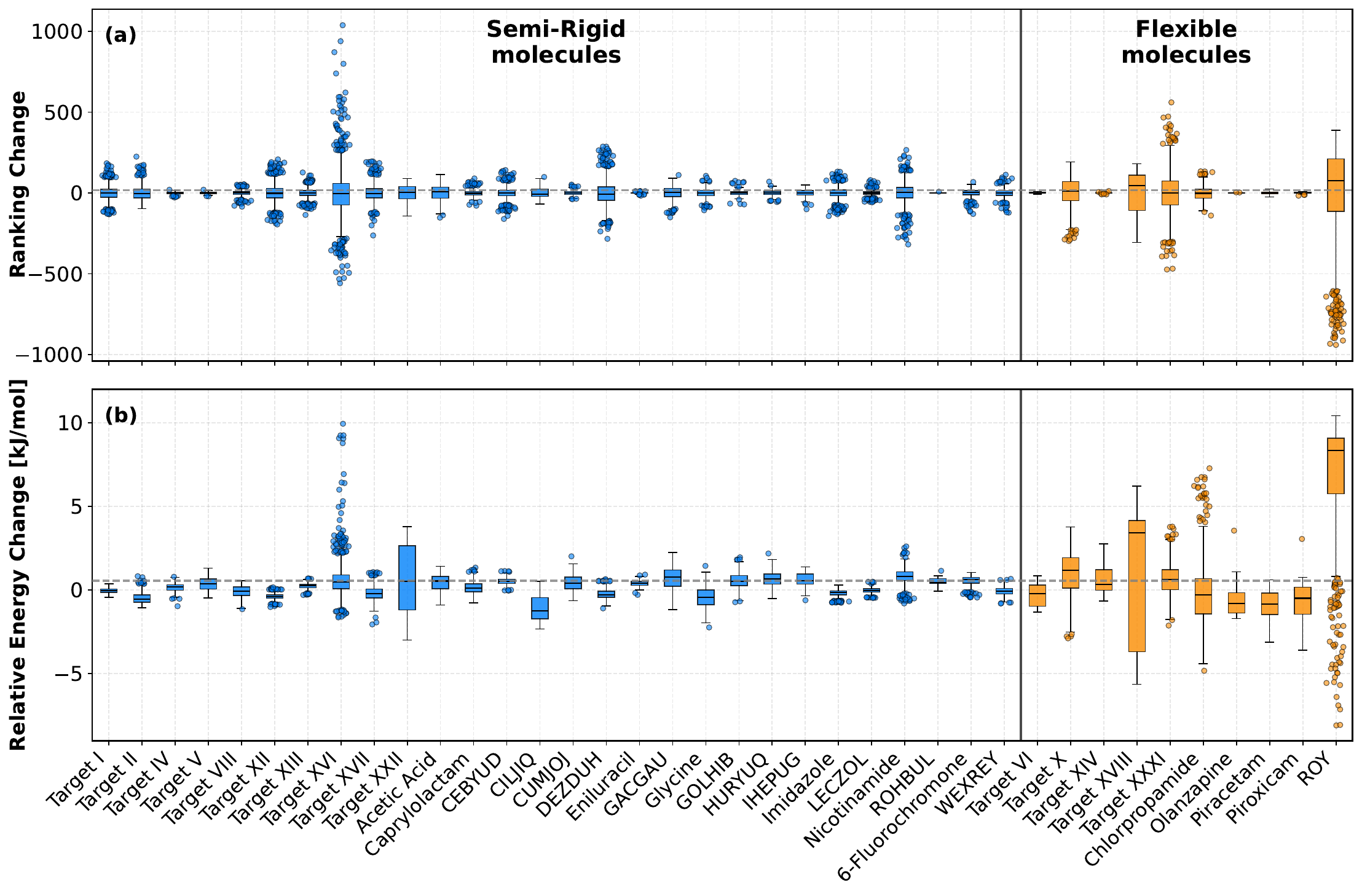}
    \caption{The distribution of (a) ranking changes and (b) relative energy changes between the lattice energy at 0 K and the lattice energy at 0 K with conformer energy corrections, obtained using UMA. The gray dashed horizontal lines show the median values.}
    \label{fig:corr_change}
\end{figure}

\newpage
 \begin{table}[h!tbp]
    \centering
    \caption{Comparison of the relative energy and ranking obtained with UMA to results reported in repositories associated with Ref.~\cite{zhou2025robust}, obtained using r$^2$SCAN-D3 at the PBE-D3 optimized geometries.
    }
    \label{tab:rank_sc}
    \begin{adjustbox}{max width=0.75\textwidth, max height=0.75\textheight, center}
    \begin{tabular}{@{}cllcccccc@{}}
    \toprule
    \multirow{2}{*}{\textbf{No.}} &
    \multirow{2}{*}{\textbf{Compound}} & \multirow{2}{*}{\textbf{Form}}  & \multicolumn{2}{c}{\textbf{UMA (0 K)}} & \multicolumn{2}{c}{\textbf{UMA+\boldmath{$\Delta$}UMA (0 K)}} & \multicolumn{2}{c}{\textbf{r$^2$SCAN-D3}~\cite{zhou2025robust}} \\
    \cmidrule(lr){4-5} \cmidrule(lr){6-7} \cmidrule(lr){8-9} 
    & & & \boldmath{$\Delta E$} \textbf{[kJ/mol]} & \textbf{Rank} & \boldmath{$\Delta E$} \textbf{[kJ/mol]} & \textbf{Rank} & \boldmath{$\Delta E$} \textbf{[kJ/mol]} & \textbf{Rank} \\ 
    \midrule
    
    \multirow{2}{*}{1} & \multirow{2}{*}{Target I} & $Z = 8$  &  0.52 & 2 & 0.00 & 1 & 1.92 & 2 \\
    && $Z =4$  & 0.00 & 1 & 0.83 & 2 & 0.00 & 1 \\
    \midrule
    4 & Target V & Exp. & 4.89 & 17 & 5.62 & 17 & 2.76 & 5\\
    \midrule
    \multirow{3}{*}{19} & \multirow{3}{*}{Glycine} & $\gamma$ &  1.12 & 3 & 0.11 & 2 & 0.00 & 1\\
    && $\alpha$ & 0.00 & 1 & 0.00 & 1 & 1.55 & 2\\
    && $\beta$ & 2.01 & 8 & 2.56 & 11 & 2.34 & 3 \\
    \midrule
    \multirow{3}{*}{25} & \multirow{3}{*}{Nicotinamide} & $\alpha$ & 0.15 & 2 & 0.00 & 1 & 0.00 & 1\\
    && $\varepsilon$ & 1.54 & 5 & 1.33 & 3 & 1.63 & 3 \\
    && $\iota$ & 2.20 & 8 & 3.68 & 11 & 4.60 & 18 \\
    \midrule

    \multirow{2}{*}{29} & \multirow{2}{*}{Target VI} & Form I & 0.00 & 1 & 0.00 & 1 & 0.00 & 1\\
    && Form III & 1.26 & 2 & 0.53 & 2 & 0.67 & 2 \\
    \midrule
    % 2 & Target X & Exp. & 1.32 & 5 &  &  &  & \\
    % \midrule
    % 3 & Target XIV & Exp. & 0.00 & 1 &  &  &  & \\
    % \midrule
    % 4 & Target XVIII & Exp. & 0.00 & 1 &   &  &  & \\
    % \midrule
    \multirow{3}{*}{33} & \multirow{3}{*}{Target XXXI} & Form A maj & 0.86 & 6 & 2.73 & 10 & 1.09 & 3 \\
    && Form A min & 1.74 & 13 & 3.17 & 23 & - & - \\
    && Form B & 1.97 & 15 & 2.72 & 9 & 5.02 & 25 \\
    \midrule
    \multirow{6}{*}{34} & \multirow{6}{*}{Chlorpropamide} & $\alpha$ & 0.00 & 1 & 0.00 & 1 & 0.88 & 6 \\
    && $\delta$ & 1.25 & 3 & 0.90 & 4 & 0.17 & 3 \\
    && $\gamma$ & 1.58 & 4 & 0.37 & 3 & 0.04 & 2 \\
    && $\varepsilon^{'}$ & 2.50 & 6 & 2.50 & 9 & 1.46 & 7 \\
    && $\varepsilon$ & 6.15 & 37 & 5.18 & 35 & 4.44 & 17 \\
    && $\beta$ & 6.42 & 44 & 4.61 & 27 & 3.97 & 15 \\
    \midrule
    \multirow{4}{*}{35} & \multirow{4}{*}{Olanzapine} & Form I & 1.68 & 4 & 0.46 & 3  & 0.25 & 2 \\
    && Form II & 2.41 & 5 & 2.31 & 5 & 2.97 & 5 \\
    && Form III & 0.00 & 1 & 1.08 & 4 & 0.71 & 3 \\
    && Form VI & 1.44 & 3 & 0.00 & 1 & 0.00 & 1 \\

    % \midrule
    % \multirow{4}{*}{Piracetam} & Form IV & 0.00 & 1 & 0.00 & 1 & - & - \\
    % & Form III & 2.25 & 3 & 1.12 & 3 & - & - \\
    % & Form II & 2.75 & 5 & 1.34 & 4 & - & - \\
    % & Form I maj & 4.91 & 10 & 4.52 & 16 & - & - \\
    \midrule
    \multirow{5}{*}{37} & \multirow{5}{*}{Piroxicam} & $\alpha_1$ & 2.89 & 4 & 0.90 & 3 & 2.26 & 5 \\
    && Form II ($\alpha_2$) & 3.85 & 10 & 1.20 & 6 & 2.47 & 9 \\
    && Form III & 4.00 & 12 & 3.95 & 14 & 1.05 & 2 \\
    && Form I & 4.17 & 14 & 2.78 & 12 & 0.00 & 1 \\
    && Form VI & 7.14 & 22 & 6.11 & 21 & 8.37 & 16 \\
    \midrule
    \multirow{10}{*}{38} & \multirow{10}{*}{ROY} & ORP & 0.00 & 1 & 4.82 & 24 & 5.44 & 10 \\
    && R & 1.27 & 4 &  2.85 & 12 & 0.75 & 2 \\
    && PO13 & 3.48 & 34 & 5.03 & 26 & 4.18 & 7 \\
    && ON & 4.26 & 63 & 5.10 & 29 & 6.07 & 13 \\
    && OP & 5.31 & 137 & 4.26 & 22 & 2.05 & 4 \\
    && Y04 & 5.96 & 197 & 3.29 & 14 & 2.80 & 6 \\
    && Y19 & 6.78 & 287 & 5.79 & 34 & 7.11 & 20 \\
    && Y & 7.14 & 336 & 0.00 & 1 & 0.00 & 1 \\
    && YT04 & 7.87 & 494 & 1.48 & 4 & 1.30 & 3 \\
    && YN & 8.35 & 583 & 2.80 & 10 & 2.51 & 5 \\
    \midrule
    \end{tabular}
    \end{adjustbox}
 \end{table} 

%% file: appendix/mlip_eval.tex
\section{UMA Performance}
\label{app:uma_eval}

\begin{table}[h!tbp]
    \small
    \centering
    \caption{Summary of the benchmark for the single point energy and the relaxation tasks, comparing UMA against PBE-D3 reference calculations at 0 K. Relative lattice energy represents the lattice energy difference between each structure and the global minimum structure as determined by each respective method. Values shown in parentheses indicate performance when using the structure with the lowest DFT energy as the global reference. In cases where no parentheses are shown, the results are unaffected by the choice of method for referencing. The bolded entries are the worst values for each metric. RMSD$_{30} < 1$ \AA\ tolerance is reported for matching.}
    \label{tab:uma_eval_results}
    \begin{adjustbox}{max width=\textwidth, max height=0.75\textheight, center}
    \begin{tabular}{@{}clcccccccc@{}}
    \toprule
    \multirow{2}{*}{\textbf{No.}} & \multirow{2}{*}{\textbf{Compound}} & \multicolumn{4}{c}{\textbf{Single Point Energy}} & \multicolumn{4}{c}{\textbf{Relaxation}} \\
    \cmidrule(lr){3-6} \cmidrule(lr){7-10}
    & & \textbf{MAE [kJ/mol]} & \textbf{R\boldmath{$^2$}} & \textbf{Spearman} & \textbf{Kendall} & \textbf{MAE [kJ/mol]} & \textbf{Spearman} & \textbf{Match} & \textbf{RMSD$_{\mathbf{30}}$ [\AA]}\\
    \midrule

    1 & Target I & 0.52 & 0.90 & 0.92 & 0.76 & 0.70 (2.62) & 0.75 & 98\% & 0.33 \\
    \midrule
    2 & Target II & 0.24 & 0.95 & 0.96 & 0.83 & 0.31 & 0.88 & 100\% & 0.18 \\
    \midrule
    3 & Target IV & 0.26 & 0.94 & 0.98 & 0.90 & 0.24 & 0.93 & 100\% & 0.14 \\
    \midrule
    4 & Target V & 0.40 & 0.89 & 0.93 & 0.81 & 0.53 & 0.88 & 100\% & 0.11 \\
    \midrule
    5 & Target VIII & 0.31 & 0.94 & 0.92 & 0.79 & 0.71 & 0.73 & 98\% & 0.21 \\
    \midrule
    6 & Target XII & 0.66 & 0.58 & 0.74 & 0.55 & 0.73 & 0.68 & 98\% & 0.20 \\
    \midrule
    7 & Target XIII & 0.38 & 0.86 & 0.92 & 0.75 & 0.55 & 0.84 & 99\% & 0.17 \\
    \midrule
    8 & Target XVI & \textbf{1.84} & \textbf{0.44} & \textbf{0.62} & \textbf{0.44} & 2.32 \textbf{(6.98)} & 0.51 & 98\% & 0.24 \\
    \midrule
    9 & Target XVII & 0.67 & 0.82 & 0.86 & 0.69 & 0.71 & 0.66 & 97\% & 0.22 \\
    \midrule
    10 & Target XXII & 0.32 & 0.96 & 0.98 & 0.93 & 0.27 & 0.98 & 100\% & 0.12 \\
    \midrule
    11 & Acetic Acid & 0.74 & 0.89 & 0.94 & 0.78 & 1.08 & 0.68 & 93\% & 0.30 \\
    \midrule
    12 & Caprylolactam & 0.20 & 0.98 & 0.96 & 0.87 & 0.54 & 0.88 & 100\% & 0.12 \\
    \midrule
    13 & CEBYUD & 1.48 (1.65) & 0.89 & 0.91 & 0.76 & 1.86 (2.02) & 0.83 & 98\% & 0.12 \\
    \midrule
    14 & CILJIQ & 0.88 & 0.84 & 0.83 & 0.66 & 1.15 & 0.73 & 100\% & 0.16 \\
    \midrule
    15 & CUMJOJ & 0.48 (0.81) & 0.91 & 0.94 & 0.79 & 0.61 (0.57) & 0.75 & 100\% & 0.16 \\
    \midrule
    16 & DEZDUH & 0.35 & 0.89 & 0.89 & 0.73 & \textbf{2.90} (3.52) & 0.71 & 99\% & 0.33 \\
    \midrule
    17 & Eniluracil  & 0.69 & 0.95 & 0.94 & 0.82 & 0.82 & 0.89 & 100\% & 0.12 \\
    \midrule
    18 & GACGAU  & 1.45 & 0.93 & 0.96 & 0.85 & 1.26 & 0.68 & \textbf{90\%} & 0.22 \\
    \midrule
    19 & Glycine  & 0.57 & 0.73 & 0.77 & 0.61 & 0.54 & 0.80 & 100\% & 0.10 \\
    \midrule
    20 & GOLHIB  & 0.53 & 0.86 & 0.77 & 0.60 & 0.50 & 0.70 & 100\% & 0.17 \\
    \midrule
    21 & HURYUQ  & 0.28 & 0.94 & 0.98 & 0.91 & 0.61 & 0.89 & 100\% & 0.16 \\
    \midrule
    22 & IHEPUG  & 0.50 & 0.86 & 0.92 & 0.77 & 0.47 & 0.84 & 100\% & 0.20 \\
    \midrule
    23 & Imidazole & 0.16 & 0.98 & 0.96 & 0.84 & 0.35 (0.45) & 0.80 & 100\% & 0.19 \\
    \midrule
    24 & LECZOL & 0.25 & 0.92 & 0.95 & 0.81 & 0.86 & 0.71 & 96\% & \textbf{0.34} \\
    \midrule
    25 & Nicotinamide  & 0.37 & 0.91 & 0.89 & 0.75 & 0.55 (2.99) & 0.84 & 98\% & 0.16 \\
    \midrule
    26 & ROHBUL & 0.49 & 0.75 & 0.85 & 0.64 & 0.79 & \textbf{0.43} & 100\% & 0.21 \\
    \midrule
    27 & 6-Fluorochromone & 0.68 (0.34) & 0.81 & 0.86 & 0.71 & 0.69 (4.32) & 0.73 & 98\% & 0.31 \\
    \midrule
    28 & WEXREY & 0.55 & 0.81 & 0.74 & 0.56 & 0.74 & 0.65 & 100\% & 0.23 \\
    \midrule
    \multicolumn{2}{c}{\textbf{MEAN}} & \textbf{0.58} \textbf{(0.59)} & \textbf{0.86} & \textbf{0.89} & \textbf{0.75} & \textbf{0.84} \textbf{(1.32)} & \textbf{0.76} & \textbf{99\%} & \textbf{0.19} \\

    \midrule
    \end{tabular}
    \end{adjustbox}
\end{table}

\newpage

\begin{table}[h!tbp]
    \ContinuedFloat
    \small
    \centering
    \caption{\textbf{(continued)} Summary of the benchmark for the single-point energy and the relaxation tasks, comparing UMA against PBE-D3 reference calculations at 0 K. Relative lattice energy represents the lattice energy difference between each structure and the global minimum structure as determined by each respective method. Values shown in parentheses indicate performance when using the structure with the lowest DFT energy as the global reference. In cases where no parentheses are shown, the results are unaffected by the choice of method for referencing. The bolded entries are the worst values for each metric. RMSD$_{30} < 1$ \AA\ tolerance is reported for matching.}
    \begin{adjustbox}{max width=\textwidth, max height=0.75\textheight, center}
    \begin{tabular}{@{}clcccccccc@{}}
    \toprule
    \multirow{2}{*}{\textbf{No.}} & \multirow{2}{*}{\textbf{Compound}} & \multicolumn{4}{c}{\textbf{Single-Point Energy}} & \multicolumn{4}{c}{\textbf{Relaxation}} \\
    \cmidrule(lr){3-6} \cmidrule(lr){7-10}
    & & \textbf{MAE [kJ/mol]} & \textbf{R\boldmath{$^2$}} & \textbf{Spearman} & \textbf{Kendall} & \textbf{MAE [kJ/mol]} & \textbf{Spearman} & \textbf{Match} & \textbf{RMSD$_{\mathbf{30}}$ [\AA]}\\
    \midrule

    29 & Target VI & 0.40 & 0.97 & 0.95 & 0.85 & 0.88 & 0.88 & 100\% & 0.13 \\
    \midrule
    30 & Target X & 0.43 & 0.92 & 0.94 & 0.79 & 0.99 & \textbf{0.73} & 96\% & 0.22 \\
    \midrule
    31 & Target XIV & 0.39 & 0.96 & 0.94 & 0.83 & 0.72 & 0.92 & 96\% & 0.09 \\
    \midrule
    32 & Target XVIII & 0.92 & \textbf{0.86} & \textbf{0.87} & \textbf{0.70} & 0.97 & 0.82 & 100\% & 0.23 \\
    \midrule
    33 & Target XXXI & \textbf{2.10} & \textbf{0.86} & 0.88 & \textbf{0.70} & \textbf{1.96} (2.84) & \textbf{0.73} & \textbf{94\%} & \textbf{0.35} \\
    \midrule
    34 & Chlorpropamide & 0.29 & 0.97 & 0.97 & 0.86 & 0.74 & 0.81 & 97\% & 0.17 \\
    \midrule
    35 & Olanzapine & 0.51 & 0.98 & 0.96 & 0.91 & 0.48 & 0.96 & 100\% & 0.14 \\
    \midrule
    36 & Piracetam & 0.43 & 0.96 & 0.96 & 0.85 & 0.51 & 0.90 & 100\% & 0.13 \\
    \midrule
    37 & Piroxicam & 0.47 & 0.96 & 0.95 & 0.85 & 0.78 & 0.92 & 100\% & 0.21 \\
    \midrule
    38 & ROY & 0.58 & 0.88 & 0.91 & 0.74 & 0.94 \textbf{(7.26)} & 0.81 & 95\% & 0.24 \\
    \midrule
    & \textbf{MEAN} & \textbf{0.65} & \textbf{0.93} & \textbf{0.93} & \textbf{0.81} & \textbf{0.90} \textbf{(1.62)} & \textbf{0.85} & \textbf{98\%} & \textbf{0.19} \\
    \midrule
    \end{tabular}
    \end{adjustbox}
\end{table}

\newpage

\begin{figure}[h!tbp]
    \centering
    \includegraphics[width=\textwidth]{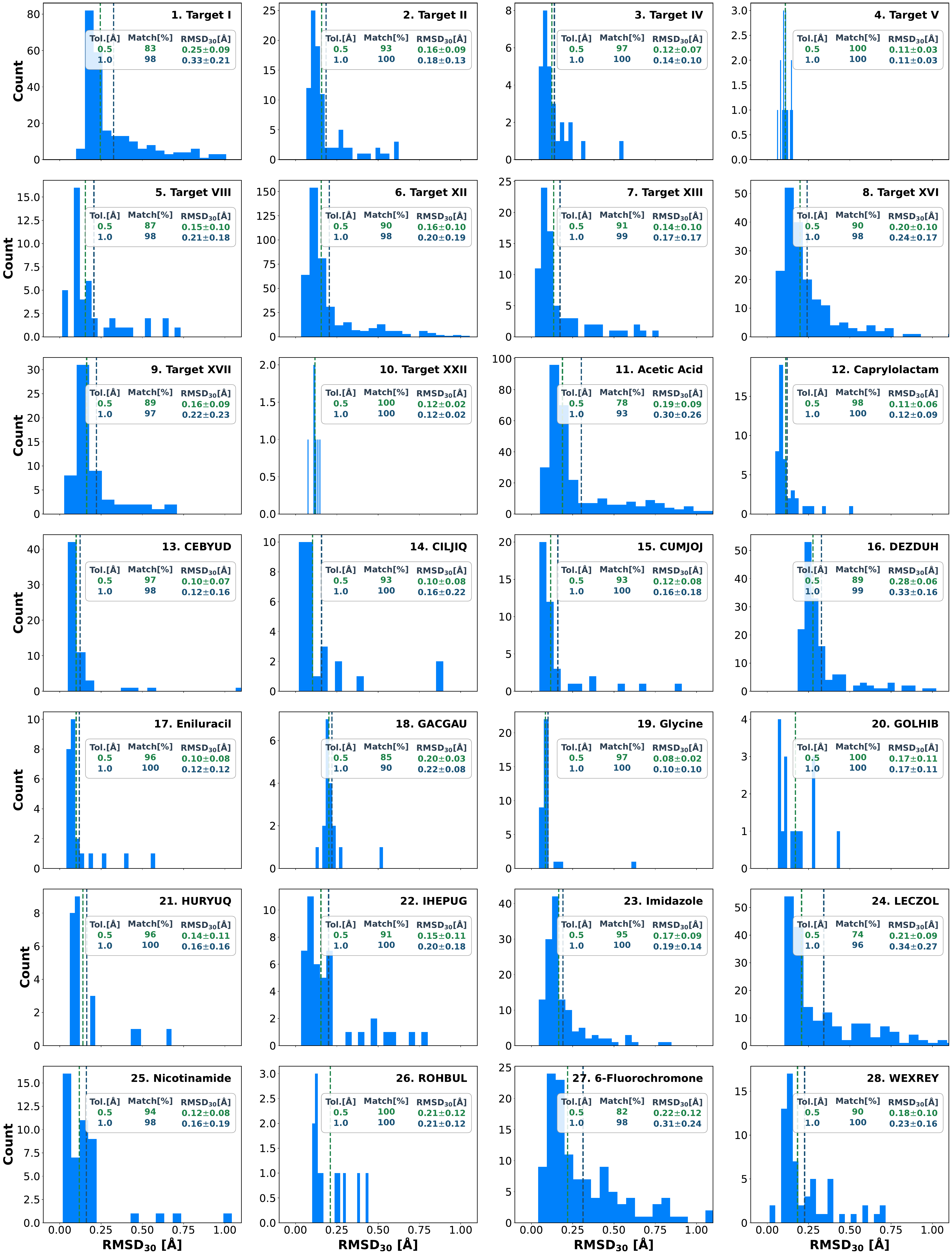}
    \caption{RMSD$_{30}$ histograms comparing the relaxed crystal structures obtained with the UMA model to those obtained with PBE-D3 relaxation for all 28 compounds in the Semi-Rigid molecules subset.}
    \label{fig:rmsd30_all}
\end{figure}

\begin{figure}[h!tbp]
    \ContinuedFloat
    \centering
    \includegraphics[width=\textwidth]{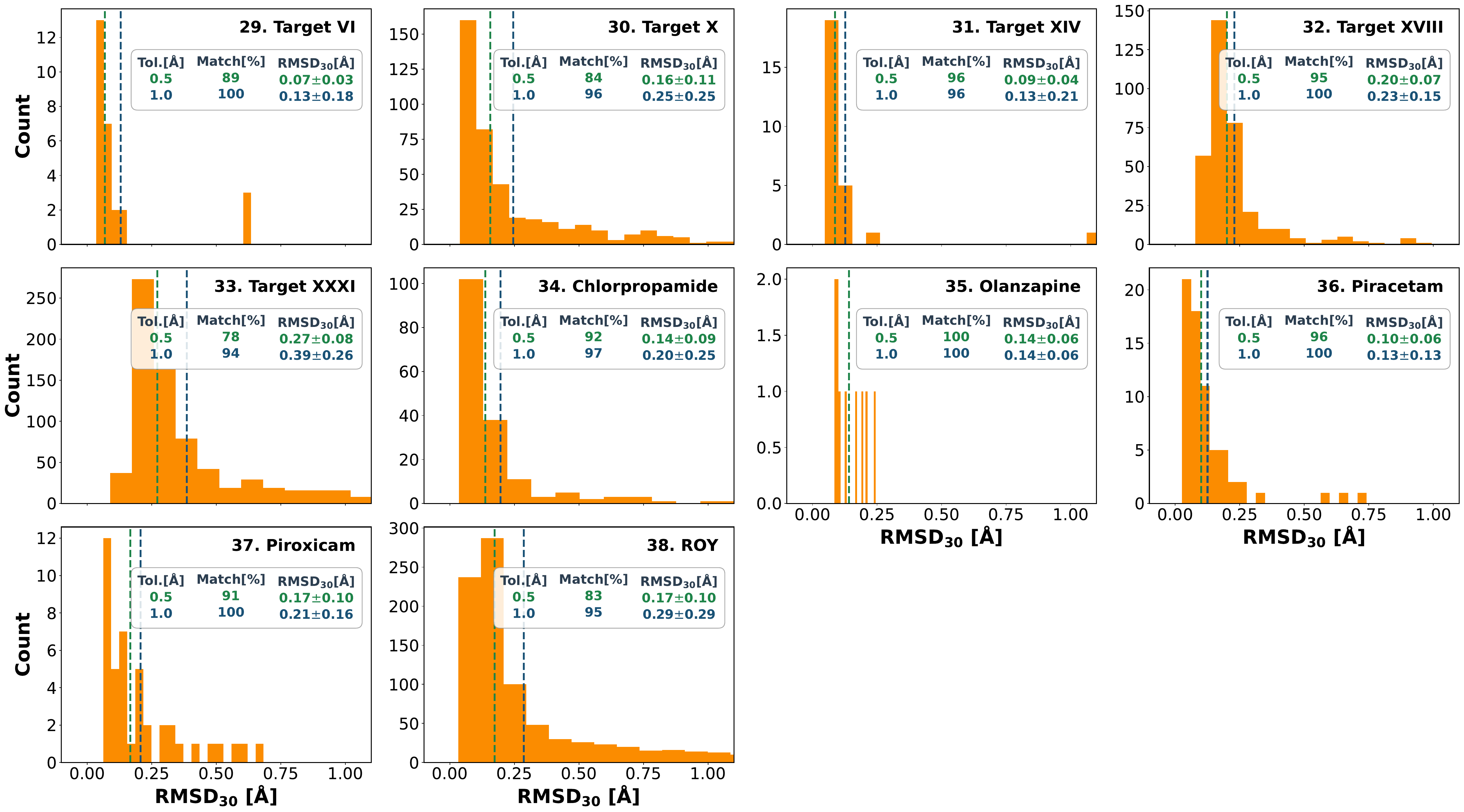}
    \caption{\textbf{(continued)} RMSD$_{30}$ histograms comparing the relaxed crystal structures obtained with the UMA model to those obtained with PBE-D3 relaxation for all 10 compounds in the Flexible molecules subset.}
\end{figure}

\begin{figure}[h!tbp]
    \centering
    \includegraphics[width=\textwidth]{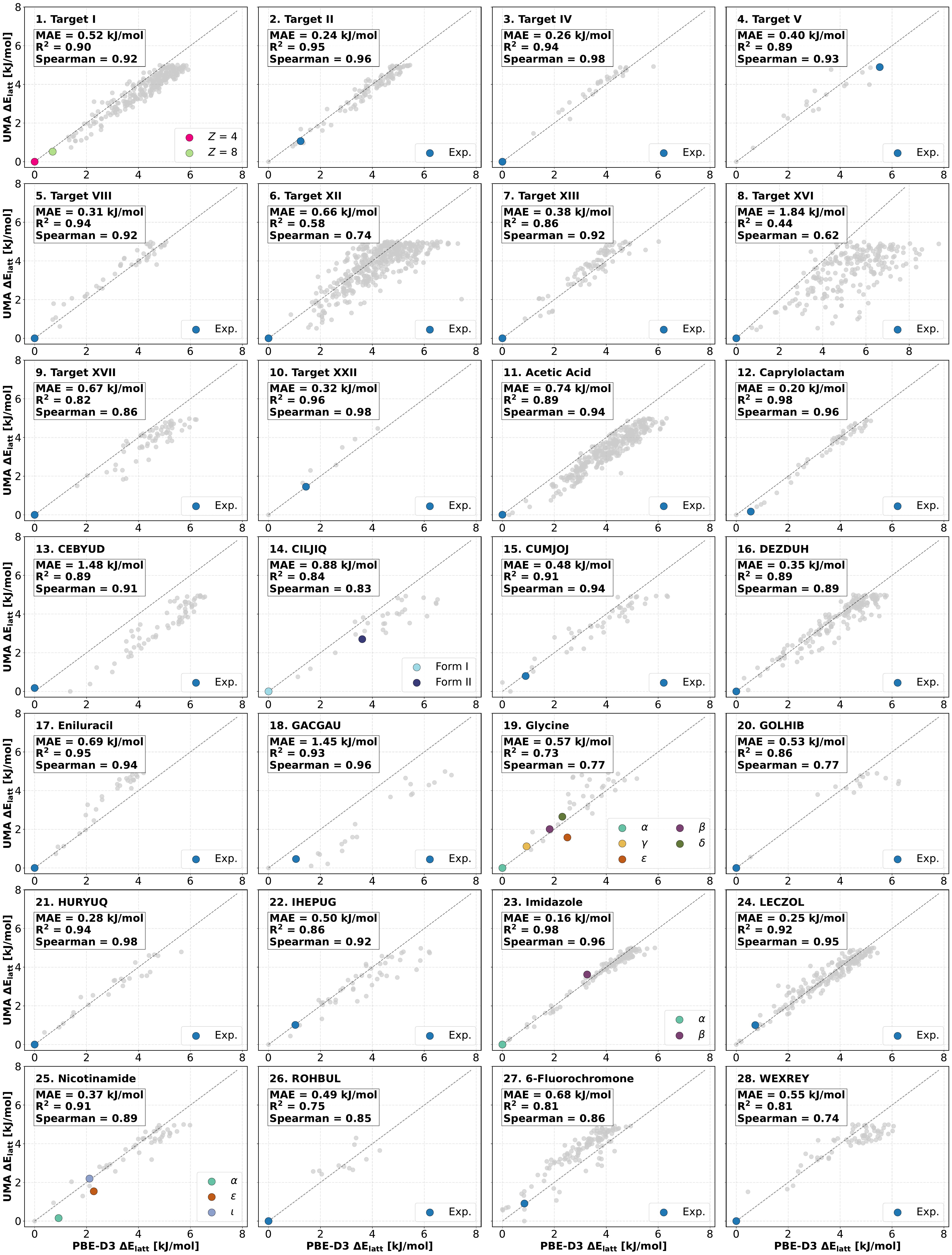}
    \caption{Parity plots of the single point energy task for all 28 compounds in the Semi-Rigid molecules subset. Relative (0~K) lattice energy represents the lattice energy difference between each structure and the global minimum structure of each method.}
    \label{fig:spe_all}
\end{figure}

\begin{figure}[h!tbp]
    \ContinuedFloat
    \centering
    \includegraphics[width=\textwidth]{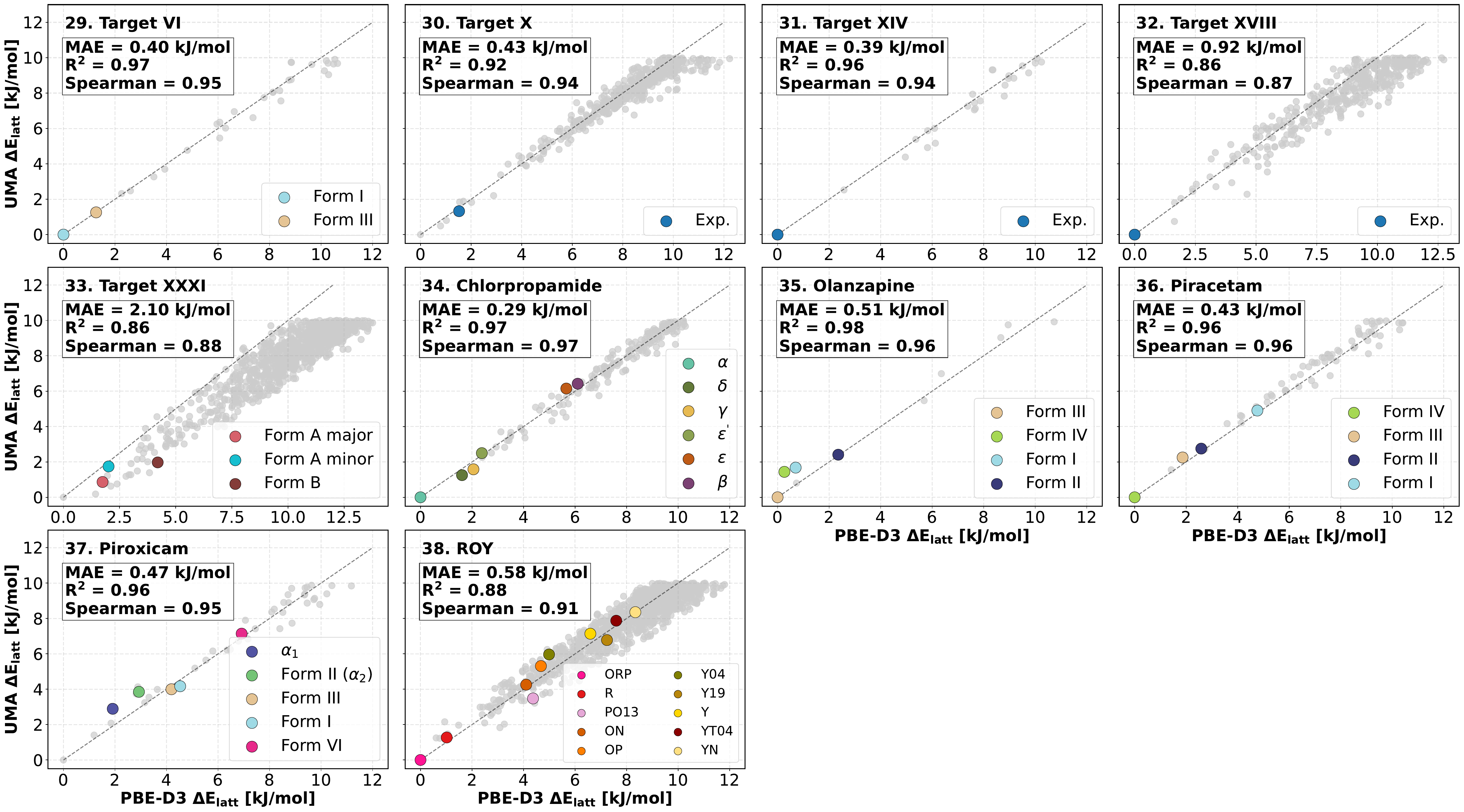}
    \caption{\textbf{(continued)} Parity plots of the single point energy task for all 10 compounds in the Flexible molecules subset. Relative (0~K) lattice energy represents the lattice energy difference between each structure and the global minimum structure of each method.}
\end{figure}

\begin{figure}[h!tbp]
    \centering
    \includegraphics[width=\textwidth]{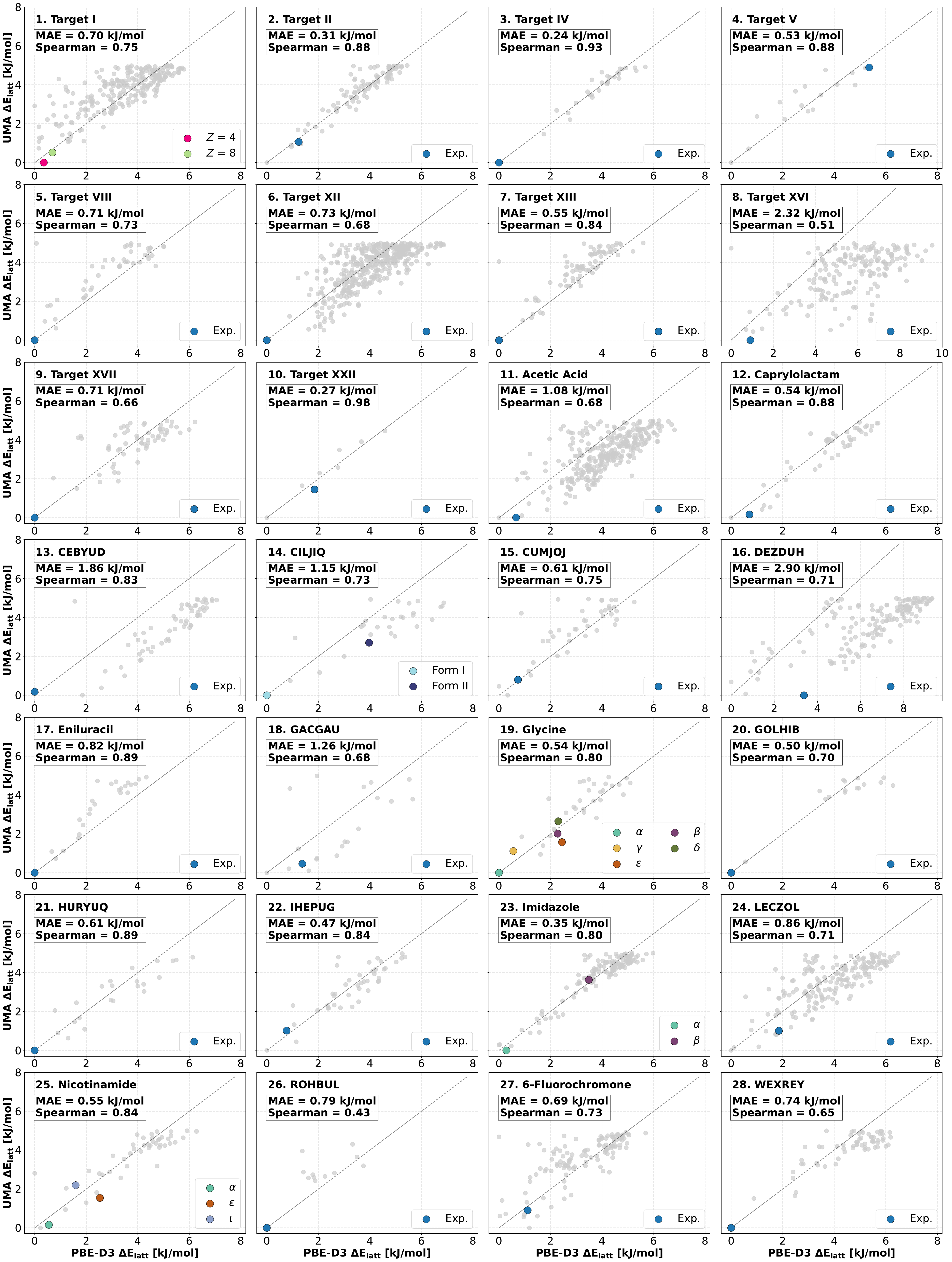}
    \caption{Parity plots for relaxation performance for all 28 compounds in the Semi-Rigid molecules subset. Relative (0~K) lattice energy represents the lattice energy difference between each structure and the global minimum of each method.}
    \label{fig:opt_all}
\end{figure}

\begin{figure}[h!tbp]
    \ContinuedFloat
    \centering
    \includegraphics[width=\textwidth]{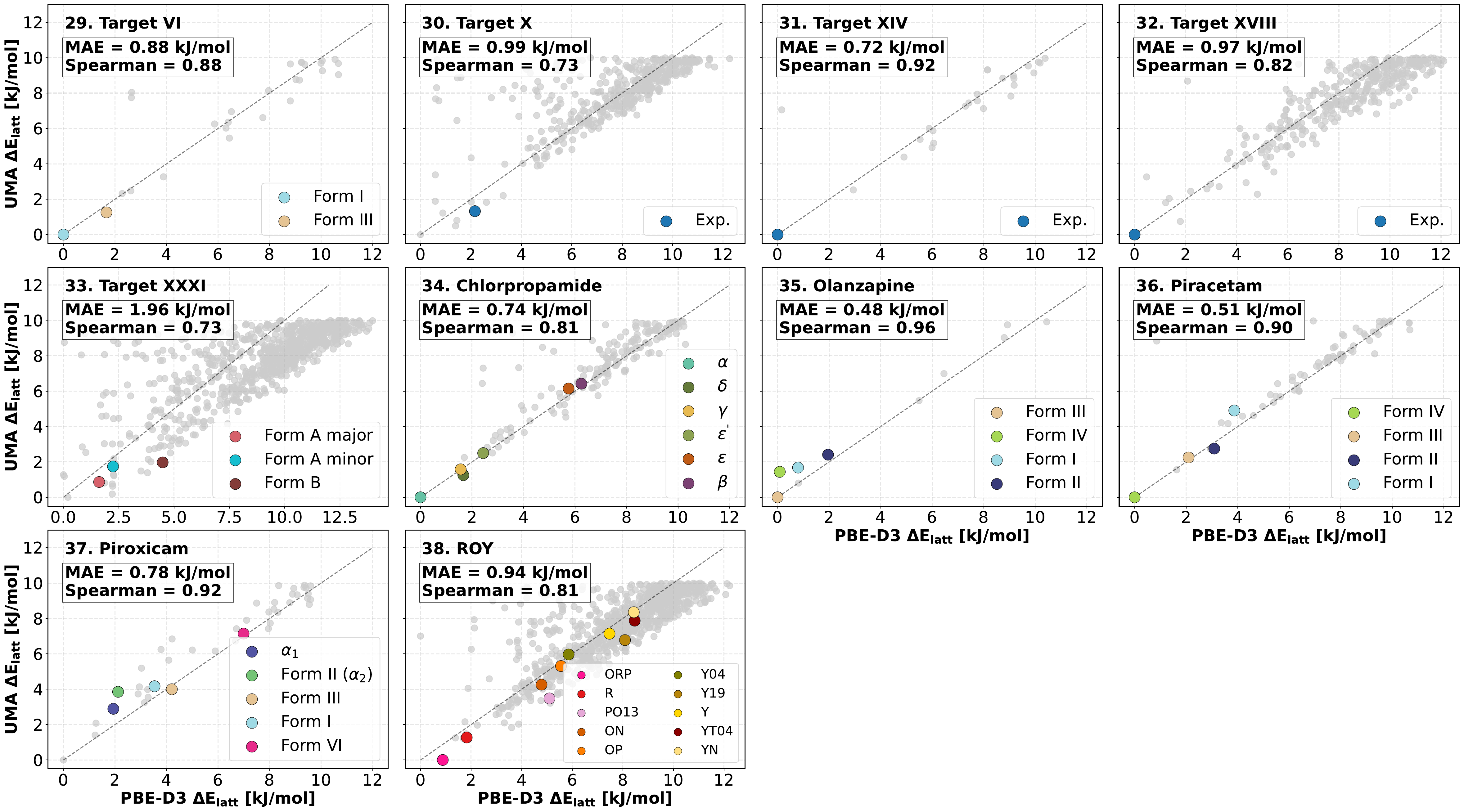}
    \caption{\textbf{(continued)} Parity plots for relaxation performance for all 10 compounds in the Flexible molecules subset. Relative (0~K) lattice energy represents the lattice energy difference between each structure and the global minimum of each method.}
\end{figure}

\begin{figure}[h!tbp]
    \centering
    \includegraphics[width=\textwidth]{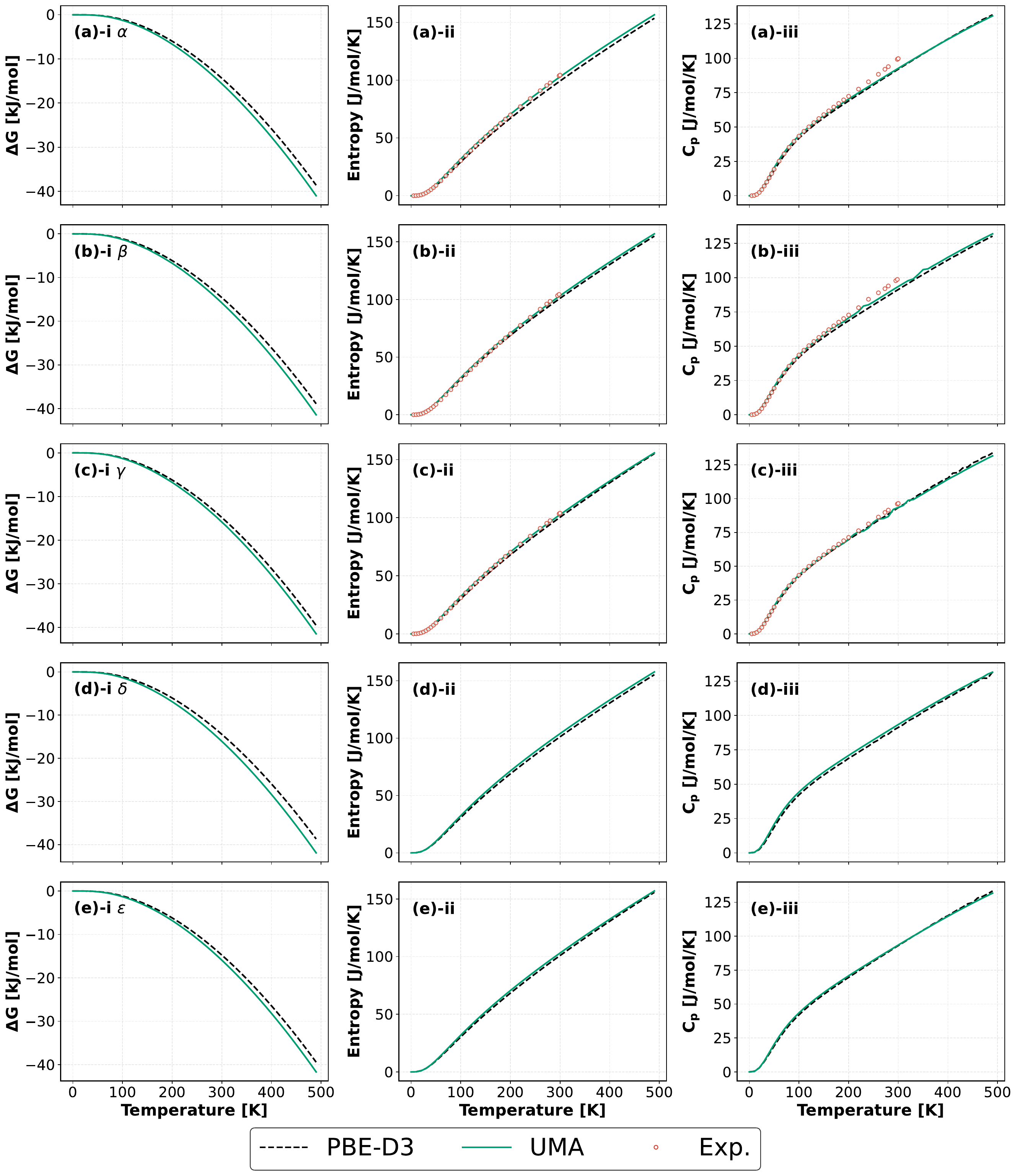}
    \caption{(i) Gibbs free energy, (ii) entropy, and (iii) heat capacity of (a) $\alpha$, (b) $\beta$, (c) $\gamma$, (d) $\delta$, and (e) $\varepsilon$-glycine polymorphs obtained from UMA, PBE-D3, and experimental data~\cite{glycine_rank3, glycine_exp_data}.}
    \label{fig:glycin_uma_fit_all}
\end{figure}

\begin{figure}[h!tbp]
    \ContinuedFloat
    \centering
    \includegraphics[width=\textwidth]{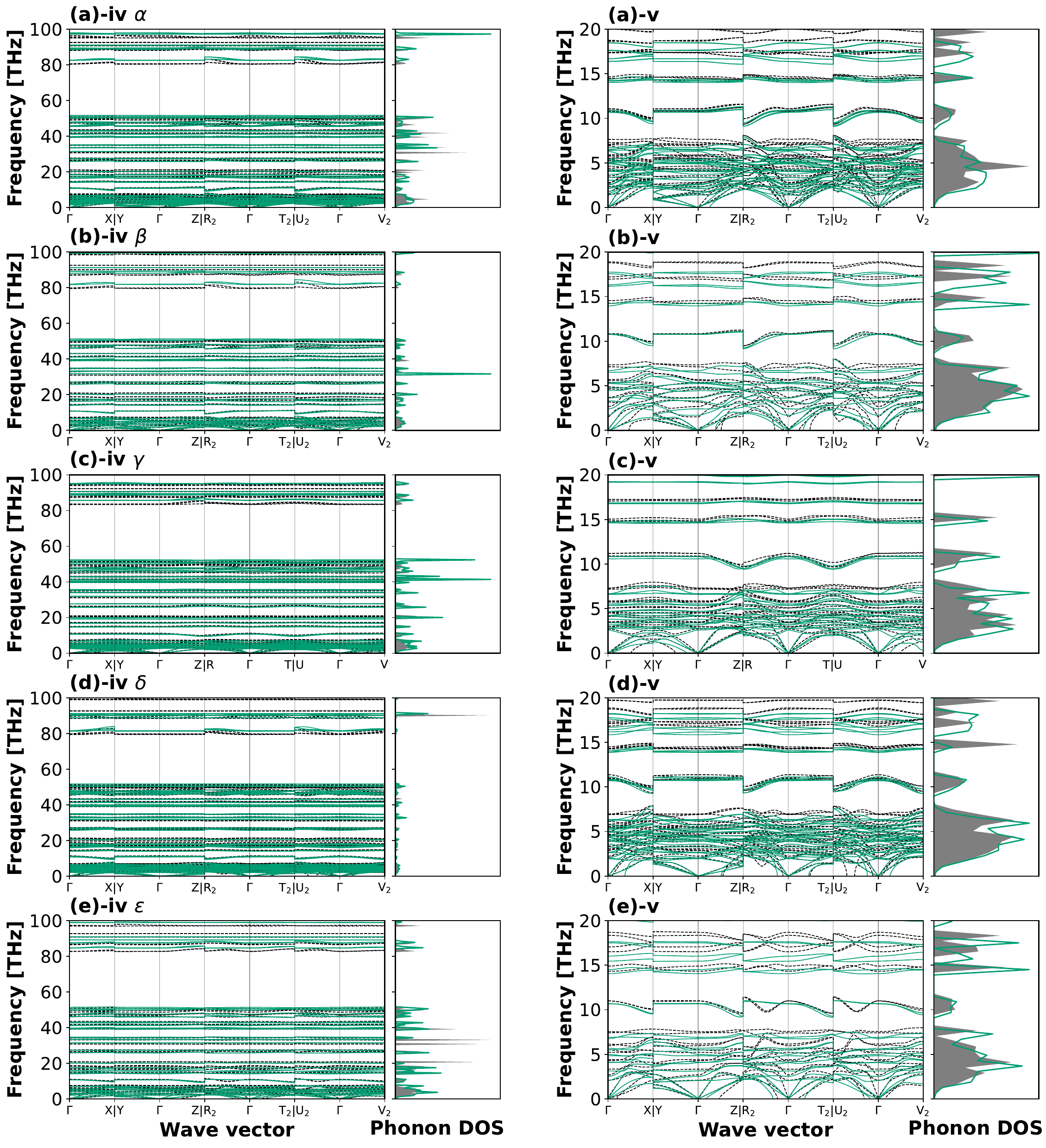}
    \caption{\textbf{(continued)}  Phonon band structures and density of states (DOS) of (a) $\alpha$, (b) $\beta$, (c) $\gamma$, (d) $\delta$, and (e) $\varepsilon$-glycine polymorphs obtained from PBE-D3 (gray) and UMA (green). (iv) Full range (0–100 THz) and (v) low-frequency region (0–20 THz).}
    \label{fig:glycin_uma_dos}
\end{figure}